\documentclass{pasj01}
\usepackage{color}
\Received{2017 February 27}
\Accepted{2017 July 15}
\Published{$\langle$publication date$\rangle$}
\begin{document}

\title{ALMA multiline observations toward the central region of NGC~613}
\author{Yusuke \textsc{Miyamoto}~\altaffilmark{1},
Naomasa \textsc{Nakai}~\altaffilmark{2,3},
Masumichi \textsc{Seta}~\altaffilmark{4}, 
Dragan \textsc{Salak}~\altaffilmark{4}, 
Makoto \textsc{Nagai}~\altaffilmark{2} and
Hiroyuki \textsc{Kaneko}~\altaffilmark{1}
}%
\altaffiltext{1}{Nobeyama Radio Observatory, NAOJ, Nobeyama, Minamimaki, Minamisaku, Nagano 384-1305, Japan}
\email{miyamoto.yusuke@nao.ac.jp}
\altaffiltext{2}{Division of Physics, Faculty of Pure and Applied Sciences, University of Tsukuba, Tsukuba, Ibaraki 305-8571, Japan}
\altaffiltext{3}{Center for Integrated Research in Fundamental Science and Technology (CiRfSE), University of Tsukuba, Tsukuba, Ibaraki 305-8571, Japan}
\altaffiltext{4}{Department of Physics, School of Science and Technology, Kwansei Gakuin University, 2-1 Gakuen, Sanda, Hyogo 669-1337, Japan}


\maketitle

\begin{abstract}

We report ALMA observations of molecular gas and continuum emission in the 90 and 350~GHz bands toward a nearby Seyfert galaxy NGC~613.
Radio continuum emissions were detected  at 95 and 350~GHz from both the circum-nuclear disk (CND) $(r\lesssim90$~pc) and a star-forming ring ($250$~pc$\lesssim r\lesssim340$~pc), and  
the 95~GHz continuum was observed to extend from the center at a position angle of $\timeform{20D} \pm \timeform{8D}$.
The archival 4.9 GHz data and our 95 GHz data show spectral indices of $\alpha\lesssim-0.6$ and $-0.2$ along the jets and in the star-forming ring; these can be produced by synchrotron emission and free--free emission, respectively. 
In addition, we detected the emission of CO(3-2), HCN(1-0), HCN(4-3), HCO$^+$(1-0), HCO$^+$(4-3), CS(2-1), and CS(7-6) in both the CND and ring. 
The rotational temperatures and column densities of molecules derived from  $J=1-0$ and $4-3$ lines of HCN and HCO$^+$ and  $J=2-1$ and $7-6$ of CS in the CND and ring were derived.
Furthermore, a non-LTE model revealed that the kinetic temperature of $T_{\rm k}=$350--550~K in the CND is higher than $T_{\rm k}=$80--300~K in the ring, utilizing the intensity ratios of  HCN, HCO$^+$, and CS.
The star-formation efficiency in the CND is almost an order of magnitude lower than those at the spots in the star-forming ring, 
while  the dominant activity of the central region is the star formation rather than active galactic nuclei.
We determined that the large velocity dispersion  of CO extending toward the north side of the CND 
and decomposing into blueshifted and redshifted features is probably explained by the effect of the radio jets.
These results strongly suggest that  the jets heat the gas in the CND, in which the feedback prevents star formation.

\end{abstract}

\section{Introduction}
\begin{figure*}[t]

 \begin{center}
  \includegraphics[width=0.9\linewidth]{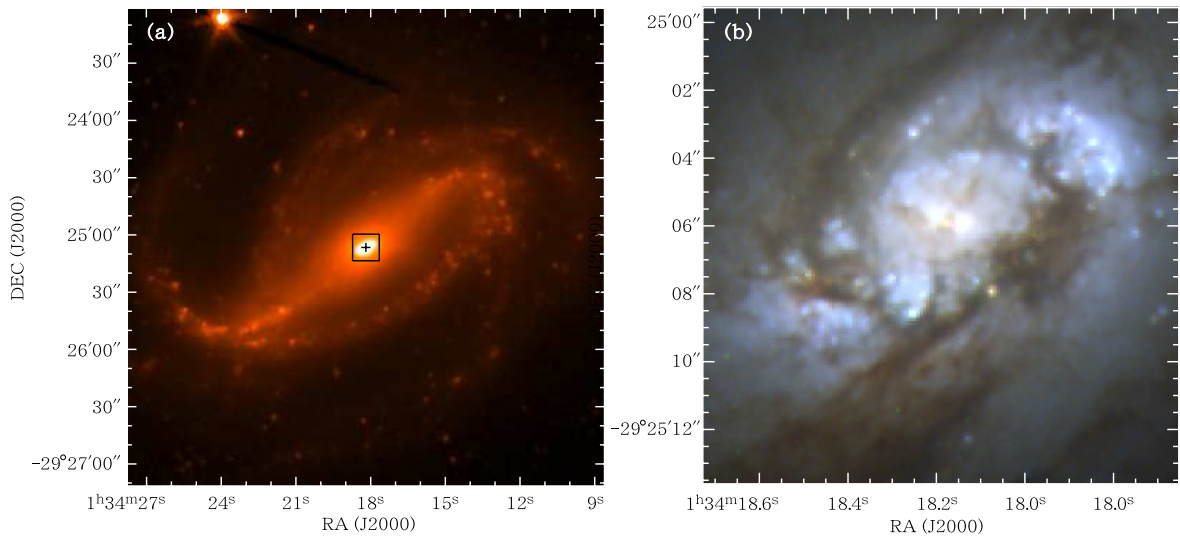}
 \end{center}
 \caption{(a)~IRAC image of NGC~613 in 3.6~$\mu$m obtained from the Spitzer Survey of Stellar Structure in Galaxies (S$^4$G, \cite{s4g}). 
(b)~HST  WFPC2 image of the center region of NGC~613. 
 The image region corresponds to the box in the IRAC image 
(left) and  ALMA images in figures~\ref{fig:cont}, \ref{fig:band3}, and \ref{fig:band7}
. }
 \label{fig:hst}
\end{figure*}
\begin{table*}
  \tbl{Parameters of NGC~613}{%
  \begin{tabular}{cccc}
      \hline
      Parameter & Value & References & this work
      \footnotemark[$\ast$]
      \\ 
      \hline
      RA (J2000.0)  & \timeform{1h34m18.235s}& \cite{simbad} & \timeform{1h34m18.1901s} 
      \\
      Dec (J2000.0)  & \timeform{-29D25'6.56"}& & \timeform{-29D25'6.601"} 
      \\
      Distance & 17.5~Mpc  & \cite{tully1988} & --- \\
      Position angle & \timeform{120D} &  \cite{rc3} &\timeform{118D} 
      (\timeform{4D})
      \\
      Inclination angle & \timeform{41D} &  \cite{rc3} & \timeform{46D} 
      (\timeform{1D})
      \\
      Systemic velocity (LSR) & $1470\pm5$~km~s$^{-1}$ & \citet{koribalski2004} &1471~km~s$^{-1}$ 
      (3~km~s$^{-1}$)
      \\
      \hline
    \end{tabular}}\label{tab:first}
     \footnotemark[$\ast$] 
     Values in parentheses represent the uncertainties.
\end{table*}

For understanding the activity and evolution of active galactic nuclei (AGN), 
it is essential to investigate its fueling process.
AGN must be fed with material available in the galactic 
disk of the host galaxy.
The galactic bar can effectively drain angular momentum of interstellar gas 
and transport the gas inward to the disk center.
The gas gathers on nearly circular orbits (x$_2$ orbits)
and forms nuclear rings ($r\sim$ a few 100~pc) \citep{binney}.
CO surveys have shown 
that central concentration of molecular gas is higher in barred galaxies than in non barred galaxies
(e.g., \cite{sakamoto}, \cite{nishiyama}, \cite{sheth2005}). 
However, the rings as dynamical barriers can prevent the gas from flowing inward \citep{regan2004}.
This is supported by the result that a correlation cannot be determined between the galactic bars and AGN activities  \citep{hunt1999}.
To discuss the fueling process of AGN, 
we must investigate the dynamics and physical properties of molecular gas
from the ring to the center via the circum-nuclear disk (CND) at $r\sim$ 1$-$$100$~pc.

X-ray radiation and jet/outflow are important feedbacks of the AGN.
X-ray radiation from AGN can effectively heat  gas in the nuclear region 
because it can penetrate large amounts of dust and gas.
Therefore, the X-ray radiation generates internal turbulence in the CND 
and constricts mass accretion to the center \citep{wada2012}.
Such an X-ray dominated region (XDR; e.g., \cite{maloney}, \cite{meijerink2005}) can contribute to the enhancement of several molecular lines; for example, HCN and its physical properties have been estimated through multiline observations of dense gas tracers (e.g., \cite{krips2008}, \cite{krips2011}). 
However, recent multiline observations have suggested that  the enhancement is explained by high-temperature chemistry (e.g., mechanical heating)  rather than XDR chemistry (e.g., \cite{izumi2013}, \cite{martin2015}, 
\cite{izumi2016}).
Jet/outflow activities from AGNs can also heavily influence the fueling process in the CND and ring. 
Recent observations toward the center of NGC 1068 with high spatial resolution ($\sim$ several 10~pc, e.g., \cite{krips2011}, \cite{garcia2014}, 
\cite{viti})
 showed that 
the HCN distribution, which traces a dense gas of $n$(H$_2) \hspace{0.3em}\raisebox{0.4ex}{$>$}\hspace{-0.75em}\raisebox{-.7ex}{$\sim$}\hspace{0.3em}10^4$ cm$^{-3}$,
is concentrated at the center. 
In addition, the molecular outflow signatures from the AGN were identified not only in the CO but also in the HCN and HCO$^+$. 
The jet can drag gas outwards on scales of $\sim100$ pc, 
and hence regulate activity of the accretion disk \citep{combes2013}.
In addition, note that shock heating by the jet is an important heating source that generates  turbulence. 
It is necessary to derive the physical properties, 
such as temperature and density of shocked gas in the CND,
to identify the heating source of the hot gas in the CND.

\begin{longtable}{*{7}{l}}
\caption{Observational log with ALMA}
\label{tab:para}
\hline
\multicolumn{7}{c}{} \\
\endfirsthead
Band  & Array &  Observation Date  & Bandpass calibrator & Flux calibrator & Phase calibrator & Observation time (on source)  \\
\hline
3 & 7~m $\times$~9 & 2014 June 5  & J0137-2430 & Uranus & J0145-2733 & 18~min \\ 
~ & 12~m $\times$~31& 2014 December 5  & J0334-4008 & Uranus & J0145-2733 & 6~min \\ 
~ & 12~m $\times$~35& 2015 April  25  & J2357-5311 & Uranus & J0145-2733 & 5~min \\
~ & 12~m $\times$~43& 2015 July 3  & J0334-4008 & Neptune & J0145-2733 & 10~min \\ 
~ & TP& 2015 September 3  & -- & -- & -- & 29~min \\ 
~ & TP& 2015 September 19  & -- & -- & -- & 61~min \\ 
\hline 
7 & 7~m $\times$~10 & 2014 June 8  & J0137-2430 & Uranus & J0145-2733 & 13~min \\
~ & 7~m $\times$~10& 2015 June  6  & J0238+1636 & Uranus & J0137-2430 & 12~min \\
~ & 7~m $\times$~12& 2015 June 8  & J2253+1608 & Uranus & J0137-2430 & 13~min \\
~ & 12~m $\times$~33 & 2014 June 14  & J0137-2430 & J0334-401 & J0145-2733 & 7~min \\
~ & 12~m $\times$~39 & 2015 June 3  & J2232+1143 & Ceres & J0145-2733 & 7~min \\
~ & TP& 2015 August 28  & -- & -- & -- & 71~min \\ 
\hline
\end{longtable}
\begin{longtable}{*{6}{l}}
\caption{Observational parameters of lines}
\label{tab:obs1}
\hline
\multicolumn{6}{}{} \\
\endfirsthead
\endlastfoot
Band  & Molecule &  Transition  & {Frequency~(GHz)\footnotemark[a]} & $E_{u}/k$~(K)\footnotemark[b] &Beam size $('')$  \\
\hline
3 & 	SiO		&	$J=$2--1	&	86.846995	&	6.25		&	0\farcs73$\times$0\farcs64 ({\it P.A.} $=\timeform{68D}$)\\
~ & 	HCN		&	$J=$1--0	&	88.631847	&	4.25		&	0\farcs72$\times$0\farcs63 ({\it P.A.} $=\timeform{68D}$)\\
~ & 	HCO$^+$	&	$J=$1--0	&	89.188526	&	4.28		&	0\farcs72$\times$0\farcs63 ({\it P.A.} $=\timeform{68D}$)\\
~ & 	CS		&	$J=$2--1	&	97.980953	&	7.05		&	0\farcs67$\times$0\farcs58 ({\it P.A.} $=\timeform{68D}$)\\
\hline\\
7 & CS		&	$J=$7--6	&	342.882857	&	65.83	&	0\farcs37$\times$0\farcs31 ({\it P.A.} $=\timeform{83D}$)\\
~ & CO		&	$J=$3--2	&	345.79599	&	33.19	&	0\farcs44$\times$0\farcs37 ({\it P.A.} $=\timeform{45D}$)\\
~ & HCN		&	$J=$4--3	&	354.505473	&	42.53	&	0\farcs43$\times$0\farcs37 ({\it P.A.} $=\timeform{40D}$)\\
~ & HCO$^+$	&	$J=$4--3	&	356.734242	&	42.80	&	0\farcs36$\times$0\farcs29 ({\it P.A.} $=\timeform{84D}$)\\
\hline
   \multicolumn{6}{@{}l@{}}{\hbox to 0pt{\parbox{130mm}{\footnotesize
 \footnotemark[a] adopted from NIST Recommended Rest Frequencies for Observed Interstellar Molecular Microwave Transitions (F. J. Lovas et al.; http://physics.nist.gov/cgi-bin/micro/table5/start.pl)
 \footnotemark[b] 
 Leiden Atomic and Molecular Database (LAMDA: \cite{lamda})
     }\hss}}
\end{longtable}

The nearby barred galaxy NGC 613 
(figure~\ref{fig:hst})
hosts a low-luminosity AGN, which is identified through mid infrared spectroscopy (Goulding \& Alexander 2009) and X-ray observations by using the XMM-Newton \citep{castangia2013}. 
NGC 613 is one of the best targets for studying the jet/outflow activities  
due to the proximity (17.5 Mpc; \cite{tully1988}), 
relatively abundant molecular gas around the center (e.g., \cite{elfhag}), 
activity shown by the radio jet \citep{hummel1987}, 
and nuclear ring with moderate inclination ($i\sim55^{\circ}$, \cite{hummel1992}). 
The basic parameters of NGC~613 are summarized in table~\ref{tab:first}.
The nuclear ring, which is clearly defined in Br$\gamma$, is likely to be disturbed 
at the north-east by the outflow from the AGN. 
The high flux ratio of [FeII]-to-Br$\gamma$ and large velocity dispersion of [FeII]  
 support the interaction between the outflow and nuclear ring \citep{falcon}. 
In contrast, the high flux of H$_2$  in the center can be explained  by the heating  due to the X-ray radiation or/and shock (e.g., \cite{maloney}, \cite{hollenbach1989}).

In this paper, we present Atacama large millimeter/submillimeter array (ALMA) cycle~2 observations toward the central region of NGC~613 at Bands~3 and 7.
The remainder of this paper is structured as follows. 
Section~\ref{sec:observation} describes the observations and data reduction.
In section~\ref{sec:result}, 
the distribution of continuum and molecular gas is presented, 
and the basic parameters of NGC~613 are determined. 
The physical conditions of the molecular gas in the central region both under local thermodynamic equilibrium (LTE) and non-LTE conditions are investigated 
and the activities for explaining the gas heating at the CND are discussed
in section~\ref{sec:discussion}. Finally, conclusions are drawn in section~\ref{sec:summary}.

\section{Observations}
\label{sec:observation}
NGC~613 was observed through the ALMA using Bands~3 and 7 receivers, as a Cycle~2 early science program (ID 2013.1.01329.S, PI: Miyamoto).
For both  Bands~3 and 7 observations, the 12-m array, Atacama compact array (ACA) and total power array (TP) were used.
The baseline lengths extended up to 1573.0~m (524.3~k$\lambda$ at 100~GHz) for Band~3 and 780.7~m (923.9~k$\lambda$ at 355~GHz) for Band~7.
The typical synthesized beams at Bands~3 and 7 were 0\farcs7 and 0\farcs4, 
corresponding to 59 and 34~pc, respectively, at the distance of the galaxy.
Single-point observations were conducted using the 12-m array and ACA for Band~3 and three-point mosaic observations were conducted using the 12-m array and ACA  for Band~7. 
These setups allowed us to image a CND of  $\sim100$~pc diameter and the surrounding star-forming ring of $\sim600$~pc  diameter \citep{boker, falcon}. 
The phase reference center was adopted to be $(\alpha_{\rm J2000.0}$, $\delta_{\rm J2000.0})=$
(\timeform{1h34m18.235s}, \timeform{-29D25'06.56''}) \citep{simbad}. 
Observations at Band~3 were conducted from June 2014 to September 2015. 
The correlators for Band~3 were configured to set three spectral windows in the lower sideband  to cover SiO(2--1) ($\nu_{\rm rest}=$86.846960~GHz), HCN(1--0)~($\nu_{\rm rest}=$88.631601~GHz), and HCO$^+$(1--0) ($\nu_{\rm rest}=$89.188526~GHz) and  two in the upper sideband to measure CS(2--1) ($\nu_{\rm rest}=$97.980953~GHz) and 100~GHz continuum, both in the 2SB dual-polarization mode.
Observations at Band~7 were cnducted from June 2014 to August 2015. 
The correlators for Band~7 were configured to set two spectral windows in the lower sideband  to cover CO(3--2) ($\nu_{\rm rest}=$345.795990~GHz) and CS(7--6) ($\nu_{\rm rest}=$342.882857~GHz) 
and  two in the upper sideband to measure HCN(4--3) ($\nu_{\rm rest}=$354.505473~GHz) and HCO$^+$(4--3) ($\nu_{\rm rest}=$356.734242~GHz), both in the dual-polarization mode.
Table~\ref{tab:para} summarizes the observational logs for this project.
Note that we could only cover the CO(3--2) line with {\it V}$_{\rm LSR}=1250$--1660~km~s$^{-1}$ because of spectral setting restrictions, 
while CO(1--0) observations with SEST, whose angular resolution was $\timeform{44''}$, revealed that the CO(1--0) velocity range 
was  {\it V}$_{\rm LSR}\sim1250$--1710~km~s$^{-1}$ \citep{elfhag}. 
The proportion of the CO(1--0) intensity at {\it V}$_{\rm LSR}=1250$--1660~km~s$^{-1}$ to the total intensity at {\it V}$_{\rm LSR}=1250$--1710~km~s$^{-1}$ is $\sim97$~\%, and the equivalent underestimation of the total flux can be expected in CO(3--2).

The data were processed using the Common Astronomy Software Application (CASA; \cite{mcmullin2007}). 
The velocity resolution of each line data obtained at different observing tracks with the 12~m array and ACA
were separately rearranged  to be 10~km~s$^{-1}$, except for the resolution of  2.5~km~s$^{-1}$ in CO(3--2). 
Each line data was then combined after subtracting continuum emission determined at the emission-free channels. 
To image the continuum emission, 
we used the flux density at the emission-free channels.
The imaging was performed using the CLEAN-algorithm in CASA.
CLEAN maps were obtained considering the Briggs weighting mode on the data with a robustness of 0.5. 
The resultant maps were $1500\times1500$~pixels with $0\farcs05$~per pixel. 
For the  line emission images,  
TP data were calibrated through flagging and the system temperature correction, and imaged independently from the 12-m array and ACA data. 
By using the Feather algorithm in CASA, the low-resolution image obtained through the TP and high-resolution image obtained through the 12-m and ACA were converted  into the gridded visibility plane and combined. 
Finally, the data was reconverted into a combined image.
The synthesized beams  of CO(4--3), $\theta=0\farcs37\times0\farcs31$ ($P.A.=\timeform{82D}$), and of HCN(4--3), $\theta=0\farcs36\times0\farcs30$ ($P.A.=\timeform{84D}$) with the 12-m array and ACA 
change to $\theta=0\farcs44\times0\farcs37$ ($P.A.=\timeform{45D}$) and $\theta=0\farcs43\times0\farcs37$ ($P.A.=\timeform{40D})$, respectively, after applying the Feather-algorithm.
The observational parameters of each line are given in table~\ref{tab:obs1}.

\section{Results}
\label{sec:result}
\subsection{Continuum emission}
\begin{figure*}[hp]
\begin{tabular}{cc}
	\begin{minipage}{0.45\hsize}
  \par
	 \begin{center}
{\bf (a)}
		  \FigureFile(70mm,150mm){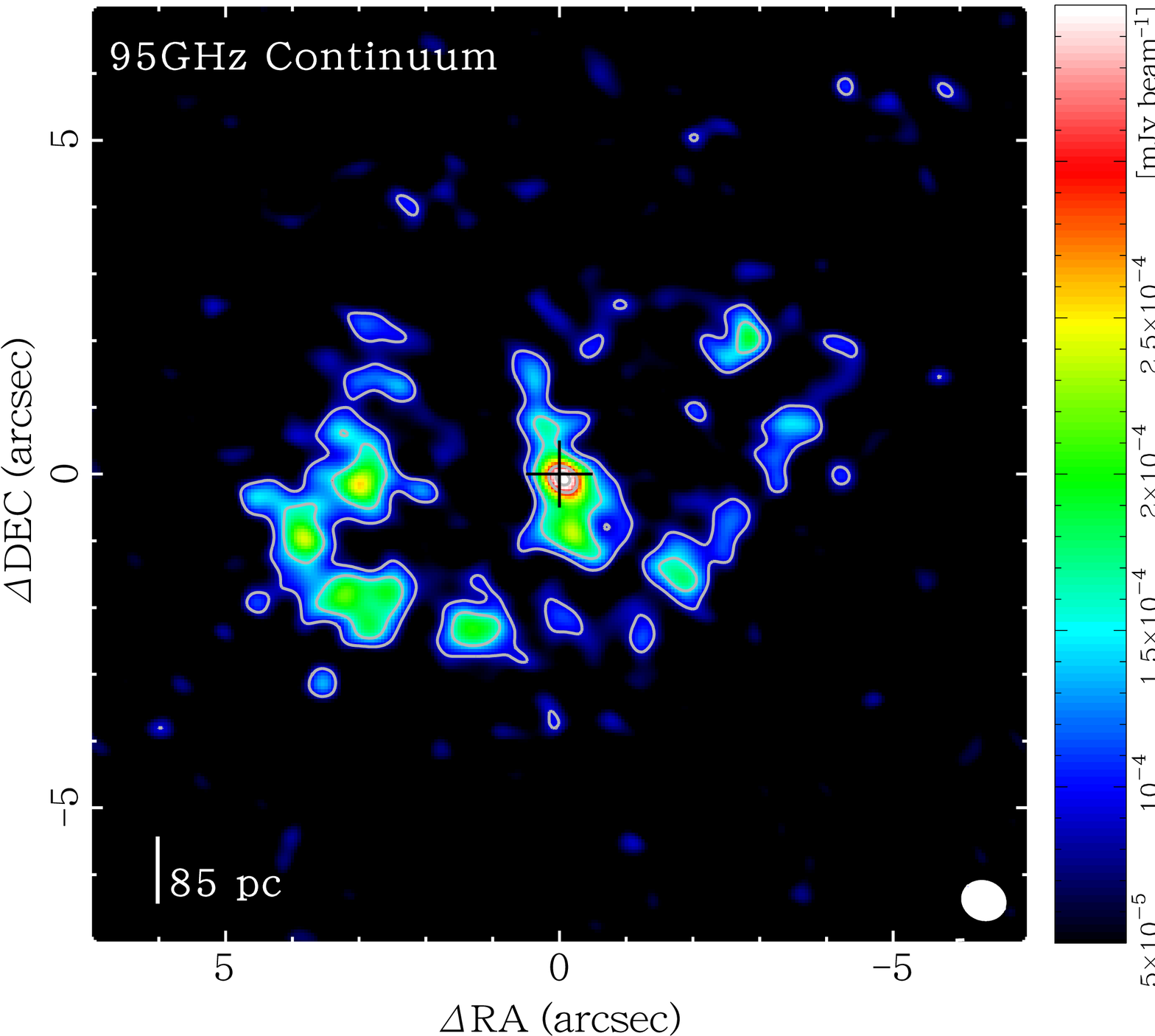}
	\end{center}
	\end{minipage}
	\begin{minipage}{0.45\hsize}
\par
	\begin{center}
{\bf (b)}
		\FigureFile(70mm,150mm){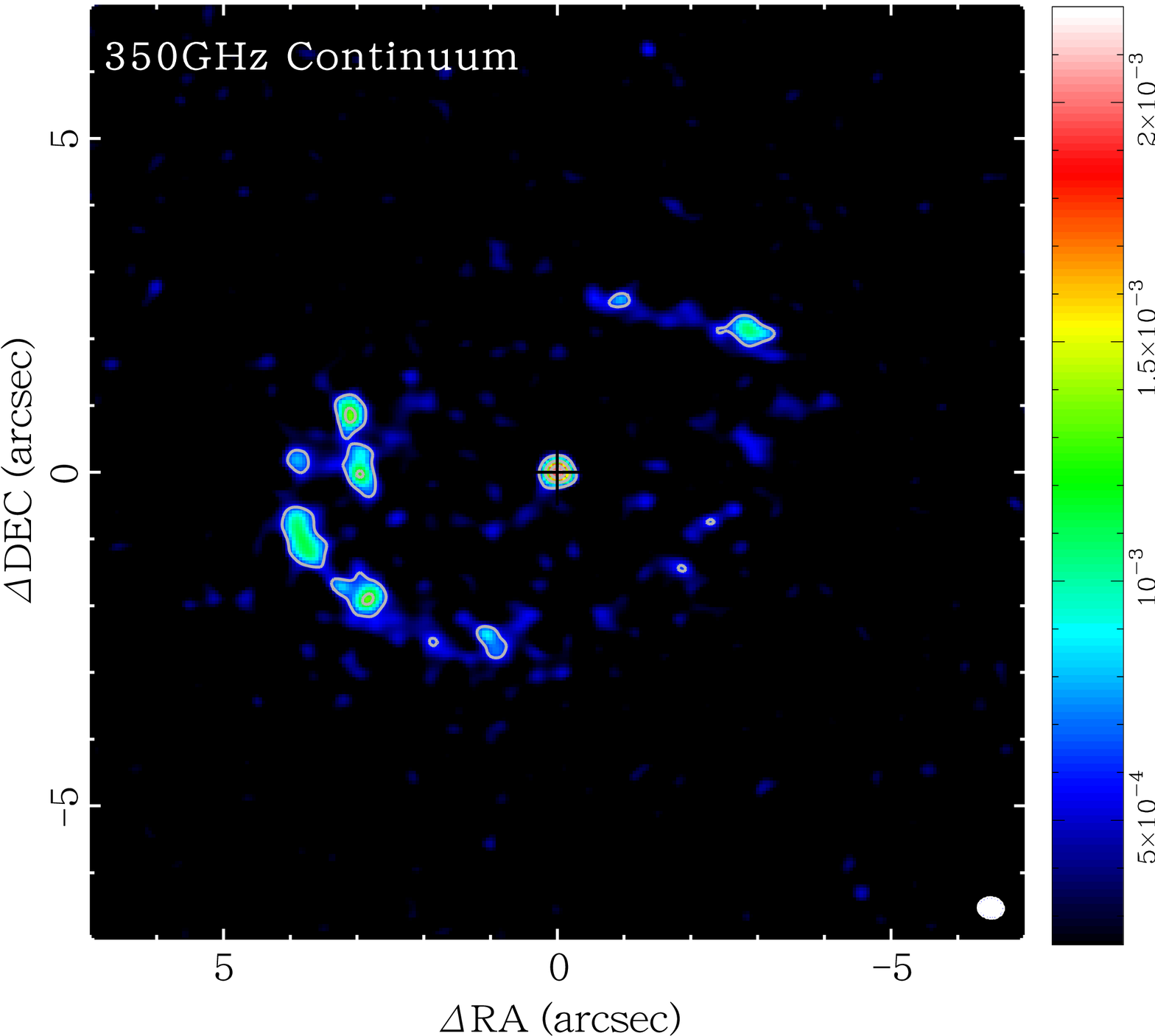}
	 \end{center}
	 \end{minipage}\\
	 \begin{minipage}{0.45\hsize}
\par

	\begin{center}
{\bf (c)}
		\FigureFile(70mm,150mm){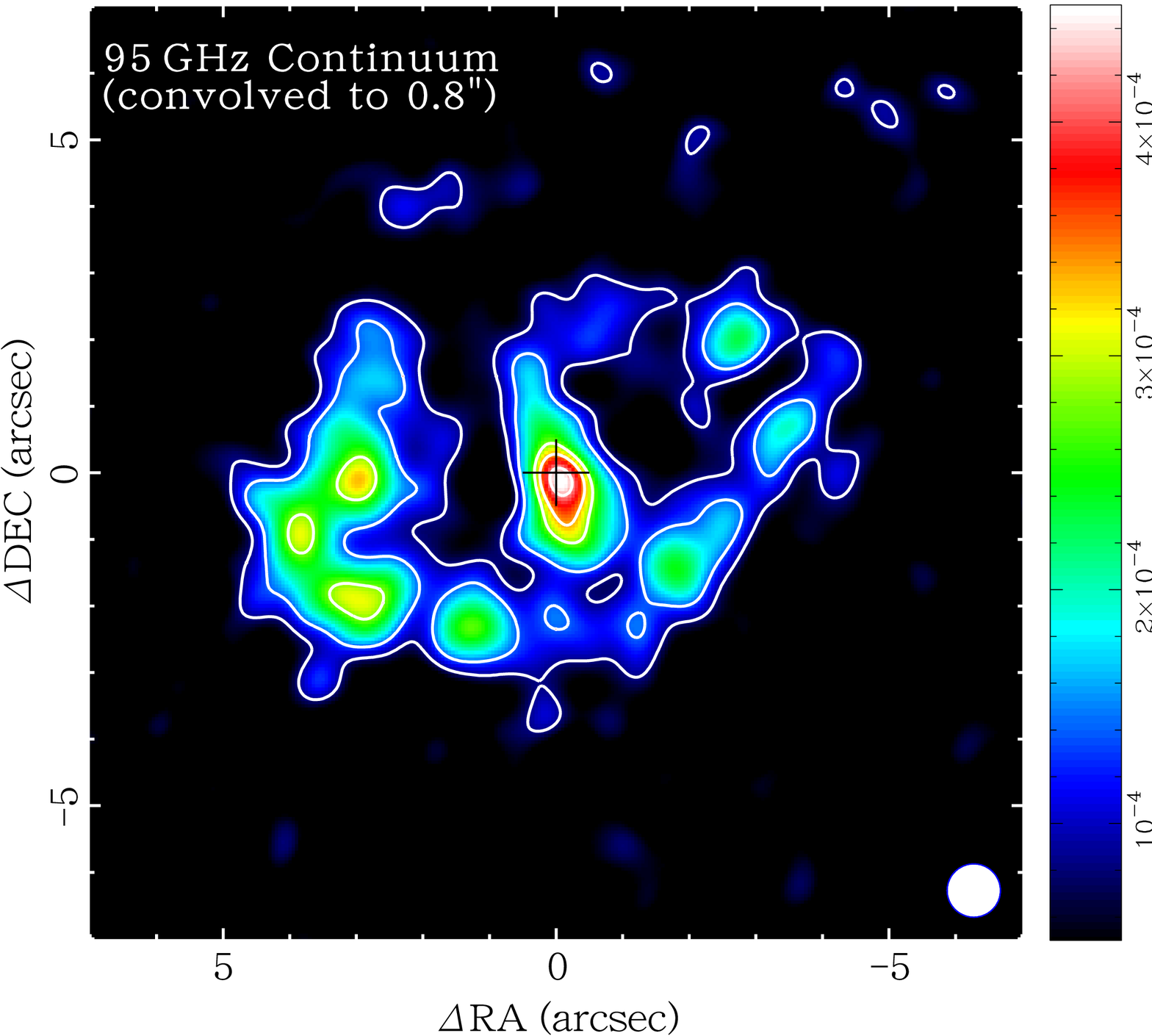}
	 \end{center}
	 \end{minipage}
	 \begin{minipage}{0.45\hsize}
\par
	\begin{center}
{\bf (d)}
		\FigureFile(70mm,150mm){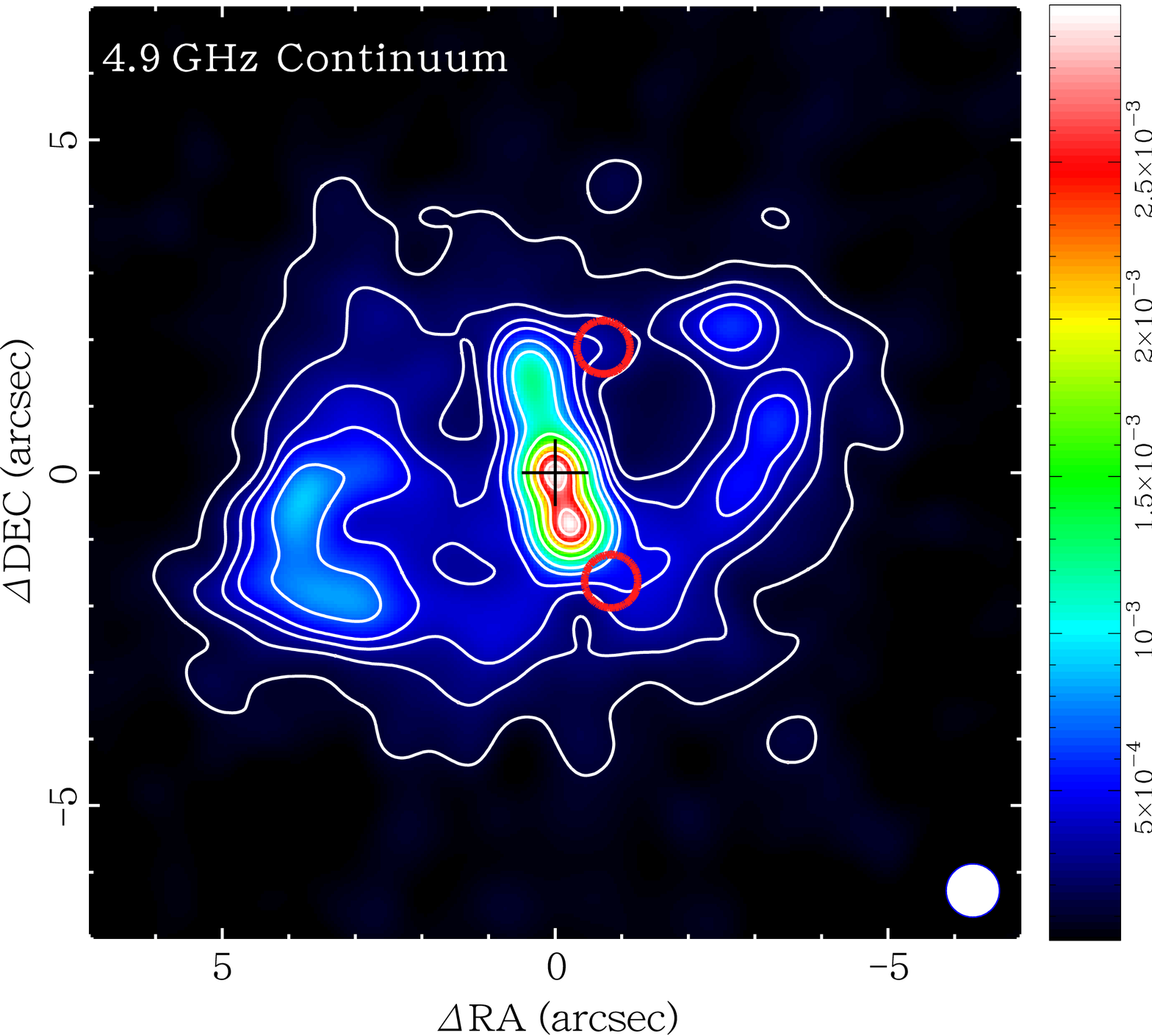}
	 \end{center}
	 \end{minipage}\\
	 
	 \begin{minipage}{0.45\hsize}
	\begin{flushleft}
{\bf (e)}
		\FigureFile(70mm,150mm){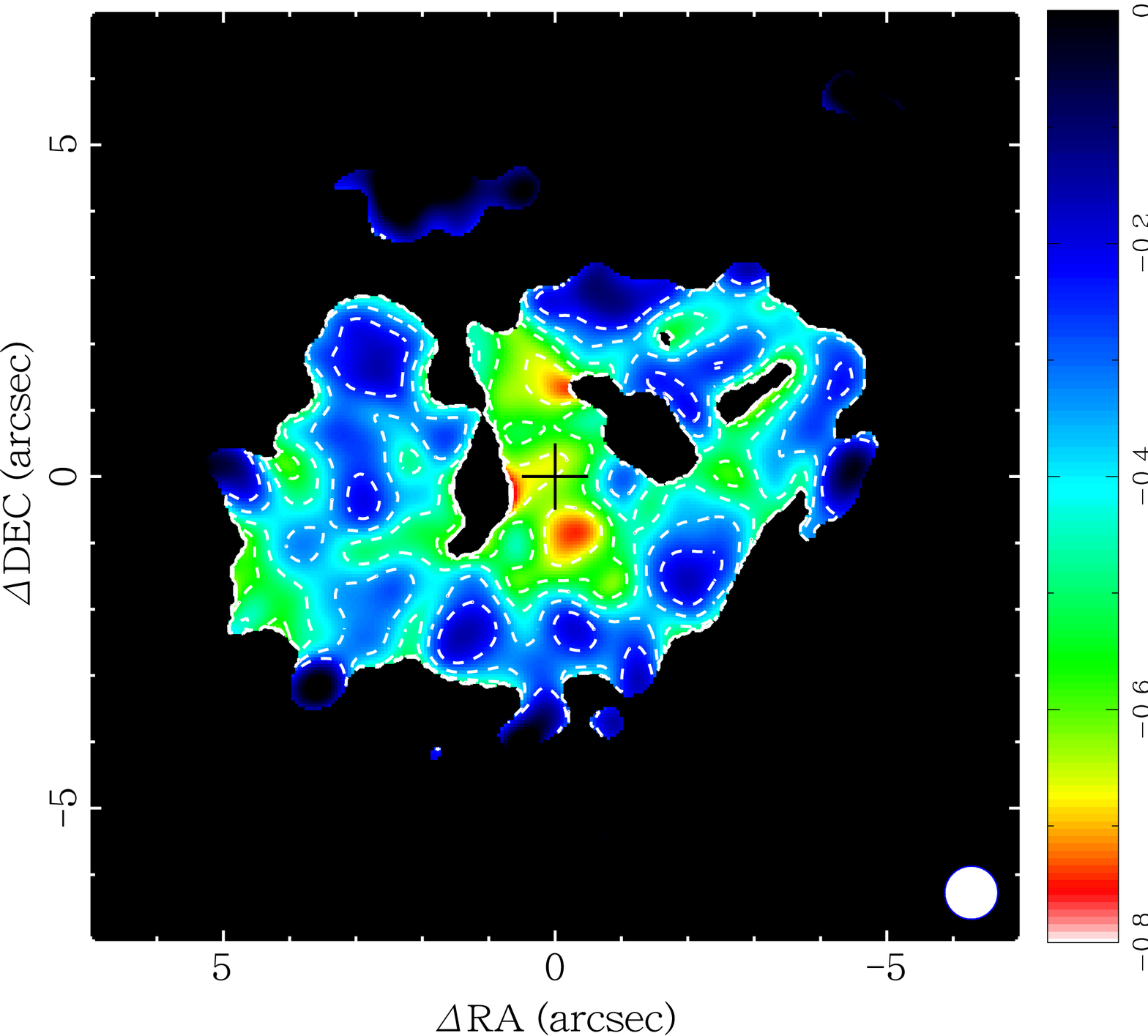}
	 \end{flushleft}
	 \end{minipage}\\
	 
 \end{tabular}
\caption{(a) 95~GHz continuum map with a resolution of $0\farcs64 \times 0\farcs56$   ({\it P.A.} $=\timeform{68D}$). The contour levels are 3$\sigma$, 5$\sigma$, 10$\sigma$, and 12$\sigma$, where 1$\sigma=30$~$\mu$Jy~beam$^{-1}$.
(b) 350~GHz continuum map with a resolution of  $0\farcs39 \times 0\farcs29$   ({\it P.A.} $=\timeform{84D}$). The contour levels are 5$\sigma$, 10$\sigma$, and 15$\sigma$, where 1$\sigma=0.12$~mJy~beam$^{-1}$.
(c) Same as panel (a) but convolved to $\theta=\timeform{0.8"}$. The contour levels are 3$\sigma$, 5$\sigma$, 10$\sigma$, 12$\sigma$, and 15$\sigma$, where 1$\sigma=28$~$\mu$Jy~beam$^{-1}$.
(d) The 4.9~GHz continuum map with a resolution of  $0\farcs8 \times 0\farcs8$. The contour levels are 5$\sigma$,10$\sigma$,15$\sigma$, 20$\sigma$, 30$\sigma$, 40$\sigma$, 60$\sigma$, 80$\sigma$, 100$\sigma$, and 120$\sigma$, where 1$\sigma=23$~$\mu$Jy~beam$^{-1}$. 
SiO is detected in positions shown in  circles.
(e) Spectral index distribution  calculated from 4.9~GHz 
(above $2\sigma$) 
and 95~GHz 
(above $2\sigma$). 
The contour levels are spectral indices of -0.15, -0.35, -0.45, -0.55, and -0.65.
}
 \label{fig:cont}
\end{figure*}

Figures~\ref{fig:cont}~(a) and (b) respectively show the 95 and 350~GHz continuum maps toward the central \timeform{14''} (1.2~kpc) region of NGC~613.
Continuum emission was detected from both the CND ($r\lesssim90$~pc) and the star-forming ring (250~pc~$\lesssim r\lesssim$~340~pc) (section~\ref{sec:mass}). 
The flux of 95~GHz continuum in the central region ($r\leq\timeform{6''}$) with only the 12-m array ($9.79\pm0.06$~mJy) is comparable to that with the 12-m array and ACA ($10.02\pm0.06$~mJy), 
indicating that the continuum emission is dominated by compact sources. 
We fit a two-dimensional Gaussian to the nuclear 350~GHz continuum source to derive the position and intrinsic size of the nucleus by using imfit in CASA.
The source size deconvolved from the synthesized beam of $0\farcs39 \times 0\farcs29$   ({\it P.A.} $=\timeform{84D}$) is  $0\farcs230 \times 0\farcs186$ ({\it P.A.} $= \timeform{147D}$), corresponding to $19.5\times 15.8$~pc at the distance of 17.5~Mpc.
The  position of the peak flux of the 350~GHz continuum ($2.180\pm0.085$~mJy~beam$^{-1}$) is at $(\alpha_{\rm J2000.0}$, $\delta_{\rm J2000.0})=$ (\timeform{1h34m18.1901s}, \timeform{-29D25'06.601''}) with an uncertainty of 
0\farcs02, 
which is consistent with the positions of the optical nucleus \citep{hummel1987} and 
an ultraluminous X-ray point source  \citep{liu2005}.
Hereafter, we adopt the peak position as the center of NGC~613. 

The 95-GHz continuum elongates from the center at a position angle of $\timeform{20D}\pm\timeform{8D}$, 
which is consistent with that of  low-frequency radio continuum of 4.9 and 14.9~GHz within the error  ({\it P.A.}$=\timeform{12D}$, \cite{hummel1992}).
We obtained the 4.9~GHz data (project AH231) from Very Large Array (VLA) archives and  reanalyzed them in a standard manner in CASA.
Figure~\ref{fig:cont}(d)  shows the 4.9-GHz continuum image  with a resolution of $0\farcs8 \times 0\farcs8$.
Figure~\ref{fig:cont}(e) shows the distribution of the spectral index $(\alpha)$ between 4.9 and 95~GHz; 
it is convolved with a resolution of  $0\farcs8$ [figure~\ref{fig:cont}(c)].
At the center, the spectral index  $\alpha\sim-0.6$.  
The index is slightly flatter than that derived from 4.9 and 14.9~GHz continuum images ($-0.8$; \cite{hummel1992}), 
suggesting the contribution of free--free emission from H\emissiontype{II} regions in addition to synchrotron emission (e.g., \cite{condon}).
The same spectral index of $\alpha\sim-0.7$ at 4.9--14.9 and 4.9--95~GHz at the southern blob 0\farcs9 from the center implies that the blob can be dominated by synchrotron emission due to the nuclear jets \citep{hummel1987}.
In contrast, the flat spectrum ($\alpha\sim-0.2$) in the ring can be produced through the free--free emission from young star-formation regions (\cite{falcon}).


\onecolumn
\twocolumn


\begin{figure*}[t]
 \begin{center}
  \includegraphics[width=\linewidth]{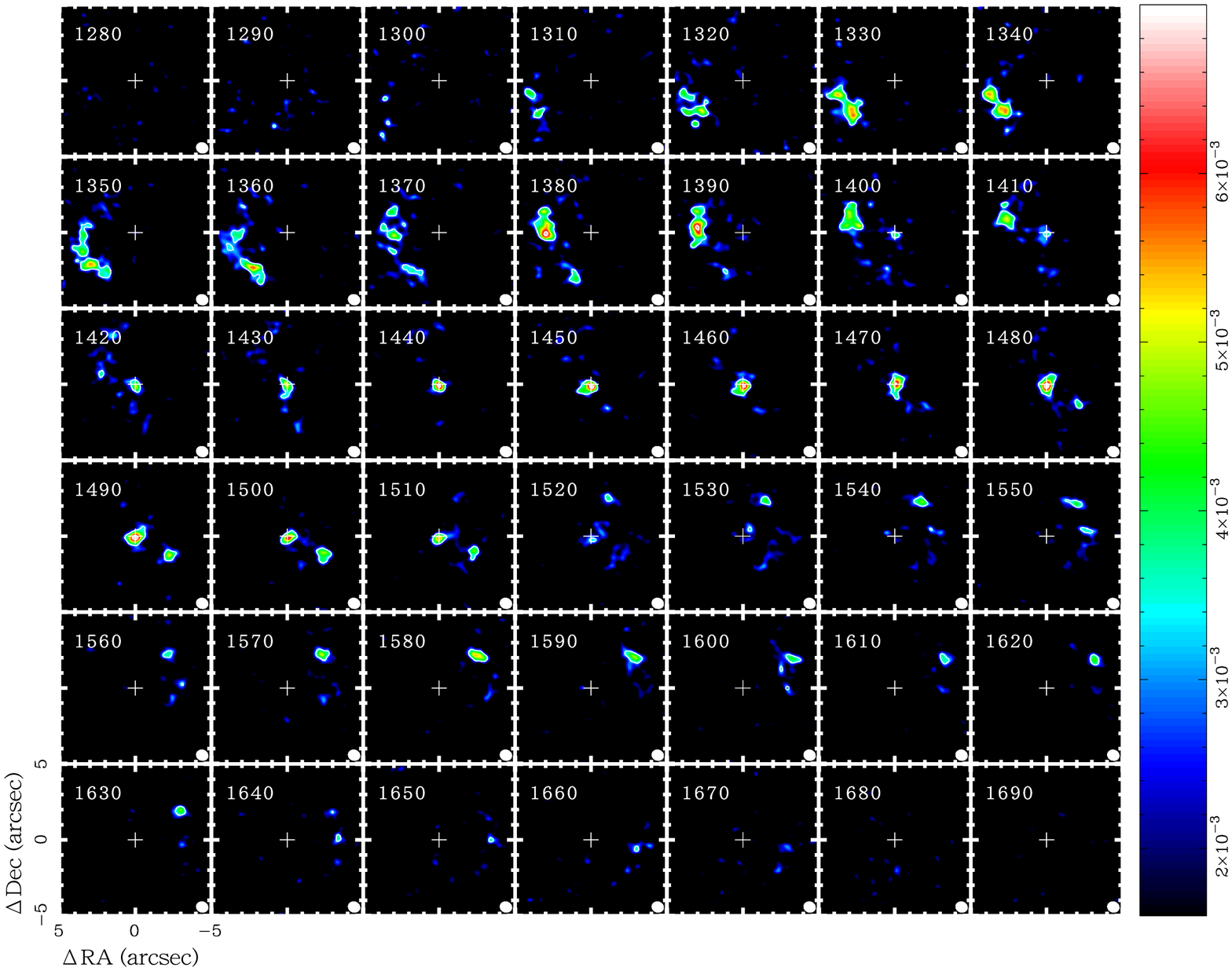}
 \end{center}
 \caption{ 
Channel maps of HCO$^+$(1--0) emission in the central $\timeform{10''}\times\timeform{10''}$ region, corresponding to $848\times848$~pc at the distance of the galaxy (17.5~Mpc).
The channels have an interval of 10~km~s$^{-1}$, and the central velocities are labeled at the upper left corner.
The synthesized beam of  0\farcs72$\times$0\farcs63 ({\it P.A.} $=\timeform{68D}$) is shown at the right bottom corner.
Crosses show the position of the galactic center.
Contours levels are 4$\sigma$ and 8$\sigma$, where 1$\sigma=0.8$~mJy~beam$^{-1}$. 
}
 \label{fig:hcop10_ch}
\end{figure*}
\begin{figure*}[t]
 \begin{center}
  \includegraphics[width=\linewidth]{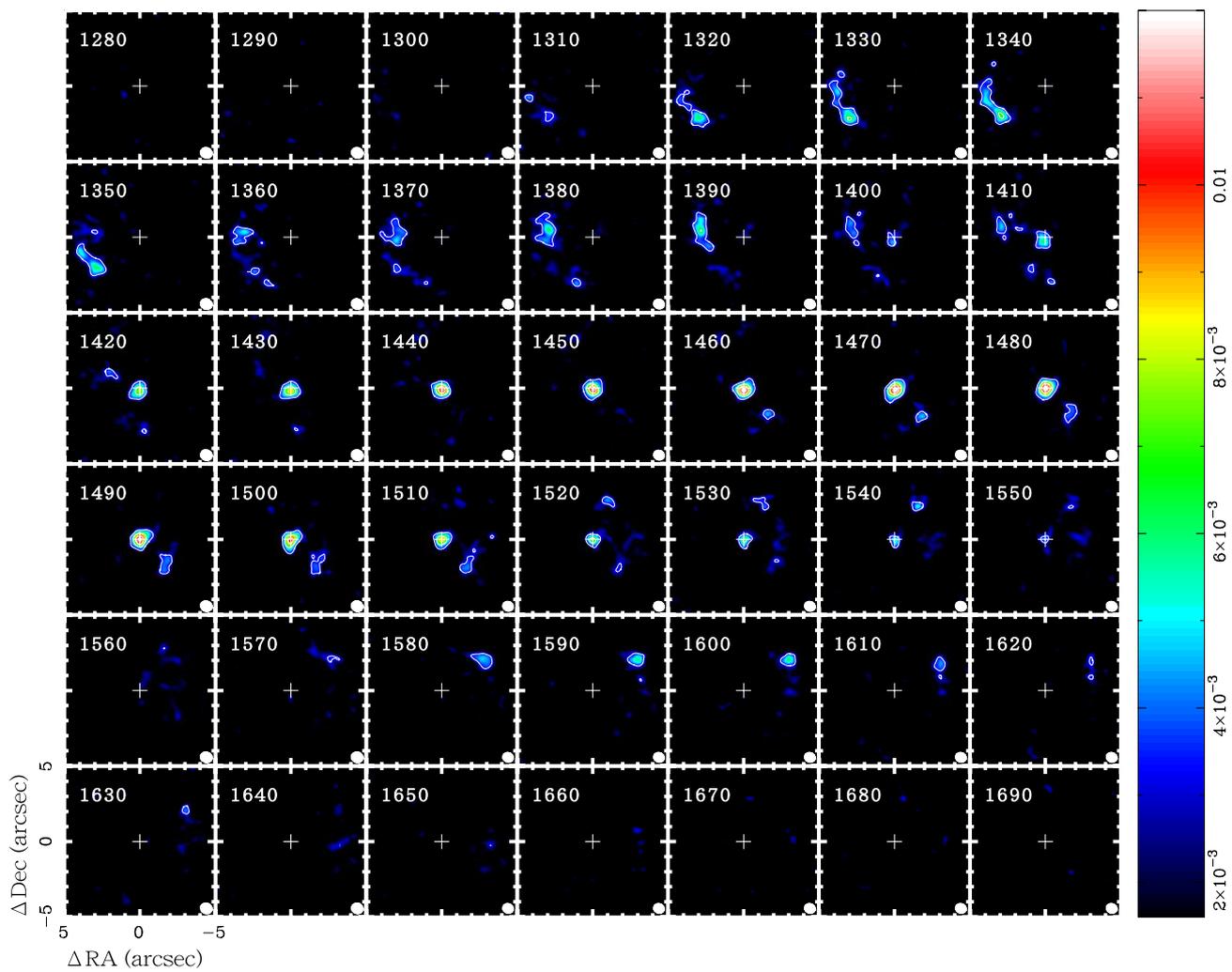}
 \end{center}
 \caption{ 
Same as figure~\ref{fig:hcop10_ch}, but for  HCN~(1--0) emission.
Contours levels are 4$\sigma$, 8$\sigma$, and 12$\sigma$, where 1$\sigma=0.8$~mJy~beam$^{-1}$. 
}
 \label{fig:hcn10_ch}
\end{figure*}

\onecolumn
\begin{figure}[p]
\begin{center}
\begin{tabular}{cc}
	\begin{minipage}{0.45\hsize}
\par
	\begin{center}
{\bf (a)}
		\FigureFile(70mm,150mm){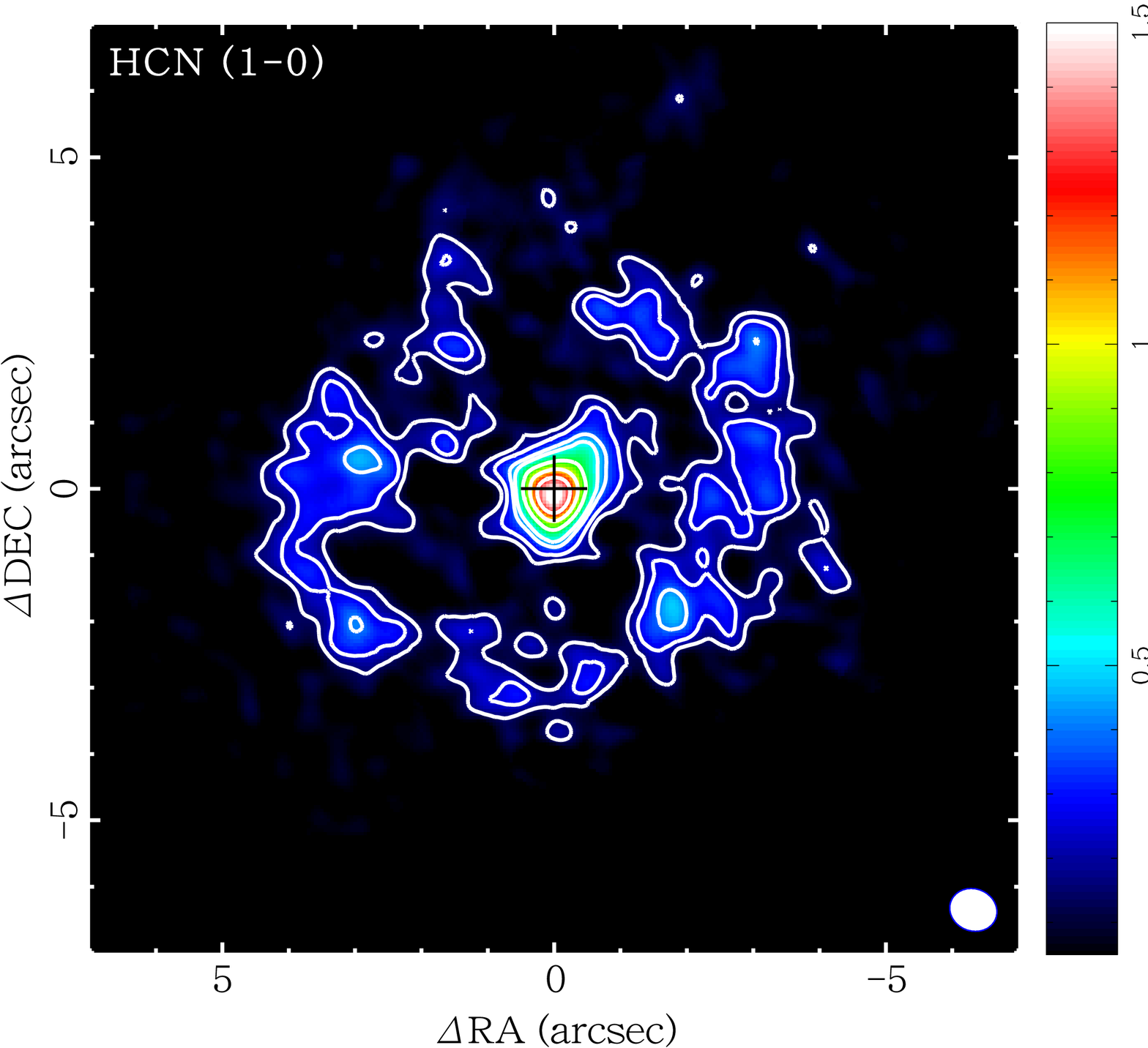}
	 \end{center}
	 \end{minipage}
	 \begin{minipage}{0.45\hsize}
\par
	\begin{center}
{\bf (b)}
		\FigureFile(70mm,150mm){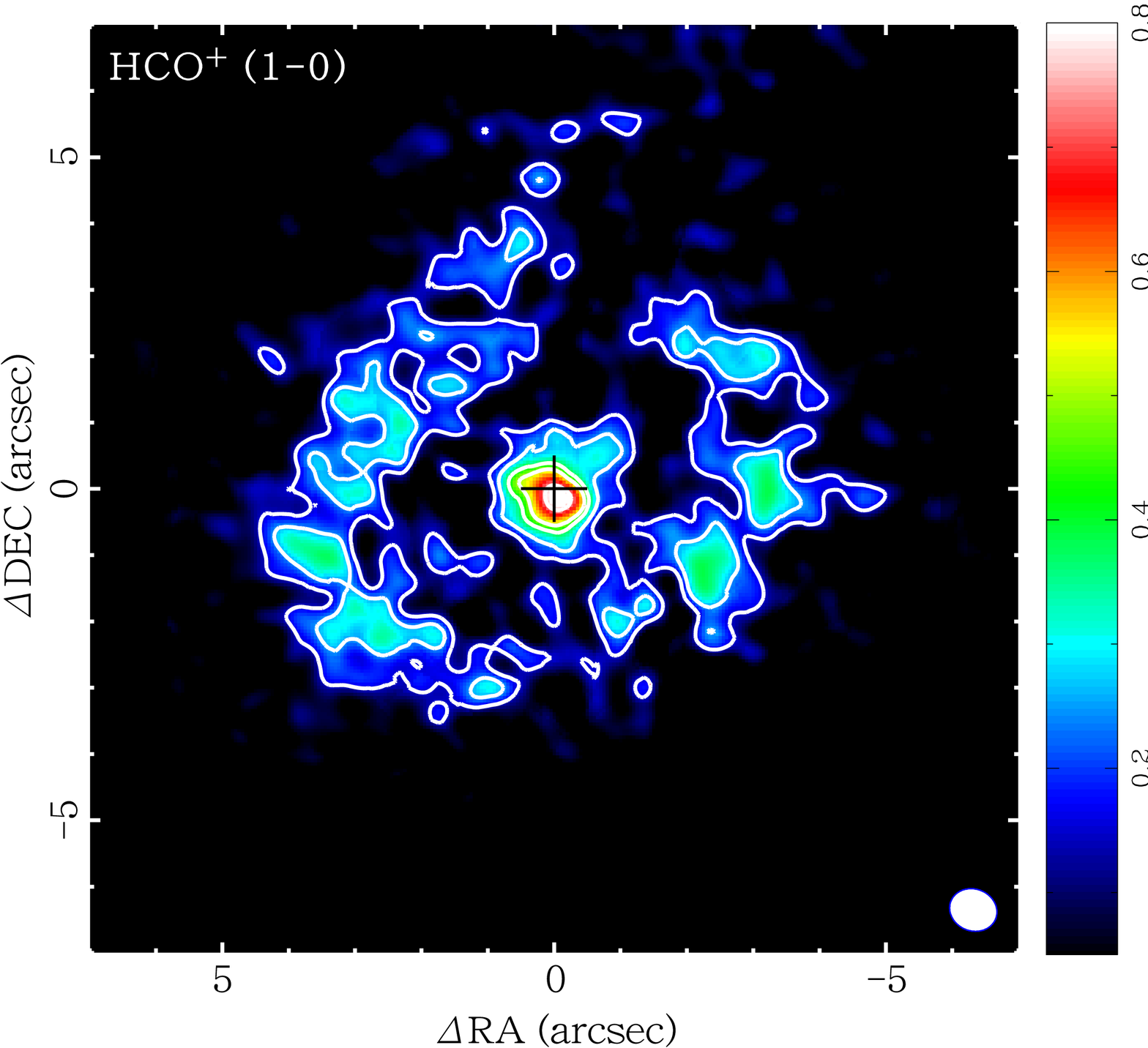}
	 \end{center}
	 \end{minipage}\\
	 \begin{minipage}{0.45\hsize}
\par
	\begin{center}
{\bf (c)}
		\FigureFile(70mm,150mm){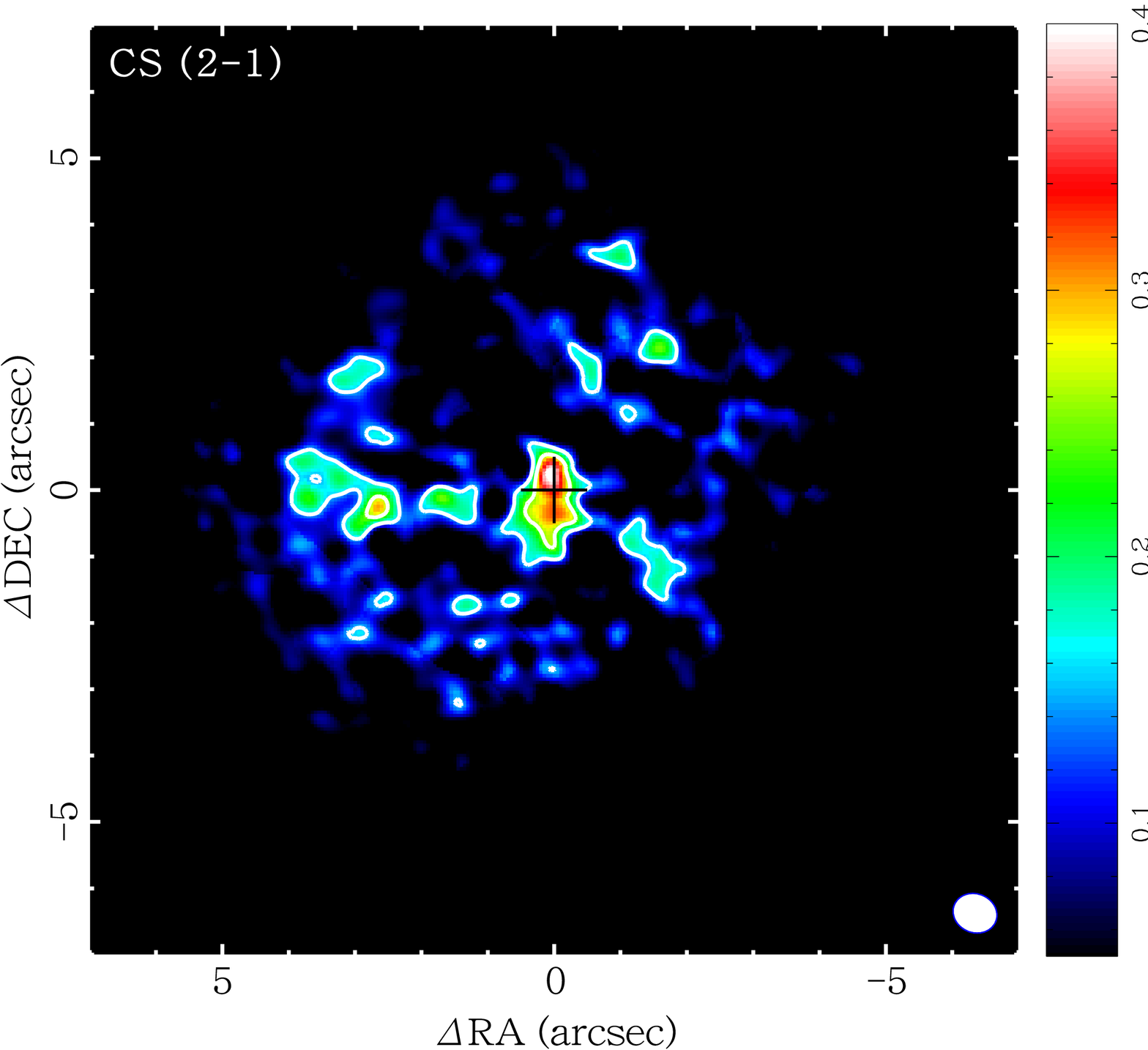}
	 \end{center}
	 \end{minipage}
	 \begin{minipage}{0.45\hsize}
\par
	\begin{center}
{\bf (d)}
		\FigureFile(70mm,150mm){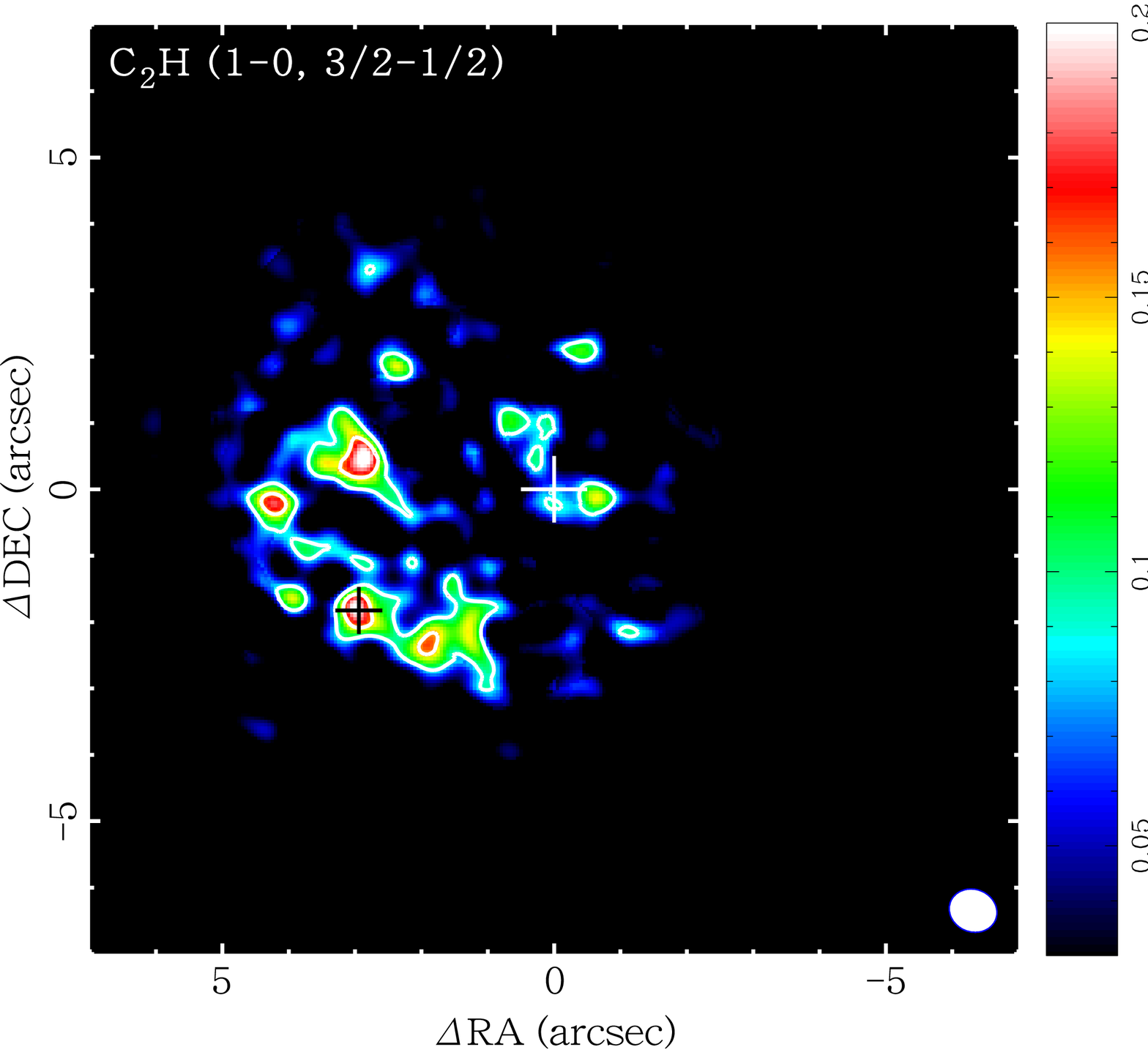}
	 \end{center}
	 \end{minipage}
 \end{tabular}
\end{center}
\caption{
Integrated intensity maps of (a) HCN(1--0), (b) HCO$^+$(1--0), (c) CS(2--1), and (d) C$_2$H(1--0, 3/2--1/2) in the central $\timeform{14"}\times\timeform{14"}$.
The synthesized beams are shown at the right bottom corner and the positions of the galactic center  is represented by crosses.
The contours are 
(a) 3, 5, 8, 10, 15, 20, 25, and 30$\sigma$ where 1$\sigma=0.05$~Jy~km~s$^{-1}$; 
(b) 3, 5, 8, 10, and 15$\sigma$ where 1$\sigma=0.05$~Jy~km~s$^{-1}$; 
(c) 3, 5, and 8$\sigma$ where 1$\sigma=0.05$~Jy~km~s$^{-1}$; and 
(d) 3 and 5$\sigma$ where 1$\sigma=0.03$~Jy~km~s$^{-1}$.
}
 \label{fig:band3}
\end{figure}
\twocolumn
\begin{figure}[t]
 \begin{center}
  \includegraphics[width=\linewidth]{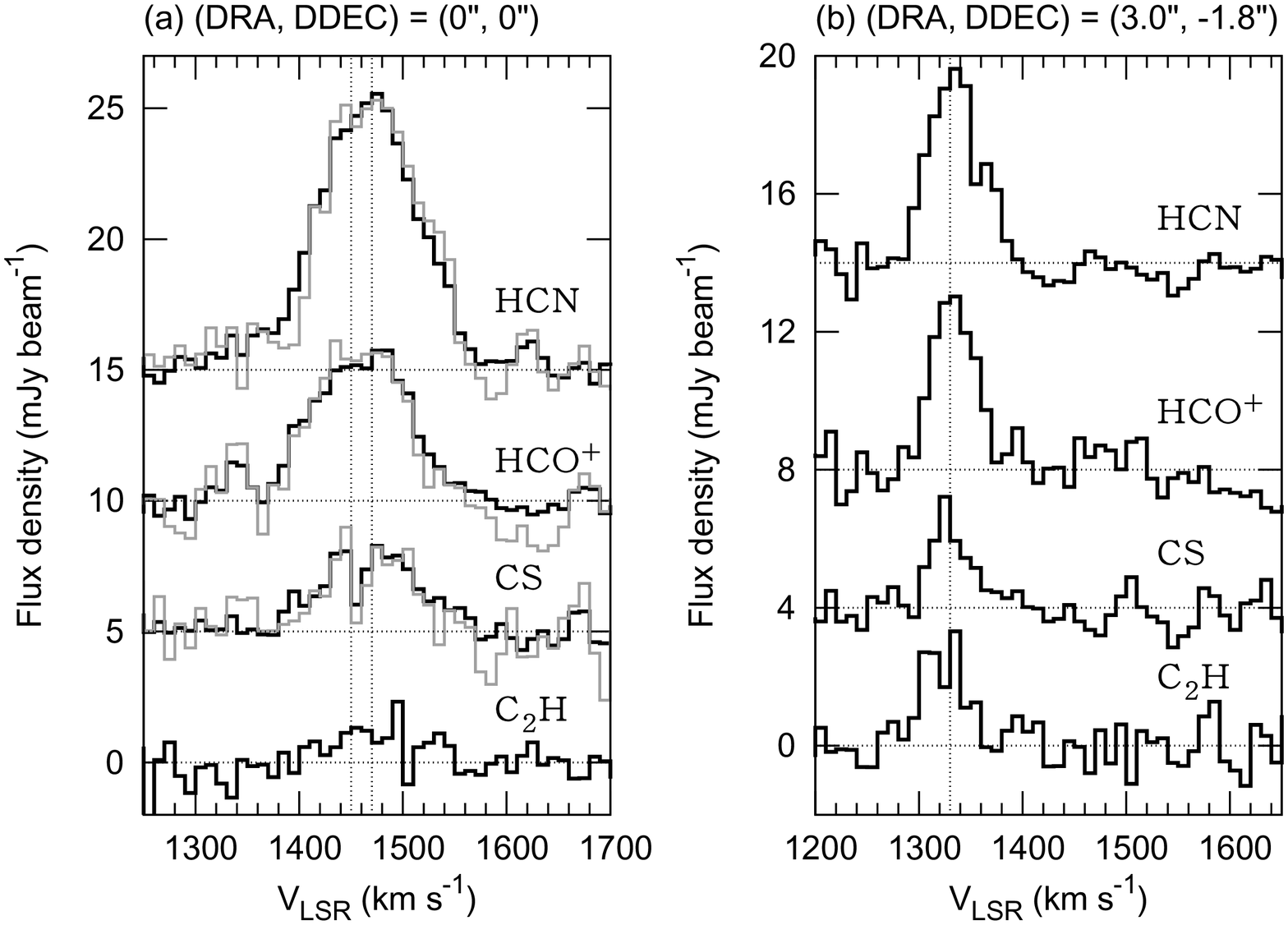}
 \end{center}
 \caption{ 
C$_2$H(1--0, 3/2--1/2), CS(2--1), HCO$^+$(1--0), and HCN(1--0) spectra at (a) the dynamical center ($\Delta \alpha$, $\Delta \delta) = (\timeform{0''}, \timeform{0''})$, and (b)  position shown by the black cross in figure~\ref{fig:band3}(d), ($\Delta \alpha$, $\Delta \delta)=(3\farcs0,-1\farcs8)$.
The gray lines in (a) are spectra derived from  only the 12-m extended array, 
providing the angular resolution of $\timeform{0.60}\times\timeform{0.52"}$, $\timeform{0.60}\times\timeform{0.51"}$, and $\timeform{0.56}\times\timeform{0.46"}$ for HCN, HCO$^+$, and CS, respectively.
The vertical lines in (a) and (b) correspond to the velocity region of the CS dip ($V_{\rm LSR}=1450$--1470~km~s$^{-1}$) and velocity of peak intensity of HCN ($V_{\rm LSR}=1330$~km~s$^{-1}$), respectively.
 }
 \label{fig:spe}
\end{figure}

\subsection{Molecular gas distribution  }
\subsubsection{Band3}
\label{sec:band3}

Figures~\ref{fig:hcop10_ch} and \ref{fig:hcn10_ch} respectively show the channel maps of HCO$^+$(1--0) and HCN(1--0) line emissions from the central $\timeform{10''}\times\timeform{10''}$ region of NGC~613, 
in which these lines were detected in the same velocity range of 1310--1660~km~s$^{-1}$. 
Figures~\ref{fig:band3}~(a)--(c) show 
the intensity maps of HCN(1--0), HCO$^+$(1--0), and CS(2--1), respectively, integrated over a velocity range of 1150--1750~km~s$^{-1}$, 
where emission-free channels were masked to derive the integrated intensities of the weak emission accurately and  enhance the contrasts.
HCN(1--0), HCO$^+$(1--0), and CS(2--1) lines were detected from both the CND and star-forming ring.
We determined that  intensities of HCN and HCO$^+$  show peaks toward the center but a dip of CS 
with $\gtrsim4\sigma$ [see figure~\ref{fig:spe}~(a)].
The dip occurs at a velocity of the peak temperature of CS(7--6) ($V_{\rm LSR}\sim1460$~km~s$^{-1}$, figure~\ref{fig:spe_band7}).
Recent observations have reported self and/or continuum absorption lines in dense gas tracers, e.g., HCN, HCO$^+$, and CS toward the nuclei of Seyfert galaxies and (ultra) luminous infrared galaxies ((U)LIRGs) (e.g., \cite{aalto2015}, \cite{scoville2015}, \cite{martin2016}, \cite{lin2016}).
Although the absorption of HCN and HCO$^+$ toward the center of NGC~613 cannot be identified from the combined data of all arrays $(\theta\sim\timeform{0.7"})$ apparently (black lines in figure~\ref{fig:spe}), by using only the 12-m extended array $(\theta\sim\timeform{0.6"})$,  
we can recognize the dip features (or plateau) at the velocity of $V_{\rm LSR}\sim1460$~km~s$^{-1}$  (gray lines in figure~\ref{fig:spe}).
These results indicate that the size of the absorbing cloud is significantly smaller than the beam size  ($\theta\sim{0.5"}=42$~pc).

\begin{figure}[t]
 \begin{center}
  \includegraphics[width=0.7\linewidth]{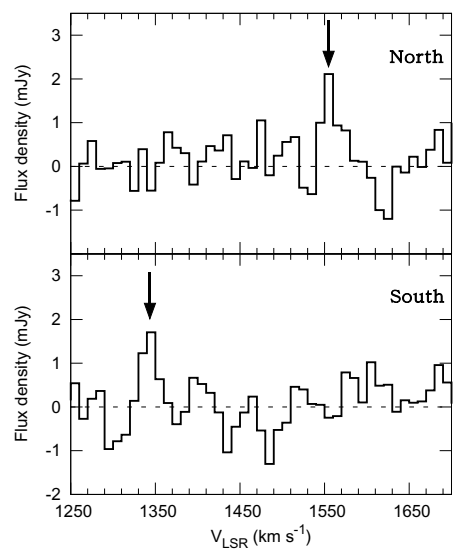}
 \end{center}
 \caption{ 
Top and bottom panels show SiO(2--1)  spectra  detected  at  the regions indicated by the northern and southern circles in figure~\ref{fig:cont}~(d), respectively.
}
 \label{fig:sio}
\end{figure}

We also detected C$_2$H(1--0, 3/2--1/2) line at 87.316925 GHz  at the east side of the ring [figure~\ref{fig:band3}~(d)]. 
The detection of C$_2$H line was confirmed by similar features of HCN(1--0), HCO$^+$(1--0), and CS(2--1) spectra [figure~\ref{fig:spe}~(b)] at a peak position of C$_2$H represented by a circle in  figure~\ref{fig:band3}~(d).
It has been reported that C$_2$H is produced in the UV irradiated cloud edges associated with massive star-forming regions \citep{beuther, meier}. 
From Br$\gamma$ observations of the central region of NGC~613, \citet{falcon} derived star-formation rates of 0.10--0.13~\MO~yr$^{-1}$ at some spots in the east side of the ring; these were clearly higher than those  in the west, 0.03--0.06~\MO~yr$^{-1}$.
In addition, we found peaks of C$_2$H located close to those of CO (3--2)  [figure~\ref{fig:band7}~(a)]. 
These results indicate that C$_2$H traces  the photodissociation region in the ring,
thus being consistent with the result of NGC~1097 \citep{martin2015}.
However, SiO(2--1), known as a tracer of shocks (e.g., \cite{martin-pintado}), is detected marginally [signal-to-noise ratio (S/N)$\sim4$; figure~\ref{fig:sio}] at  the edge of the radio jets, 
as shown by circles in figure~\ref{fig:cont}~(d), 
probably indicating the existence of shock regions related to  the jets.
However, for a detailed discussion, it is necessary to improve data quality.

\begin{figure*}[h]
 \begin{center}
  \includegraphics[width=\linewidth]{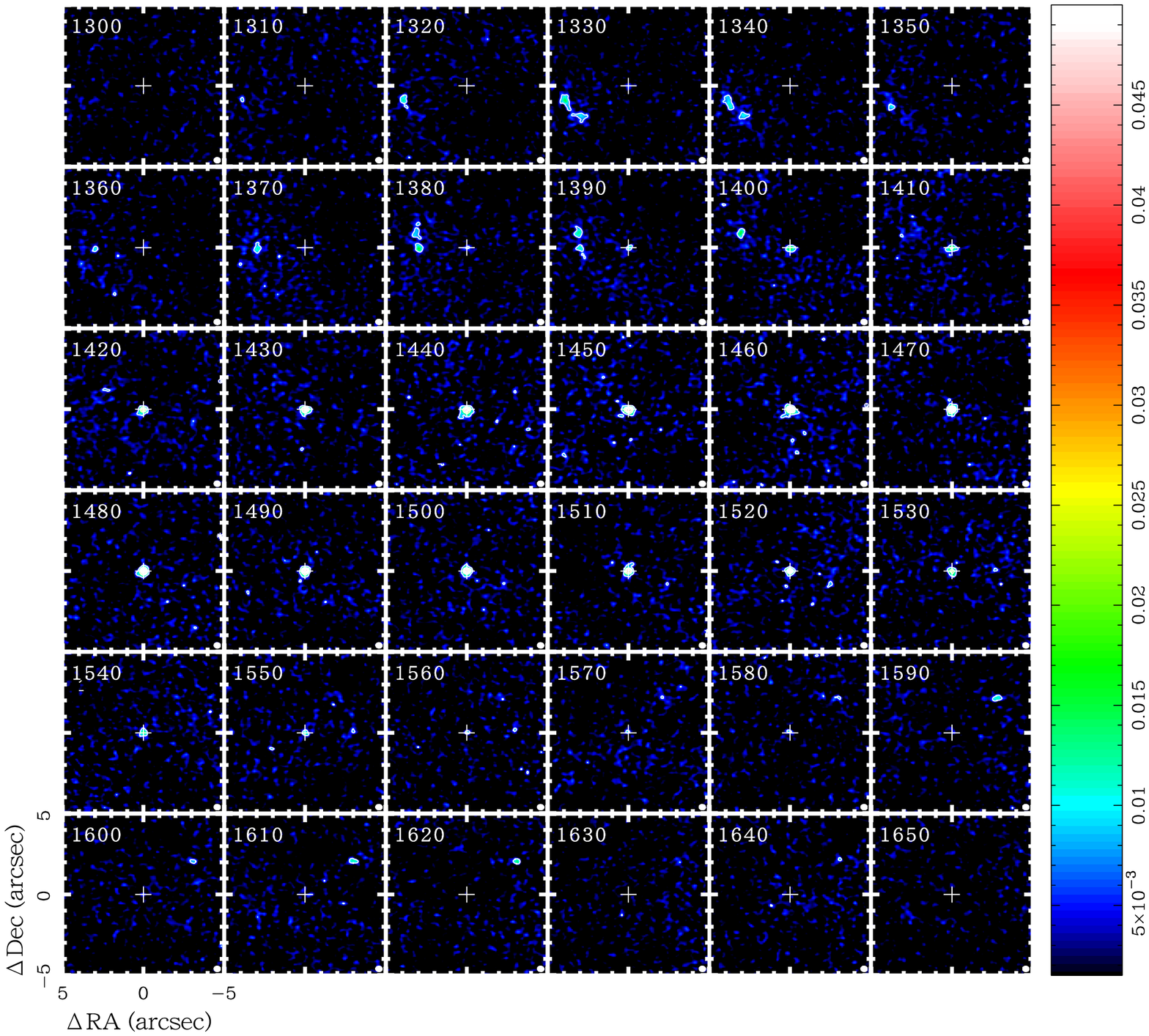}
 \end{center}
 \caption{ 
Channel maps of HCO$^+$(4--3) emission in the central $\timeform{10''}\times\timeform{10''}$ region, corresponding to $848\times848$~pc at the distance of the galaxy (17.5~Mpc).
The channels have an interval of 10~km~s$^{-1}$ and the central velocities are labeled at the upper left corner.
The synthesized beam of  0\farcs36$\times$0\farcs29 ({\it P.A.} $=\timeform{84D}$) is shown at the right bottom corner.
Crosses show  the position of the dynamical center.
Contours levels are 4$\sigma$, 10$\sigma$, 15$\sigma$, 20$\sigma$ and 25$\sigma$ where 1$\sigma=1.5$~mJy~beam$^{-1}$. 
 }
 \label{fig:hcop43_ch}
\end{figure*}

\subsubsection{Band7}
\label{sec:band7}
\begin{figure*}[t]
 \begin{center}
  \includegraphics[width=\linewidth]{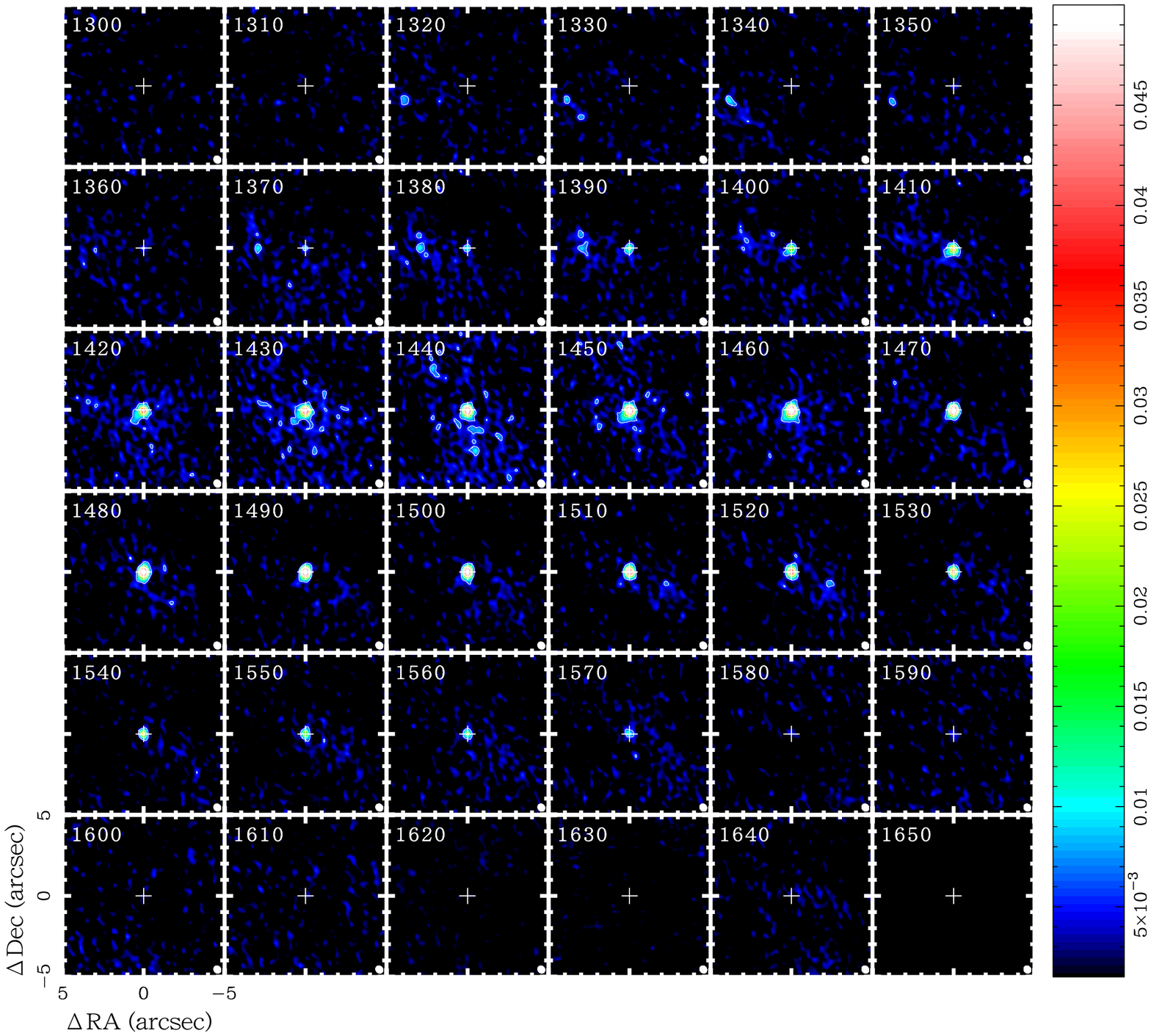}
 \end{center}
 \caption{ 
 Channel maps of HCN(4--3) emission in the central $\timeform{10''}\times\timeform{10''}$ region, corresponding to $848\times848$~pc at the distance of the galaxy (17.5~Mpc).
The channels have an interval of 10~km~s$^{-1}$, and the central velocities are labeled at the upper left corner.
The synthesized beam of  0\farcs43$\times$0\farcs37 ({\it P.A.} $=\timeform{40D}$) is shown at the right bottom corner.
Crosses show  the position of the dynamical center.
Contours levels are 4$\sigma$, 10$\sigma$, 15$\sigma$, 20$\sigma$, and 25$\sigma$, where 1$\sigma=1.7$~mJy~beam$^{-1}$. 
}
 \label{fig:hcn43_ch}
\end{figure*}

Figures~\ref{fig:hcop43_ch} and \ref{fig:hcn43_ch} show the channel maps of the HCO$^+$(4--3) and HCN(4--3) line emissions, respectively, from the central $\timeform{10''}\times\timeform{10''}$ region of NGC~613.
As the critical densities of $J=4$--3 transitions of HCO$^+$ and HCN ($n^{\rm crit}_{\rm H_2}\sim 10^{6-7}$~cm$^{-3}$) are several orders of magnitude higher than those of $J=1$--0 ($n^{\rm crit}_{\rm H_2}\sim 10^{4-5}$~cm$^{-3}$), 
the distribution of the higher transitions is much compact than those of the lower transitions, 
and the HCO$^+$(4--3) and HCN(4--3) emissions at the west side of the star-forming ring are weaker than those at the east side where star formation is active \citep{falcon}. 

Figures~\ref{fig:band7}~(a)--(d) show the velocity-integrated intensity maps  in the same manner as Band~3  but the velocity range of 
1250--1660~km~s$^{-1}$ for CO(3--2) and 1250--1650~km~s$^{-1}$ for HCN(4--3).
The high velocity components of HCN(4--3) ($V_{\rm LSR}\gtrsim1630$~km~s$^{-1}$) are regarded as negligible (figure~\ref{fig:hcn43_ch}).
HCN(4--3), HCO$^+$(4--3), and CS(7--6) were detected in the CND and a part of the star-forming ring. 
In contrast, CO(3--2) was detected in the CND and star-forming ring and  along the galactic bar. 
The southern galactic bar is weaker than the northern; this  may be attributed to the limited high velocity range of CO(3--2).

\subsection{Basic parameters of NGC~613}
\subsubsection{Systemic velocity, position angle, inclination angle, and rotation curve}
\label{sec:param}
\begin{figure}[t]
 \begin{center}
  \includegraphics[width=\linewidth]{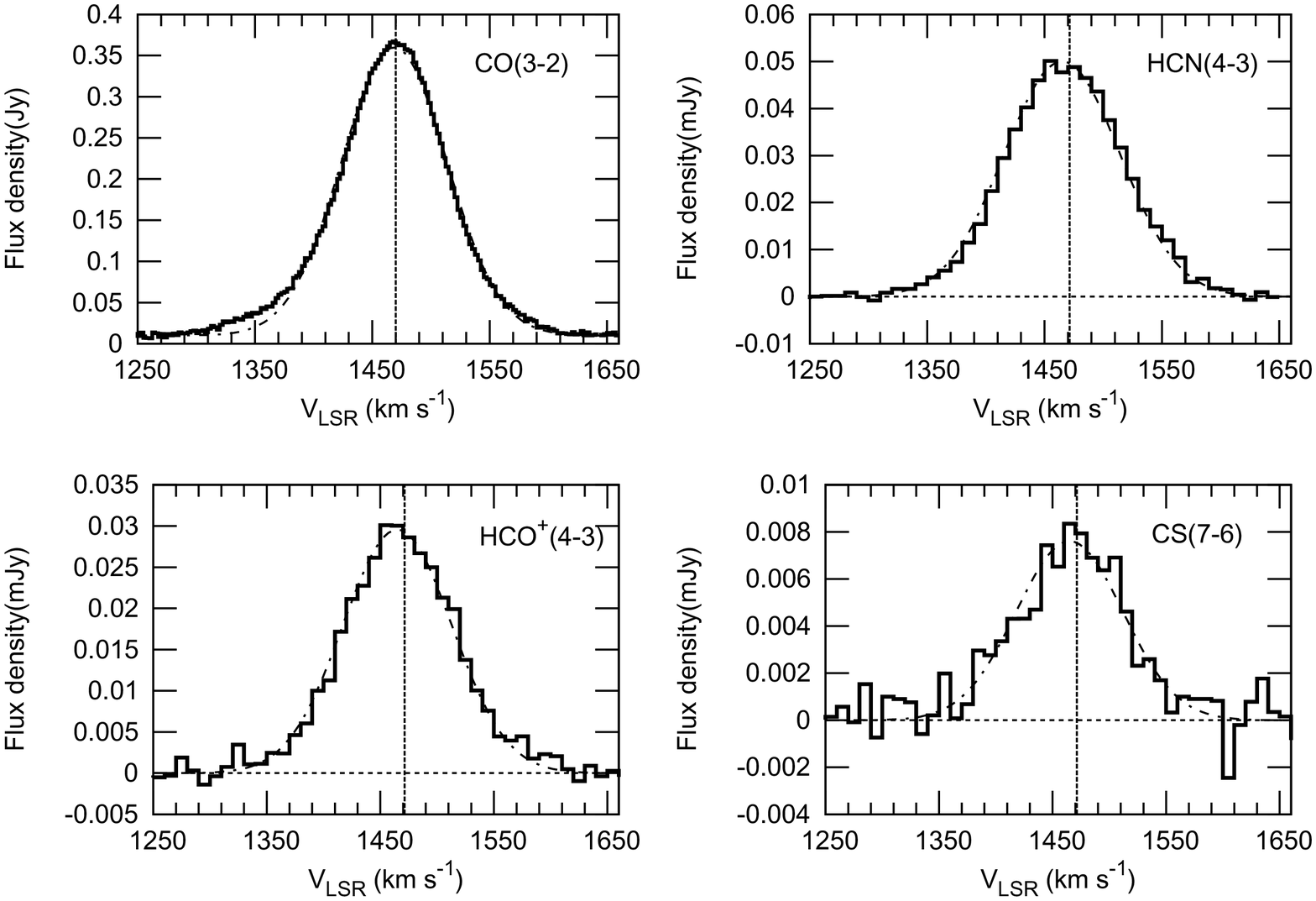}
 \end{center}
 \caption{ CO(3--2), CS(7--6), HCN(4--3), and HCO$^+$(4--3) emission line profiles toward the center of NGC~613. 
The corresponding aperture is $0\farcs5$  (42.4~pc). 
Gaussian fits to the lines are overlaid with the broken lines.
The vertical dot lines are the central velocities of $V_{\rm LSR}=1471$~km~s$^{-1}$, derived from the Gaussian fitting to the spectra.
 }
 \label{fig:spe_band7}
\end{figure}

Figure~\ref{fig:spe_band7} shows the spectra of CO(3--2), CS(7--6), HCN(4--3), and HCO$^+$(4--3) toward the center of NGC~613, $(\alpha_{\rm J2000.0}$, $\delta_{\rm J2000.0})=$ (\timeform{1h34m18.190s}, \timeform{-29D25m06.60s}).
By fitting the Gaussian to these  line profiles, we derived the central velocity  of $V_{\rm LSR}=1471.3\pm0.3$~km~s$^{-1}$, which was consistent with the systemic velocity determined by \citet{koribalski2004}, that is, $V_{\rm sys}=1470\pm5$~km~s$^{-1}$  in radio definition  with respect to LSR.

\onecolumn
\begin{figure}[p]
\begin{tabular}{cc}
	\begin{minipage}{0.45\hsize}
\par
	\begin{center}
{\bf (a)}
		\FigureFile(70mm,150mm){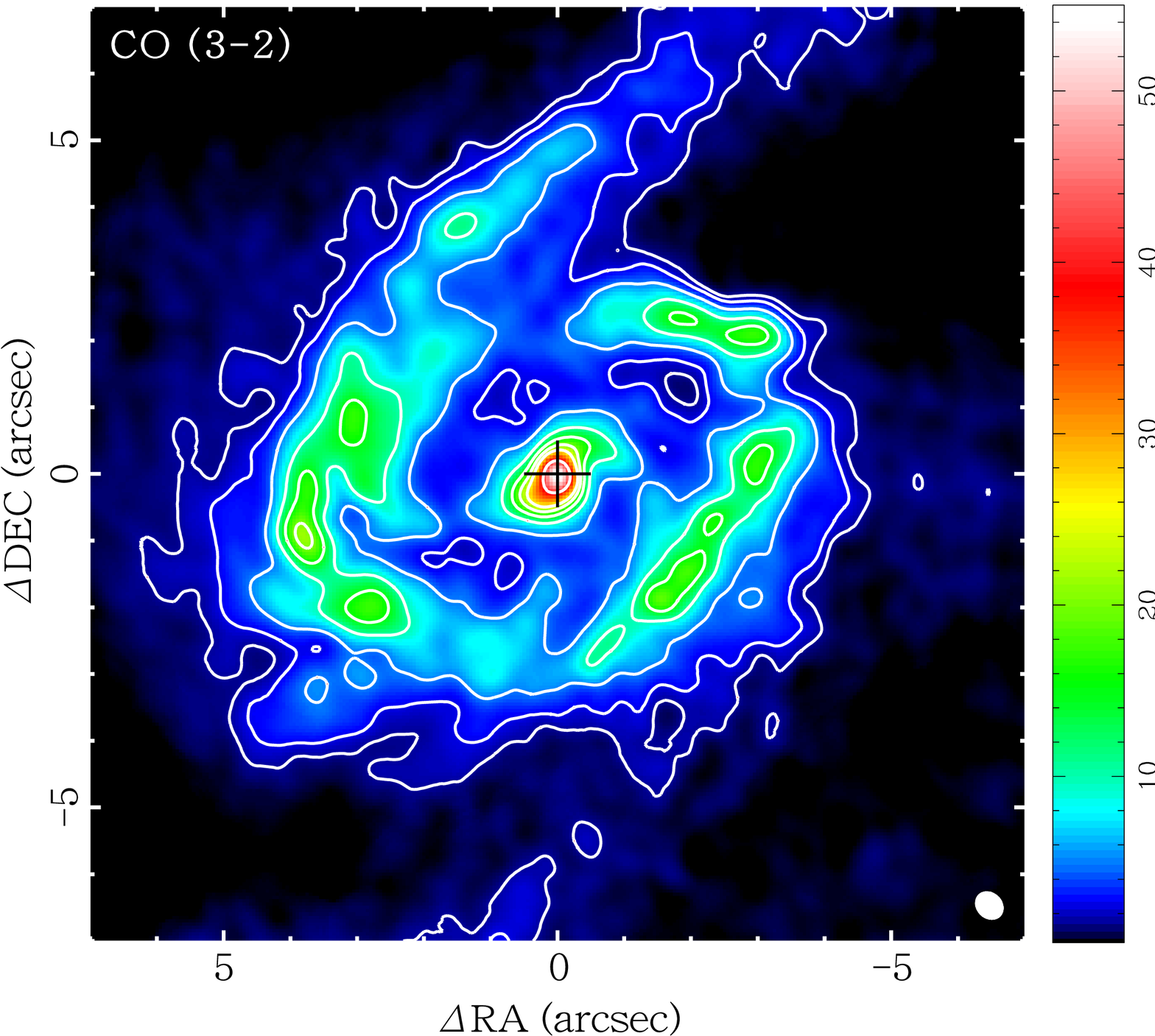}
	 \end{center}
	 \end{minipage}
	 \begin{minipage}{0.45\hsize}
\par
	\begin{center}
{\bf (b)}
		\FigureFile(70mm,150mm){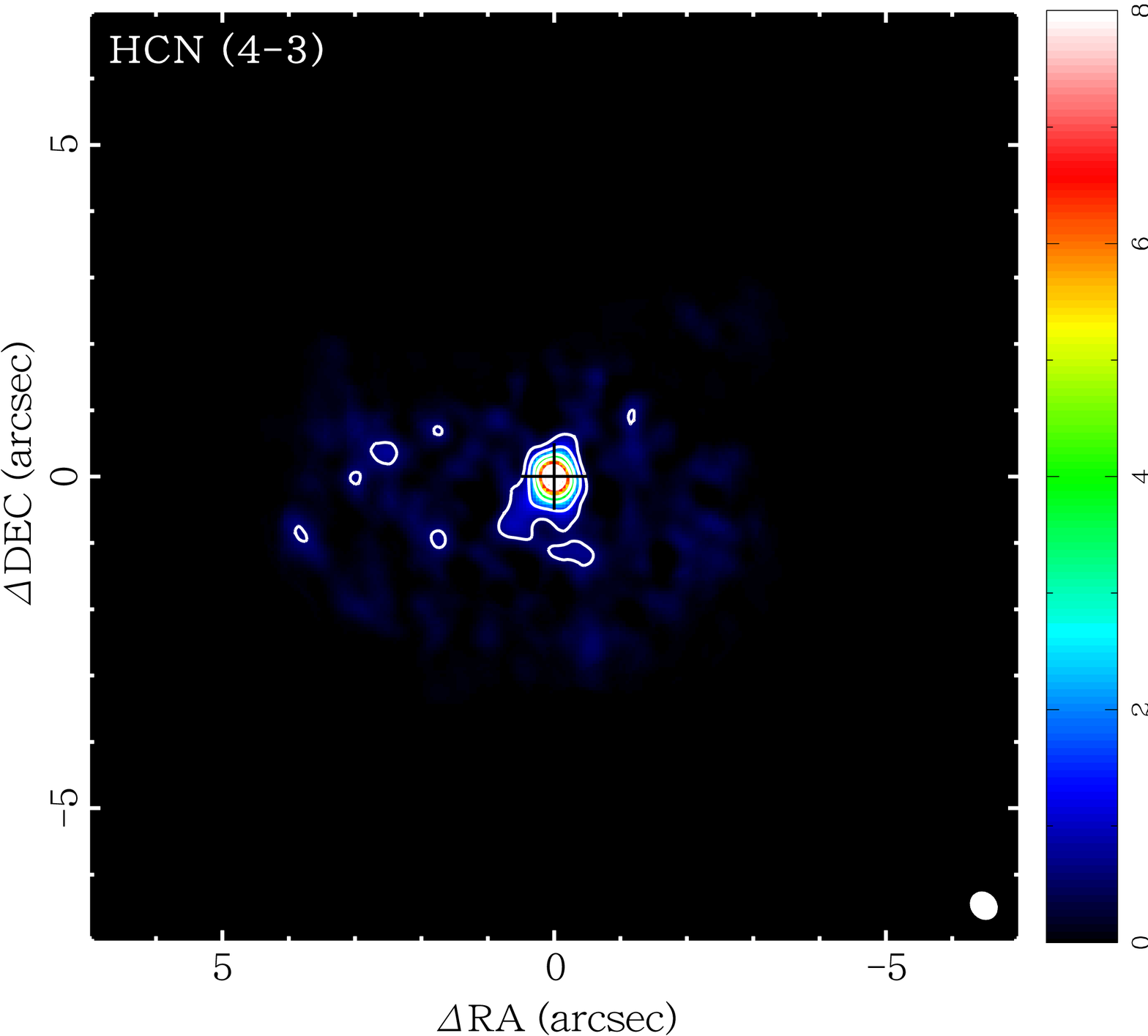}
	 \end{center}
	 \end{minipage}\\
	 \begin{minipage}{0.45\hsize}
\par
	\begin{center}
{\bf (c)}
		\FigureFile(70mm,150mm){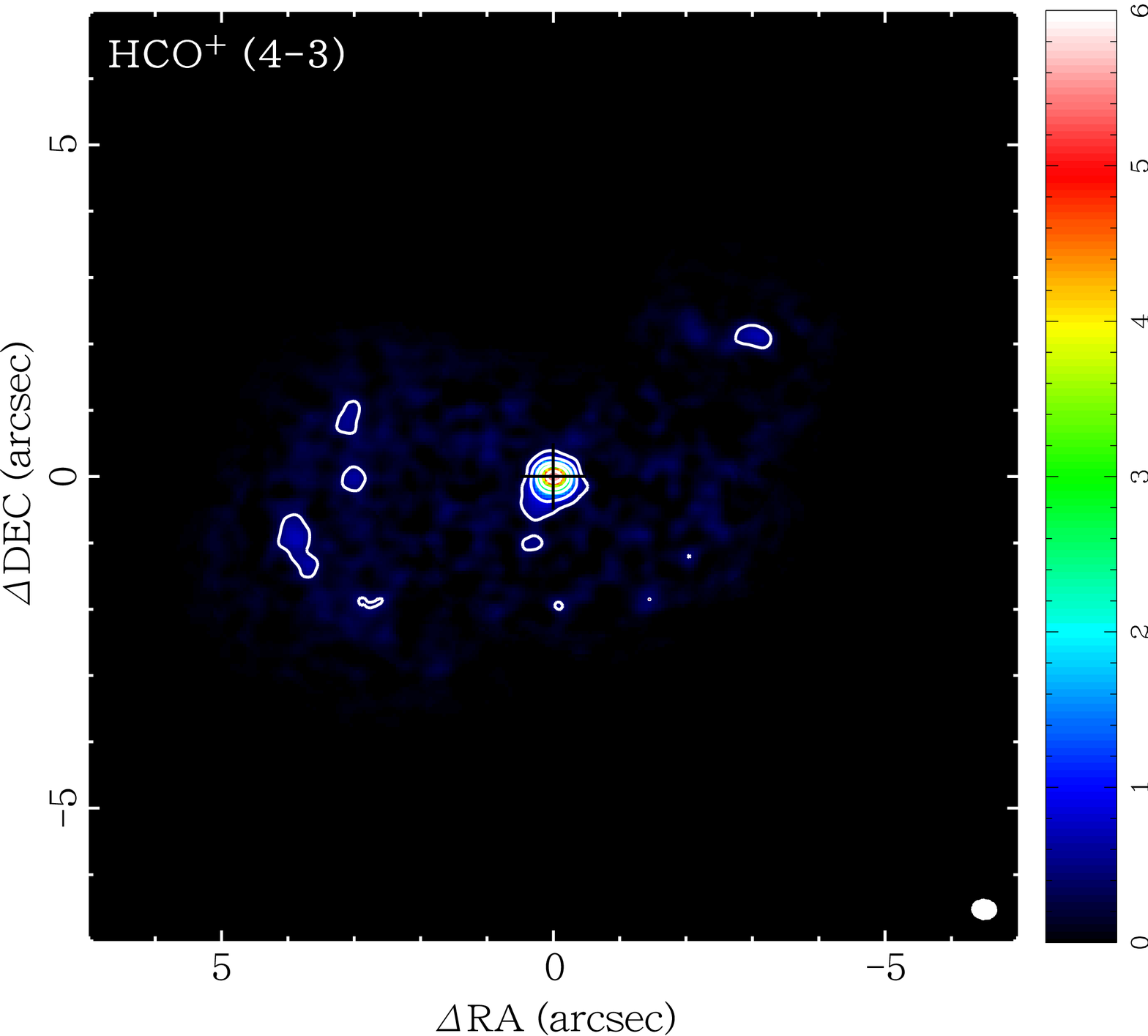}
	 \end{center}
	 \end{minipage}
	 \begin{minipage}{0.45\hsize}
\par
	\begin{center}
{\bf (d)}
		\FigureFile(70mm,150mm){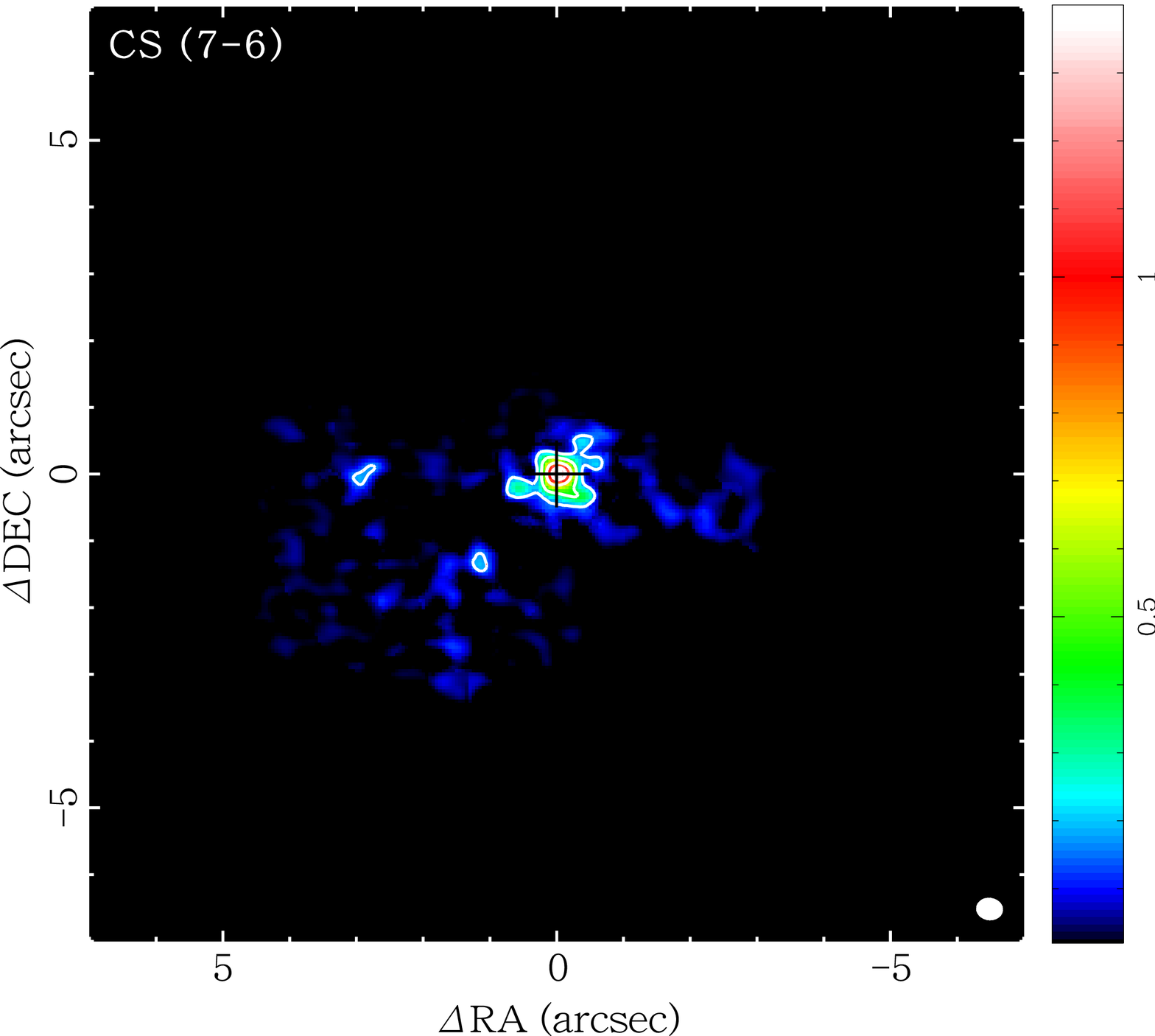}
	 \end{center}
	 \end{minipage}
 \end{tabular}
\caption{
Integrated intensity maps of (a) CO(3--2), (b) HCN(4--3), (c) HCO$^+$(4--3), and (d) CS(7--6) in the central $\timeform{14"}\times\timeform{14"}$.
The synthesized beams are shown at the right bottom corner, and the positions of the galactic center  shown by crosses.
The contours are 
(a) 3, 5, 10, 20, 30, 40, 50, 80, and 110$\sigma$, where 1$\sigma=0.5$~Jy~km~s$^{-1}$; 
(b) 4, 10, 20, 30, 50, 70, and 80$\sigma$, where 1$\sigma=0.1$~Jy~km~s$^{-1}$; 
(c) 4, 10, 20, 30, and 50$\sigma$, where 1$\sigma=0.1$~Jy~km~s$^{-1}$, and 
(d) 4$\sigma$ where 1$\sigma=0.03$~Jy~km~s$^{-1}$.
}
 \label{fig:band7}
\end{figure}
\twocolumn

\begin{figure*}[h]
 \begin{center}
  \includegraphics[width=\linewidth]{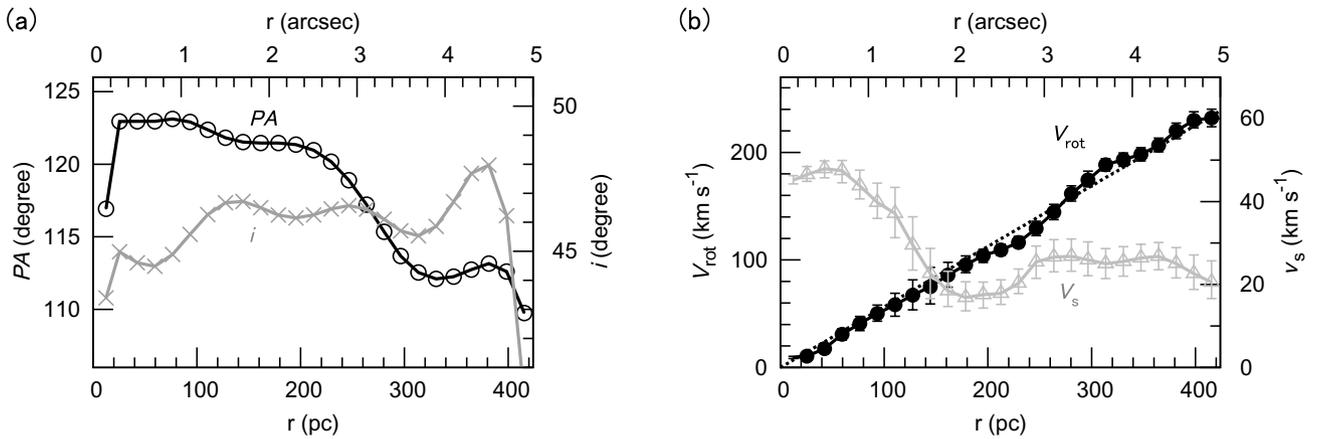}
 \end{center}
 \caption{ 
Radial variation of (left) the position angle of the major axis, inclination angle, (right) rotational velocity, and velocity dispersion of the central region of NGC~613 ($r\leq\timeform{5"}$). 
These values were kinematically determined using $^{\rm 3D}$Barolo  \citep{barolo}.
 }
 \label{fig:param}
\end{figure*}

We kinematically determined  the basic parameters of NGC~613, 
such as the systemic velocity $(V_{\rm sys})$, inclination angle $(i)$, position angle of the major axis $(\theta_{\rm maj})$, rotational velocity $(V_{\rm rot})$, and velocity dispersion $(v_{\rm s})$,  by using the program of 3D-based analysis of rotating objects from line observations ($^{\rm 3D}$Barolo) that fits a 3D tilted-ring to spectroscopic data cubes  \citep{barolo}.
The $^{\rm 3D}$Barolo can estimate the basic parameters, such as the dynamical center $(x_0, y_0)$,  $V_{\rm sys}$,  $i$,  $\theta_{\rm maj}$,  $V_{\rm rot}$, and  $v_{\rm s}$, by comparing the model with the observed data at each ring of radius $r$ and width $w$, assuming that the ring has a constant circular velocity $(V_{\rm rot})$ depending only on $r$.
The data cube of CO(3--2) with the velocity resolution $\Delta v=$2.5~km~s$^{-1}$ was utilized for this procedure.
The central region of NGC~613 $(r\leq\timeform{5"})$ was divided into 25 annuli with a width of $\Delta r=0\farcs2\sim \theta_{\rm maj}/2$, where $\theta_{\rm maj}$ is the synthesized beam size along its major axis, centered on the galactic center.
The scale-height of the disk is estimated as $h_z=v_s^2/2\pi G \Sigma_{\rm H_2}$, where $v_s$ is the gas velocity dispersion, $G$ is the gravitational constant, and $\Sigma_{\rm H_2}$ is the molecular gas surface density (section \ref{sec:mass}).
For $v_s=25$~km~s$^{-1}$ and $\Sigma_{\rm H_2}=200$\MO~pc$^{-2}$ in the ring, the scale height $h_z\approx150$~pc.
By adopting the  dynamical center of NGC~613 $(\alpha_{\rm J2000.0}$, $\delta_{\rm J2000.0})=$ (\timeform{1h34m18.190s},\timeform{-29D25m06.60s})  
and  scale-height $h_z=150$~pc,
we derived  $V_{\rm sys}$, $\theta_{\rm maj}$, $i$, $V_{\rm rot}$, and $v_s$ as functions of the radius by $^{\rm 3D}$Barolo.
In all the fitting steps, we applied a weighting function of $|\cos\theta|$, where $\theta=0$ corresponds to the major axis of the galaxy $(X)$, to minimize the error caused by the points close to the minor axis $(Y)$.
The resultant $V_{\rm sys}$ of $1471\pm3$~km~s$^{-1}$ is consistent with the central velocities of CO(3--2), CS(7--6), HCN(4--3), and HCO$^+$(4--3) spectra, and therefore we regarded $V_{\rm sys}=1471$~km~s$^{-1}$ as the systemic velocity of the galaxy.

\begin{figure*}[h]
 \begin{center}
  \includegraphics[width=\linewidth]{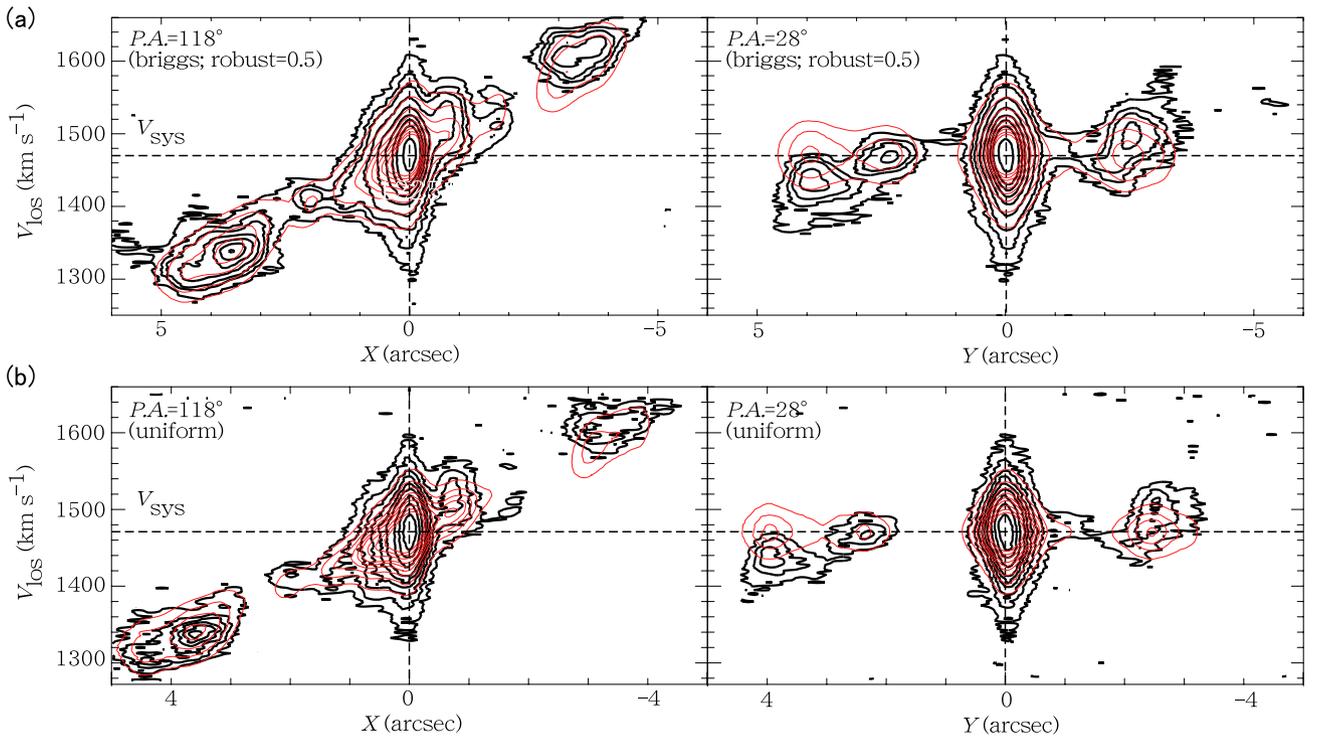}
 \end{center}
 \caption{ (a) PV diagrams of CO(3--2) along the major axis $(X)$ {\it P.A.}$=\timeform{118D}$ and minor axis $(Y)$ {\it P.A.}$=\timeform{28D}$  for the Briggs weighting with a robustness of 0.5 in CLEAN.
Contour levels are 0.02, 0.04, 0.06, 0.1, 0.15, 0.2, 0.25, 0.3, and 0.4~Jy~beam$^{-1}$. 
Red contours show the PV diagram of best fits model obtained using $^{\rm 3D}$Barolo.
(b) Same as panel (a) but for the uniform weighting in CLEAN $(\theta_{\rm FWHM}=0\farcs33\times0\farcs29)$.
Contour levels are 0.025, 0.05, 0.075, 0.1, 0.125, 0.15, 0.2, 0.25, and 0.3~Jy~beam$^{-1}$.
 }
 \label{fig:pv_diagram}
\end{figure*}

Figure~\ref{fig:param}(a) shows the results of $\theta_{\rm maj}$ and $i$ derived through this method.
The position  and inclination angles in the outer region than at $r\approx4\farcs5$ change abruptly because of low S/N of the data.
In addition, the values at the innermost region $(r<0\farcs1)$ cannot be determined rigidly because the size is smaller than that of the beam.
For simplicity,  $\theta_{\rm maj}=\timeform{118D}\pm\timeform{4D}$ and $i=\timeform{46D}\pm\timeform{1D}$, which are mean values between the radii $r=0\farcs2$ and $4\farcs5$, 
are regarded as the position and  inclination angles, 
although the systematic radial variances of these angles can be due to the warp or misalignment between the CND and ring.
The position and inclination angles are consistent with the result of \citet{rc3} ($\theta_{\rm maj}=\timeform{120D}$) and the value of $i=\timeform{47D}$ adopted by \citet{burbidge}, respectively.

For the fixed values of $\theta_{\rm maj}=\timeform{118D}$ and $i=\timeform{46D}$, 
the rotation velocity and velocity dispersion were derived (figure~\ref{fig:param}(b)). 
The velocity dispersion of $v_s\approx50$~km~s$^{-1}$ at the CND decreases to $v_s\approx20$~km~s$^{-1}$ with the radius and then  increases to $v_s\approx25$~km~s$^{-1}$ at the ring 
($r\sim\timeform{3"-4"}$).
The rotation curve shows the rigid rotation as a function of $V_{\rm rot}=49\pm1$~[km~s$^{-1}$~arcsec$^{-1}$]~$\times~r$ at $r>0\farcs5$, 
while the position--velocity diagram along the major axis shows that the velocity rises steeply in the central region ($r < 0\farcs5 \approx 42$~pc; figure~\ref{fig:pv_diagram}).
It is difficult to derive the high-rotation velocity separately from the velocity dispersion 
because of the limited beam width of $0\farcs44\times0\farcs37$ for the Briggs weighting with a robustness of 0.5 [figure~\ref{fig:pv_diagram}(a)] and even for  the uniform weighting [$0\farcs32\times0\farcs28$;  figure~\ref{fig:pv_diagram}(b)]. 
To resolve the degeneracy and clarify the variance of the rotation velocity, 
observations with higher angular resolutions would be needed.
If the rotational velocity is $V_{\rm rot}=100$~[km~s$^{-1}$]~$\sin(\timeform{46D})$ at $r=\timeform{0.1"}(=8.5$~pc),   
the central dynamical mass becomes $M=3.8\times10^7~\MO$, which is comparable to the supermassive black hole mass of NGC~613 $(\sim10^7~\MO)$ estimated by 
\citet{beifiori} and \citet{davis2014}

The red contours 
superimposed on the PV diagram by the data cube of CO(3--2) (black contour) 
in figure~\ref{fig:pv_diagram} are the best fit model with $^{\rm 3D}$Barolo along $\theta_{\rm maj}=\timeform{118D}$ (major axis) and $\theta_{\rm min}=\timeform{28D}$ (minor axis) for $i=\timeform{46D}$.  
The model can trace the ring at $|X|\sim\timeform{3"}$--\timeform{5"} 
but not the high-velocity components in the CND at $|X|<\timeform{0.5"}$.
In the PV diagram along the minor axis, a bridging feature is observed between the CND and the southern part of the ring [$Y\sim(-1\farcs0$) -- ($-2\farcs3$)]. 
Considering that 
the gas motion in NGC~613 is counterclockwise  and the southern part of NGC~613 is on the near side \citep{burbidge},
the components with a higher velocity than $V_{\rm sys}$ at $Y\sim\timeform{1"}-\timeform{2"}$ may be associated with the outflow from the CND.  
The velocity shifts of $v>V_{\rm sys}$ ($Y\sim\timeform{-1"}$) and $v<V_{\rm sys}$  ($Y\sim\timeform{-2"}$)  indicate the inflowing  and outflowing gas, respectively, 
whereas 
it is difficult to explain this velocity shift based on the coplanar gas motion in the galactic disk.
At the point  of the velocity shift $(Y\sim-1\farcs5)$, any particular gas distribution cannot be found (figure~\ref{fig:band7}(a));  
however, the point corresponds to the edge of the radio jets (figure~\ref{fig:cont}(d)).
In addition, an extent of strong [FeII] emission was reported from the center to around the velocity shift point \citep{falcon}.
The velocity shift is presumed to diffuse outflowing gas over a wide-range direction such as a spherical bubble (e.g., \cite{wagner2011}).
The velocity offset ($v>V_{\rm sys}$) at the outer ring $(|Y|\gtrsim\timeform{2.5"})$ is caused by the noncircular (inflow) motion of the galactic bar because the 
velocity difference from systemic velocity $(\Delta V\sim30$~km~s$^{-1})$ is obviously larger than the effect of the uncertainty of the inclination $(\Delta =\pm\timeform{1D})$ on the rotation velocity $(\sim6\% 
\approx7~{\rm km~s^{-1}})$.

\subsubsection{Mass and radial distribution of molecular gas}
\label{sec:mass}

\begin{figure}[t]
 \begin{center}
  \includegraphics[width=\linewidth]{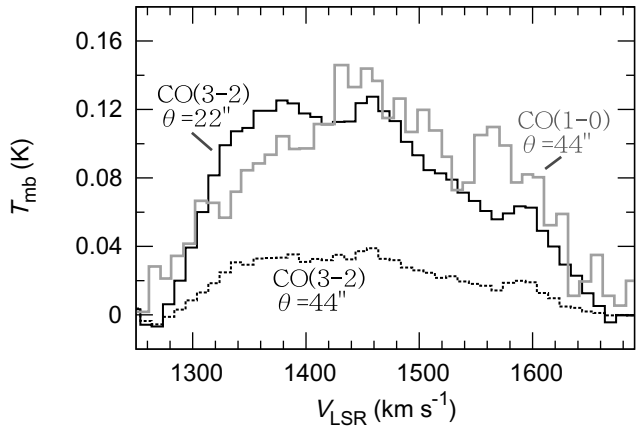}
 \end{center}
 \caption{ Comparison among the spectra of CO(3--2) with ALMA TP array and CO(1--0) with SEST (gray, \cite{elfhag}) toward the central region of NGC~613.
The spectra of CO(3--2) are convolved with the beam of $\theta_{\rm FWHM=}\timeform{44"}$ (black dash) and \timeform{22"} (black solid).
 }
 \label{fig:co_spe}
\end{figure}

\begin{figure}[t]
 \begin{center}
  \includegraphics[width=\linewidth]{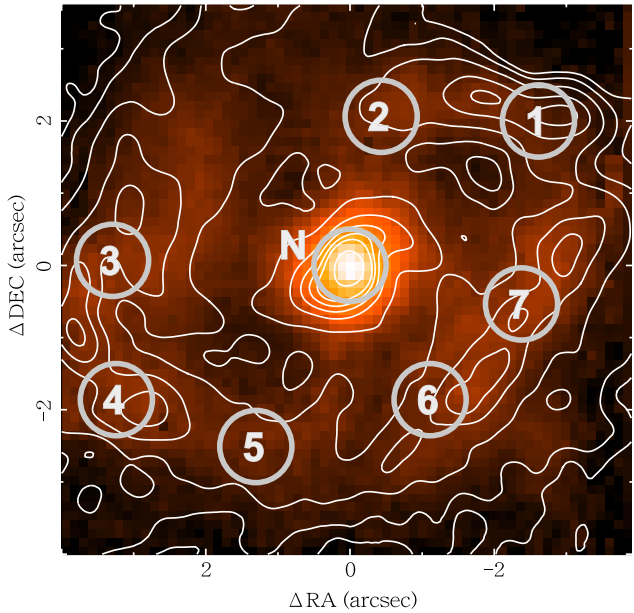}
 \end{center}
 \caption{ VLT H$_2$ map (color; \cite{falcon}) overlaid with contours of CO(3--2) integrated intensity (figure~\ref{fig:band7}(a)).
The peak positions of these maps matched each other.
The circles mark the positions of the nucleus and spots~1 through 7 denoted by \citet{falcon}. }
 \label{fig:h2}
\end{figure}

Figure~\ref{fig:co_spe} shows the spectra of CO(3--2) with ALMA TP array and of CO(1--0) with SEST for comparison \citep{elfhag}. 
The integrated intensities of CO(3--2) over the velocity range of  1280--1680~km~s$^{-1}$ within the beam of $\theta_{\rm FWHM}=\timeform{44"}$ and \timeform{22"}  are $I_{\rm CO32}=8.5\pm0.4$ and $31.2\pm0.9$~K~km~s$^{-1}$, respectively, and that of CO(1--0) with $\theta_{\rm FWHM}=\timeform{44"}$ is $I_{\rm CO10}=33.7\pm0.8$~K~km~s$^{-1}$.
For the same beam size ($\theta_{\rm FWHM}=\timeform{44"}$) in both CO(1--0) and CO(3--2), 
 the ratio of CO(3--2) to CO(1--0) is $R_{32/10}=0.25\pm0.01$, which is consistent with the values of  $R_{32/10}=$0.2--0.7 toward the centers of 28 nearby galaxies observed through a beamwidth of \timeform{21"} \citep{mauersberger} .
However, for the central region of $r\leq\timeform{5"}$
(=420~pc), 
the $R_{32/10}$ could be underestimated 
because it increases toward the center (e.g., \cite{dumke}, \cite{muraoka}, \cite{tsai2012}, \cite{vlahakis}, \cite{garcia2014}).
\citet{matsushita2004} showed 
$R_{32/10}\geq1.9\pm0.2$ at the nucleus of the Seyfert 2 galaxy M51,  
whose 
IR luminosity in 8--1000~$\mu$m comparable to that of NGC~613 \citep{sanders2003}, 
averaged over the beamsize of $4\farcs2\times3\farcs4$.
The intensity ratio of M51 ($R_{32/10}=0.8\pm0.1$) derived from  $I_{\rm CO10}=53.4\pm4.3$~K~km$^{-1}$ over $\theta_{\rm FWHM}=\timeform{44"}$ 
\citep{young1995} and $I_{\rm CO32}=44.4\pm2.8$~K~km$^{-1}$  over $\theta_{\rm FWHM}=\timeform{22"}$ \citep{mao2010} 
is slightly smaller than the intensity ratio of NGC~613  ($R_{32/10}=0.93\pm0.03$) using CO(1--0) over $\theta_{\rm FWHM}=\timeform{44"}$ and CO(3--2) over $\theta_{\rm FWHM}=\timeform{22"}$,
although the distance of M51 ($\sim8.4$~Mpc; \cite{feldmeier}) is approximately a half that of NGC~613.
We adopted the $R_{32/10}\sim2$ as the upper limit
in the central region of NGC~613 ($r\leq\timeform{5"}$).
The molecular mass of $3.0\times10^8$~$\MO$ in the central region of NGC~613 ($r\leq\timeform{22"}$) can be derived from 
the CO(1--0) intensity of $I_{\rm CO}=33.7$~K~km~s$^{-1}$ 
by adopting the CO-to-H$_2$ conversion factor of $X_{\rm CO}=0.5\times10^{20}$ cm$^{-2}$~(K~km~s$^{-1}$)$^{-1}$ \citep{bolatto2013}. 
The total flux within the region of 
$r\leq\timeform{5"}$ is 
$S_{\rm CO32}=2.15\times10^3$~Jy~km~s$^{-1}$, 
which corresponds to $M_{\rm H_2}=7.2\times10^7$
--$5.8\times10^8$~$\MO$,
given by 
(e.g., \cite{kohno2002}, \cite{bolatto2013})
\begin{eqnarray}
\left(\frac{M_{\rm H_2}}{\MO}\right) 
&=& 2.2\times10^2
\frac{1}{R_{32/10}}
\left(\frac{\lambda}{{\rm 0.87~mm}}\right)^2
\left(\frac{S_{\rm CO32}}{\rm Jy~km~s^{-1}}\right)\nonumber \\
&\times&
\left(\frac{D}{\rm Mpc}\right)^2 \left[\frac{X_{\rm CO}}{0.5\times10^{20}~{\rm cm}^{-2}~{\rm (K~km~s^{-1})}^{-1}}\right]
\label{eq:mass}
\end{eqnarray}
for 
$R_{32/10}=0.25$--2 with
galaxy distance $D=17.5$~Mpc.


\begin{figure}[t]
 \begin{center}
  \includegraphics[width=\linewidth]{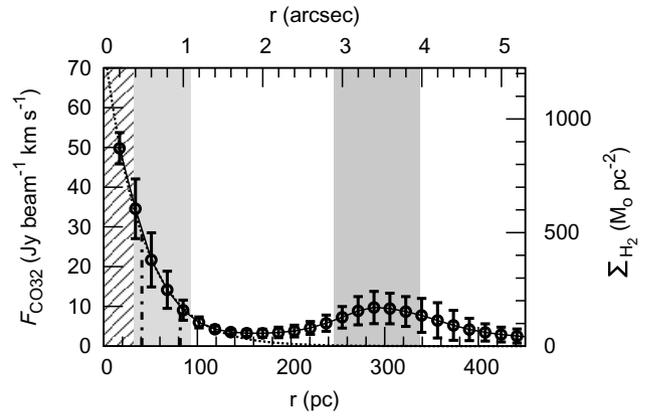}
 \end{center}
 \caption{ 
The solid line presents the radial distribution of CO(3-2) emission of NGC~613. 
The CO-integrated intensities are azimuthally averaged with the correction for the inclination of \timeform{46D}.
The right-hand axis shows the surface densities 
for $R_{32/10}=2$
calculated using equation~(\ref{eq:surface}).
The dotted line presents a fitted exponential profile of $\Sigma_{\rm H_2}=1330\exp(-r/r_{\rm e})$~[\MO~pc$^{-2}$] ($r_{\rm e}=41$~pc), 
at $r\leq100$~pc.
The vertical broken lines show the radii $r_{\rm e}$ and $2r_{\rm e}$.
The diagonal hatched region $(r\leq0.75r_{\rm e})$, light gray region $(0.75r_{\rm e}\leq r\leq2.25r_{\rm e})$, and dark gray region  $(6r_{\rm e}\leq r\leq8.25r_{\rm e})$ are defined as the inner CND, outer CND, and star-forming ring, respectively.
}
 \label{fig:rad}
\end{figure}

By using the integrated intensity of CO(3--2)  $(F_{\rm CO32})$, galaxy inclination $i$, beamwidths $\theta_{\rm maj}$ and $\theta_{\rm min}$, and conversion factor $X_{\rm CO}$,
the face-on surface density of molecular gas is given  by
\begin{eqnarray}
\left(\frac{\Sigma_{\rm H_2}}{\MO~{\rm pc}^{-2}} \right) &=& 8.2~\cos i ~
\frac{1}{R_{32/10}}~
\left(\frac{\lambda}{{\rm 0.87~mm}}\right)^2~
\left(\frac{\theta_{\rm maj}}{\rm arcsec}\right)^{-1}\nonumber \\
&\times&
\left(\frac{\theta_{\rm min}}{\rm arcsec}\right)^{-1} \left(\frac{F_{\rm CO32}}{\rm Jy~beam^{-1}~km~s^{-1}}\right) \nonumber \\
&\times&
\left[\frac{X_{\rm CO}}{0.5\times10^{20}~{\rm cm}^{-2}~{\rm (K~km~s^{-1})}^{-1}}\right].
\label{eq:surface}
\end{eqnarray}
Figure~\ref{fig:rad} shows the radial distribution of $\Sigma_{\rm H_2}$ 
for $R_{32/10}=2$, 
which was derived by averaging into annuli with width of $0\farcs2$ ($=17$~pc). 
Position angle $PA=\timeform{118D}$ and inclination angle $i=\timeform{46D}$ were adopted as the parameters of the ellipses.
The radial distribution $\Sigma_{\rm H_2}$ shows an exponential decrease (dotted line) of $\Sigma_{\rm H_2}=1330\exp(-r/r_{\rm e})$~[\MO~pc$^{-2}$], where $r_{\rm e}=0\farcs48(=41$~pc) at $r\leq100$~pc.
We define the regions $r\leq 0\farcs36(=0.75 r_{\rm e})$, $0\farcs36(=0.75r_{\rm e} )\leq r\leq 1\farcs08(=2.25r_{\rm e})$, and $2\farcs88(=6r_{\rm e})\leq r\leq 3\farcs96(=8.25r_{\rm e})$ as 
the inner CND, outer CND, and star-forming ring, respectively.

\section{Discussion}
\label{sec:discussion}
\subsection{LTE analysis}
\label{sec:lte}

\begin{table*}
  \tbl{Resultant parameters from rotational diagrams.}{%
  \begin{tabular}{cccc}
\hline
& Inner CND & Outer CND & Star forming ring \\
\hline
$T_{\rm rot}$ (HCN) [K] &	$11.6\pm{1.8}$ &	$10.6\pm3.7$ &	$8.3\pm2.7$	\\
$T_{\rm rot}$ (HCO$^+$) [K] &	$12.9\pm2.6$ &	$11.2\pm4.1$ &	$8.7\pm2.6$	\\
$T_{\rm rot}$ (CS) [K] &	$17.5\pm2.9$ &	$15.9\pm7.1$ &	$11.7\pm3.7$	\\
$N$(HCN) [$\times 10^{13}$cm$^{-2}$] &		$7.7\pm0.9$ &	$4.0\pm1.1$ &	$1.3\pm0.4$	\\
$N$(HCO$^+$) [$\times 10^{13}$cm$^{-2}$] &	$2.2\pm{0.3}$ &	$1.3\pm{0.3}$ &	$0.7\pm{0.2}$	\\
$N$(CS) [$\times 10^{13}$cm$^{-2}$] &		$2.2\pm{0.3}$ &	$1.3\pm{0.5}$ &	$0.6\pm{0.2}$	\\
$N$(HCN)$/N$(HCO$^+$) &	$3.5\pm{0.7}$ &	$3.1\pm{1.1}$ &	$1.9\pm{0.8}$	\\
$N$(HCN)$/N$(CS) 		&	$3.5\pm{0.7}$ &	$3.1\pm{1.5}$ &	$2.2\pm{1.1}$	\\
\hline
\end{tabular}}
\label{tab:diagram}
  \end{table*}

\begin{figure*}[t]
 \begin{center}
  \includegraphics[width=\linewidth]{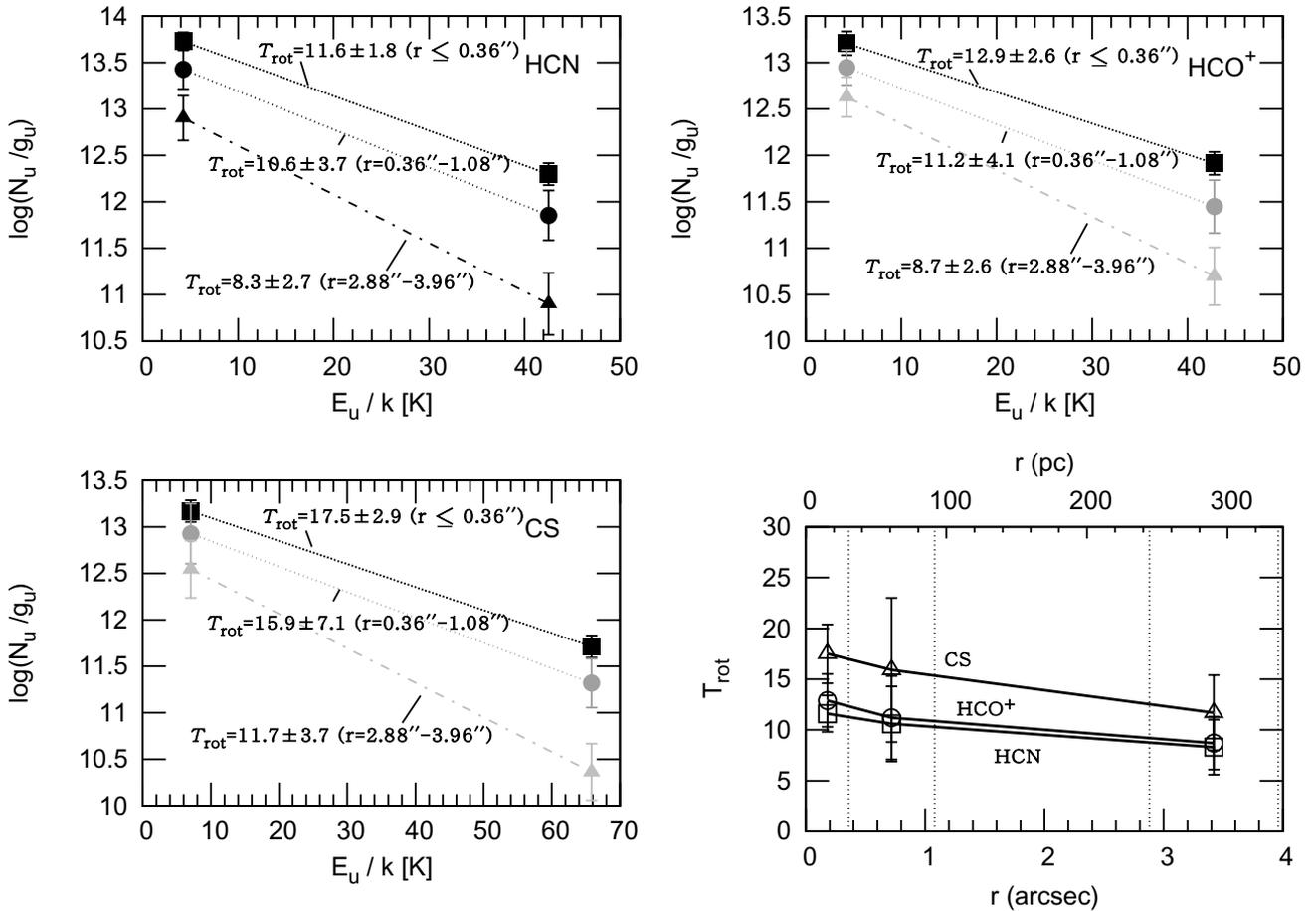}
 \end{center}
 \caption{ 
Rotational diagrams of the HCN (top left), HCO$^+$ (top right), and CS (bottom left) measurements, 
and the radial variation of the resultant rotational temperature of the molecules (bottom right).
In each rotational diagram, the squares, circles, and triangles present the region of the inner-CND, outer-CND, and star-forming ring, respectively.
}
 \label{fig:diagram}
\end{figure*}

The rotational temperatures ($T_{\rm rot}$) and column densities ($N_{\rm mol}$) of HCN, HCO$^+$, and CS can be estimated from the rotational diagrams, assuming optically thin LTE. 
The rotational diagram is the  logarithmic plot of the normalized column densities with the statistical weight of the level as a function of energy levels corresponding to the transitions 
(cf. \cite{izumi2013}).
To construct the rotational diagram of each observed molecule, i.e., HCN, HCO$^+$, and CS, 
the complete data in Bands~3 and 7 were convolved to the beam size of $0\farcs72\times0\farcs72$.
As described in figure~\ref{fig:rad}, we divided the central region of NGC~613 into three regions: the inner CND ($r \leq 0\farcs36$),  outer CND ($0\farcs36\leq r \leq 1\farcs08$), and star-forming ring ($2\farcs88\leq r \leq 3\farcs96$).
The resultant diagrams of HCN, HCO$^+$, and CS are shown in figure~\ref{fig:diagram}, and 
the derived rotational temperatures and column densities of the molecules in the inner and outer CNDs and in the star-forming ring are summarized in table~\ref{tab:diagram}.
While the rotational temperatures and column densities of all molecules gradually decrease with the distance from the nucleus, 
the temperature of CS is slightly higher than those of HCN and HCO$^+$ (bottom-right panel of figure~\ref{fig:diagram}). 
Considering that the temperature of CS is high not only in the inner CND but also in the outer region, 
the absorption of CS(2--1) in the CND (figure~\ref{fig:spe}) does not play a major role in the temperature.
Note that $T_{\rm rot}$ and $N$ were derived by fitting only two transitions with approximately one order of magnitude energy differences.
By utilizing  three transition lines ($J=1$--0, 3--2, and 4--3) of HCN and HCO$^+$, 
\citet{izumi2013} suggested  the presence of two temperature gas components in the CND of NGC~1097: one is responsible for the low-$J$ ($J\leq3$) emission and the other for the high-$J$ ($J\geq 3$) emission.
In addition, observations toward the center of nearby galaxies in multiple transitions of CS by using a single dish have presented the existence of two temperature gas components
(e.g., \cite{bayet}, \cite{aladro}).
CS(7--6), whose energy level is  higher than those of HCN(4--3) and HCO$^+$(4--3), may trace the higher temperature components selectively. 
To examine the presence of two temperature components and the effect of the warm gas on the higher rotational temperature of CS, it is necessary  to obtain the data of adjoining transitions.

\subsection{The HCN/HCO$^+$ and HCN/CS ratios}
\label{sec:ratio}
\begin{figure}[t]
\begin{tabular}{cc}
	\begin{minipage}{0.45\hsize}
\par
	\begin{center}
{\bf (a)}
		\FigureFile(70mm,150mm){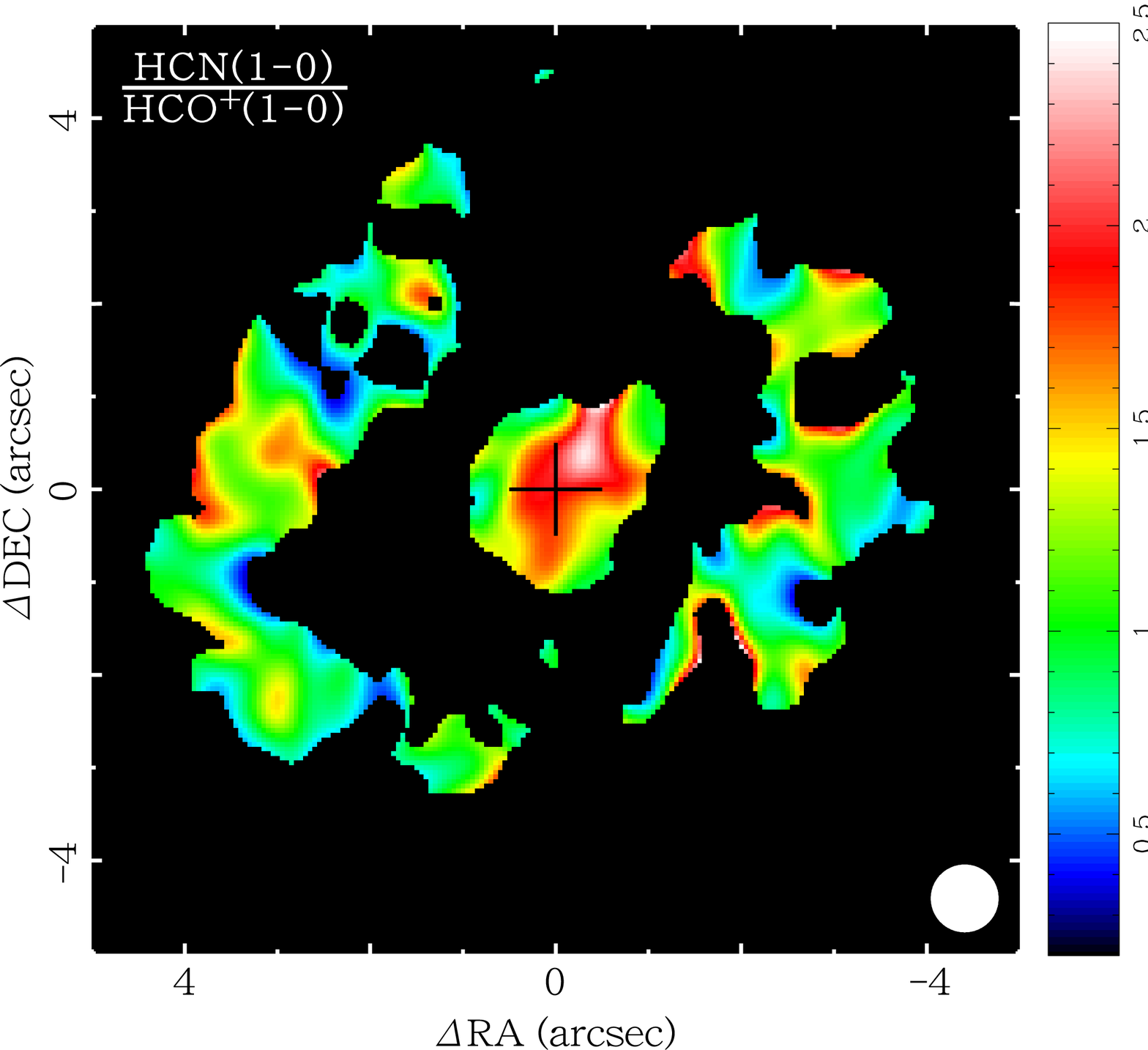}
	 \end{center}
	 \end{minipage}
	 \begin{minipage}{0.45\hsize}
\par
	\begin{center}
{\bf (b)}
		\FigureFile(70mm,150mm){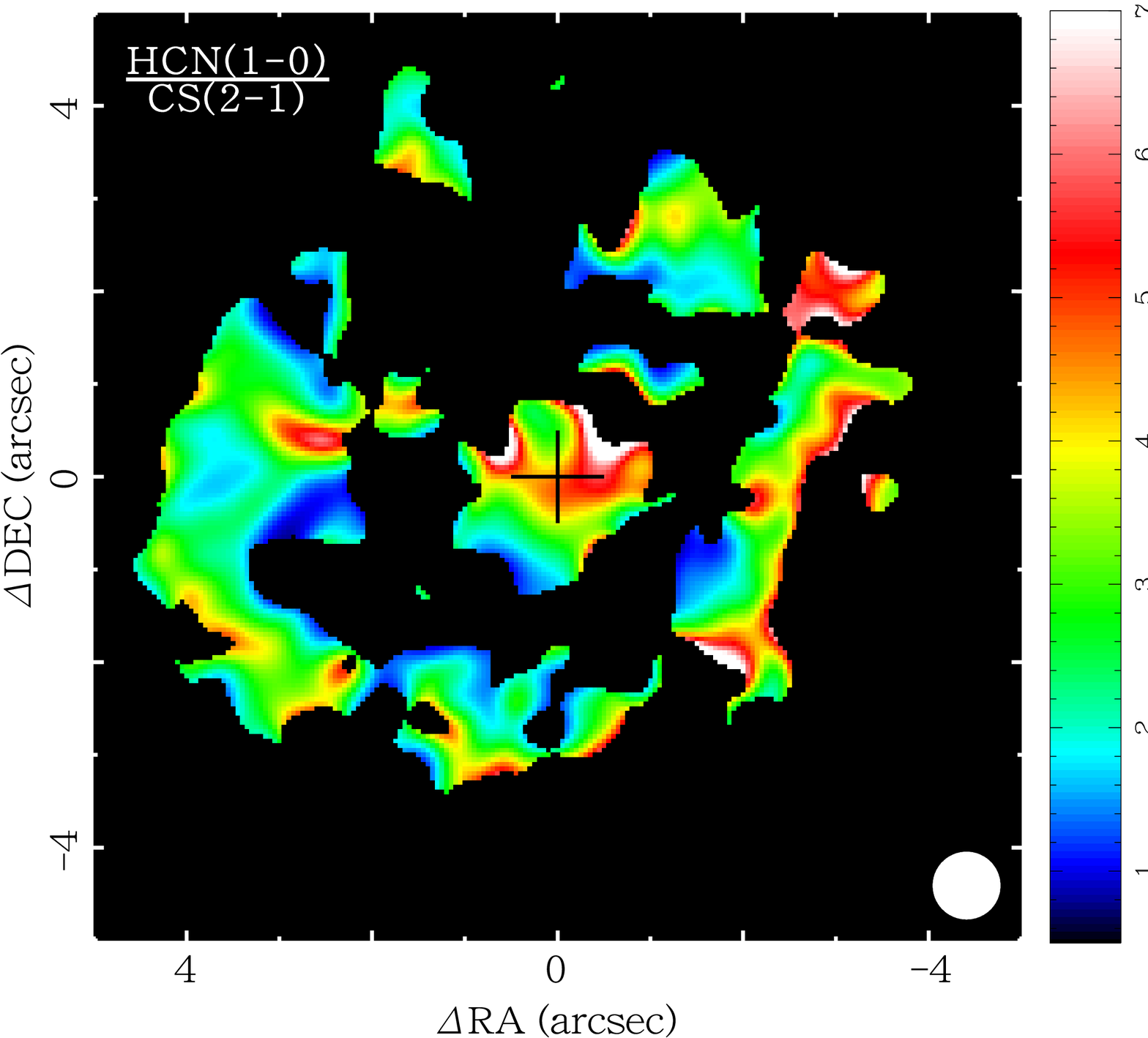}
	 \end{center}
	 \end{minipage}\\
 \end{tabular}
\caption{
Integrated intensity ratios of (a) HCN(1--0)-to-HCO$^+$(1--0) and (b)  HCN(1--0)-to-CS(2--1). 
The individual intensity maps are convolved with a beam of 0\farcs72, and 
values only above 4$\sigma$ in the individual  maps have been used. 
}
 \label{fig:ratio}
\end{figure}

Figure~\ref{fig:ratio} shows the integrated intensity ratio maps of HCN(1--0)/HCO$^+$(1--0) and HCN(1--0)/CS(2--1).
We convolved each integrated intensity map to a beam of $0\farcs72$ and
only picked points above a 4~$\sigma$ level in each integrated intensity map.
Both the ratios of  HCN/HCO and HCN/CS in the center region are higher than those in the ring, 
even considering the high ratio at the edge of the ring affected by the low S/N.
However, note that the peak of the ratio is offset from the center to the northern west ($r\sim0\farcs5-0\farcs6$). 
The peak position likely corresponds to a region of high velocity dispersion (section~\ref{sec:outflow} and figure~\ref{fig:out}). 

By dividing the central region of NGC~613 into an annuli with a width of $\Delta r=0\farcs72$ from the galactic center ($r\leq0\farcs36$),  
we investigated the variances of HCN(1--0)/HCO$^+$(1--0) and HCN(1--0)/CS(2--1) ratios as a function of the radius.
The parameters used to determine the ellipses were {\it P.A.}$=\timeform{118D}$ and $i=\timeform{46D}$.
The ratios of the integrated intensities  were azimuthally averaged  in the inner CND ($r\leq0\farcs36$), outer CND ($0\farcs36 \leq r \leq 1\farcs08$), and star-forming ring ($2\farcs88 \leq r \leq 3\farcs96$). 
The ratio of HCN/HCO$^+=1.9\pm0.1$ at the inner CND decreases gradually with the radius,  $1.6\pm0.2$ in the outer CND and $1.3\pm0.4$ in the ring.
Similarity,  HCN/CS shows radial decrease, 
$4.6\pm0.5$ in the inner CND, $4.1\pm1.5$ in the outer CND, and $2.9\pm0.7$ in the ring.
It is expected that 
the optical depths of these lines 
are high ($\tau\gtrsim$1; e.g., \cite{jimenez}), as supported by the dips of the spectra at the center (figure~\ref{fig:spe}). 
For the optically thick case, as the critical density is lower than its nominal value by a factor equal to the optical depth (\cite{shirley}, \cite{jimenez}), 
these lines can be thermalized.
Under these conditions, the intensity ratios  mainly depend on the abundance ratio. 
Actually, the variations of the abundance ratios of $N$(HCN)/$N$(HCO$^+$) and $N$(HCN)/$N$(CS) from the inner CND to the ring (table~\ref{fig:diagram}) can be found to correspond to the variation of the  intensity ratios.

\begin{figure}[h]
\begin{tabular}{cc}
	\begin{minipage}{0.45\hsize}
\par
	\begin{center}
{\bf (a)}
		\FigureFile(70mm,150mm){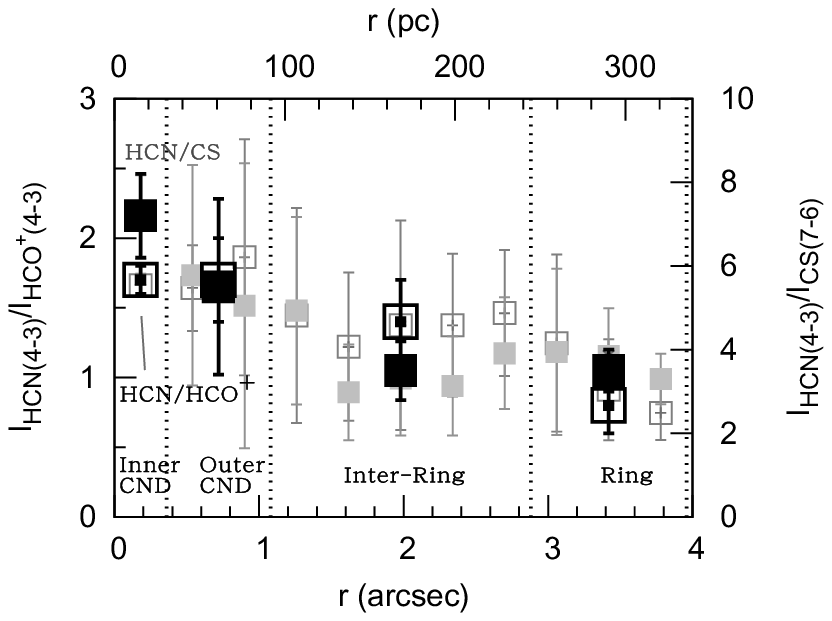}
	 \end{center}
	 \end{minipage}
	 \begin{minipage}{0.45\hsize}
\par
	\begin{center}
{\bf (b)}
		\FigureFile(70mm,150mm){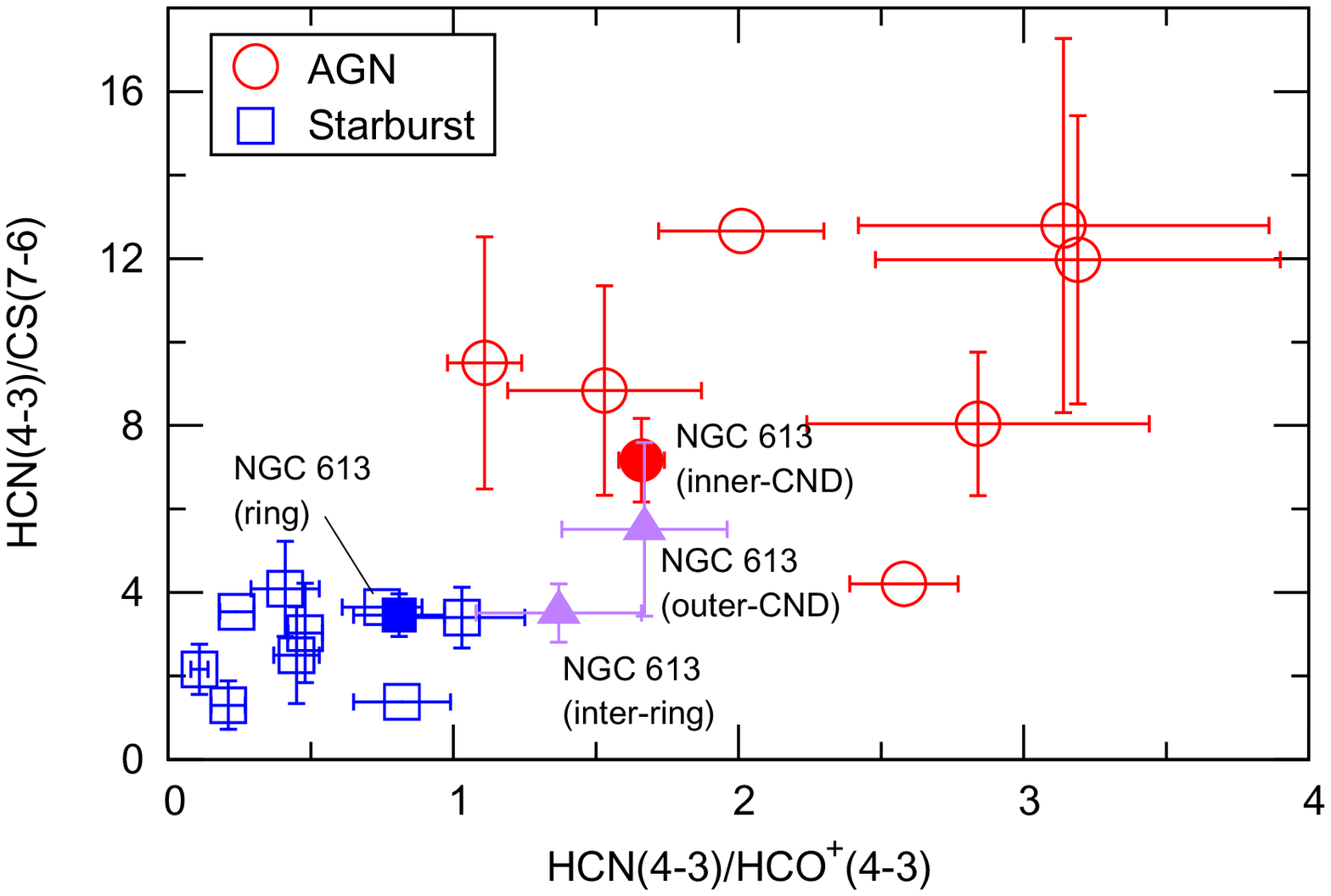}
	 \end{center}
	 \end{minipage}\\
 \end{tabular}
\caption{
(a) The ratios of the integrated intensities of  HCN/HCO$^+$ (open gray square) and HCN/CS (filled gray square) in annuli with a width of $\Delta r=0\farcs36$.
Open and filled black squares represent 
azimuthal averages of the integrated intensity ratios of HCN/HCO$^+$ and HCN/CS, respectively, in the inner CND ($r\leq0\farcs36$), outer CND ($0\farcs36 \leq r \leq 1\farcs08$), star-forming ring ($2\farcs88 \leq r \leq 3\farcs96$), and inter-ring ($1\farcs08 \leq r \leq 2\farcs88$).
Vertical dot lines correspond to $r=0\farcs36$, $1\farcs08$, $2\farcs88$ and $3\farcs96$. 
(b) Diagram of HCN(4--3)/HCO$^+$(4--3) and HCN(4--3)/CS(7--6) integrated intensity ratios. The open circles and  squares indicate  AGNs and starbust galaxies shown in figure~1 of \citet{izumi2016}.
The filled circles and  squares show the ratios in the inner-CND and star-forming ring of NGC~613, respectively.
The filled triangles represent the ratios in the outer-CND and inter-ring for comparison.
}
 \label{fig:izumi}
\end{figure}

For Band~7 data, 
we divided the central region of NGC~613 into annuli with widths of $\Delta r=0\farcs36$  from the  center ($r\leq0\farcs18$). 
The integrated intensities were convolved to a beam of $0\farcs36$.
In a manner similar to Band~3, 
the ratios of the integrated intensities  were azimuthally averaged  in the inner CND, outer CND, star-forming ring, and the intermediate region between the CND and ring (the inter-ring; $1\farcs08 \leq r \leq 2\farcs88$). 
Figure~\ref{fig:izumi}(a) shows the variances of the HCN(4--3)/HCO$^+$(4--3) and HCN(4--3)/CS(7--6) ratios as a function of the radius.
The ratios of both HCN/HCO$^+$ and HCN/CS in the CND decrease gradually with the radius, as is the case in Band~3.  
We plotted the HCN(4--3)/HCO$^+$(4--3) and HCN(4--3)/CS(7--6) ratios  in each region on  
the submillimeter-HCN diagram developed by \citet{izumi2016} [figure~\ref{fig:izumi}(b)].
From the inner CND through outer CND,  the HCN(4--3)/CS(7--6) ratio slightly decreases in spite of the constant ratio of HCN(4--3)/HCO$^+$(4--3). 
The ratios of HCN(4--3)/HCO$^+$(4--3) and  HCN(4--3)/CS(7--6) in the outer CND and inter-ring locate at the border between the AGN and starburst regimes, and the ratios in the  star-forming ring are consistent with those of starburst galaxies. 
The relatively low HCN/HCO$^+$ in the ring could be caused by the HCO$^+$ enhancement 
because HCO$^+$ is abundant in strong UV fields, i.e., massive star formation (e.g., \cite{bayet2011, meijerink2011}). 
By adopting the column densities of CS and HCN derived in section~{\ref{sec:lte}} and a line width of $\sim200$~km~s$^{-1}$ (figure~\ref{fig:spe_band7}), 
we estimated the ratio of line-of-sight molecular column density to the velocity width as $N_{\rm CS}/dV (\approx N_{\rm HCN}/dV)\sim2\times10^{11}$~(cm$^{-2}$~km~s$^{-1})$. 
Under these conditions, 
it is expected that both CS(7--6) and HCO$^+$ are optically thin ($\tau<<1$), 
and the variations of the integrated intensity ratios of HCN(4--3)/CS(7--6) and HCN(4--3)/HCO$^+$(4--3) in the regions  depend on the kinetic temperature $T_{\rm k}$,  volume density of H$_2$ $n_{\rm H_2}$ and abundance ratio of CS and HCO$^+$ relative to HCN \citep{izumi2016}. 

\subsection{Non-LTE analysis of HCN, HCO$^+$, and CS}
\label{sec:non-lte}
We investigated the difference of the physical parameters ($T_{\rm kin}$, $n_{\rm H_2}$, and $N_{\rm H_2}$) in the CND ($r\lesssim0\farcs36$) and ring ($2\farcs88\lesssim r \lesssim3\farcs96$) by utilizing the intensity ratios of  HCN, HCO$^+$, and CS through the non-LTE radiative transfer code RADEX  \citep{vandertak}.
RADEX uses an escape probability approximation to solve the non-LTE excitation, 
assuming that all lines radiate from the same region.
We adopted a uniform spherical geometry 
($dV=15$~km~s$^{-1}$)
and evaluated the minimum residuals between the observed line and modeled intensity ratios.
Intensity ratio is described as 
$R^{\rm mol}_{J_{\rm u}J_{\rm l}/J'_{\rm u}J'_{\rm l}}=I^{\rm mol}_{J_{\rm u}J_{\rm l}}/I^{\rm mol}_{J'_{\rm u}J'_{\rm l}} $, 
where $J_{\rm u}$ and $J_{\rm l}$ are the upper and lower rotational transitions and 
$I^{\rm mol}_{J_{\rm u}J_{\rm l}}$ is the integrated intensity of a given molecule at transition ($J_{\rm u}\rightarrow J_{\rm l}$), 
as shown by \citet{krips2008}.
In our analysis, we convolved the individual integrated intensity maps with a resolution of 0\farcs72, and 
used the averaged intensity ratios of 
$R^{\rm HCN}_{43/10}=0.44\pm0.07$, 
$R^{\rm HCN/HCO^+}_{43/43}=1.41\pm0.24$,
and $R^{\rm HCN/CS}_{43/76}=5.70\pm0.97$ at the CND,
and 
$R^{\rm HCN}_{43/10}=0.10\pm0.05$, 
$R^{\rm HCN/HCO^+}_{43/43}=0.75\pm0.44$, 
and $R^{\rm HCN/CS}_{43/76}=4.62\pm2.60$  in the ring.
All line specifications (spectroscopic data and collisional excitation rates) were taken from the Leiden Atomic and Molecular Database (LAMDA: \cite{lamda}). 
For our non-LTE analysis, 
we varied the gas kinetic temperature within a range of $T_{\rm k}=10$--700~K with steps of 
$\Delta T_{\rm k}=10$~K, 
the H$_2$ column density of $N_{\rm H_2}=10^{18}$--$10^{25}$~cm$^{-2}$ with steps of $\Delta N_{\rm H_2}=10^{1.0}$~cm$^{-2}$, 
and gas density of $n_{\rm H_2}=10^3$--$10^9$~cm$^{-3}$ with steps of $\Delta n_{\rm H_2}=10^{0.25}$~cm$^{-3}$. 
Our LTE analysis  shows the comparable abundance of HCO$^+$ and CS with respect to HCN (table~\ref{tab:diagram});  
it is compatible with the results in the central region of NGC~1097 \citep{martin2015}, 
and hence we assumed that the relative abundance ratios [HCN]/[HCO$^+$] are the same as [HCN]/[CS]. 
We considerably changed the initial parameters of the relative abundance ratios,  
${\rm [HCN]/[CS]}$ and ${\rm [HCN]/[HCO^+]}$ to 1, 5, 10, 20, 30, 40, and 50, 
where ${\rm [HCO^+]/[H_2]=1.0\times10^{-9}}, 5.0\times10^{-9}$ (e.g., \cite{paglione}, \cite{martin2006}), 
and the background temperature $T_{\rm bg}=2.7,10,20,30,40,50$~K.

\begin{figure}[h]
\begin{tabular}{cc}
	\begin{minipage}{0.45\hsize}
\par
	\begin{center}
{\bf (a)}
		\FigureFile(70mm,150mm){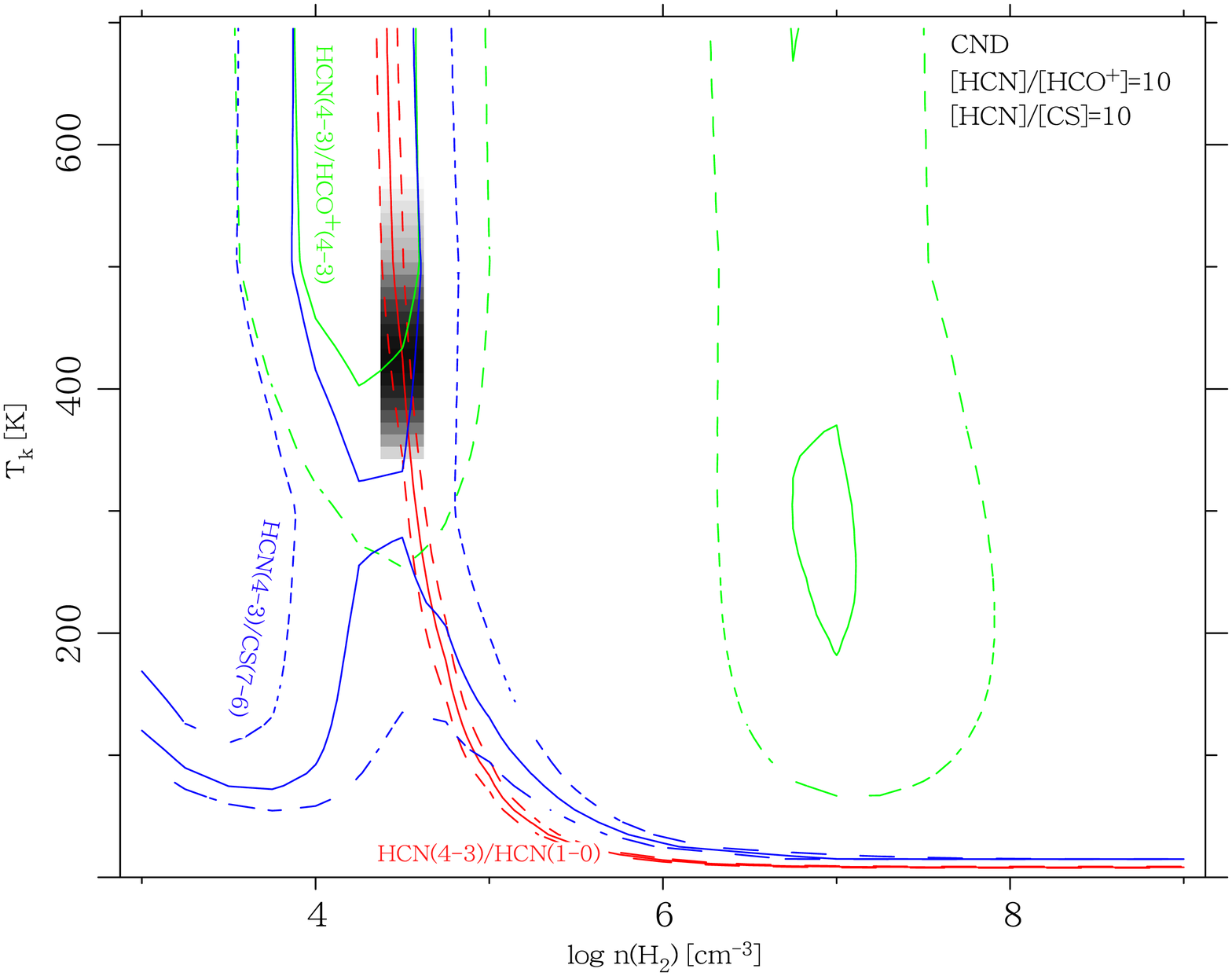}
	 \end{center}
	 \end{minipage}
	 \begin{minipage}{0.45\hsize}
\par
	\begin{center}
{\bf (b)}
		\FigureFile(70mm,150mm){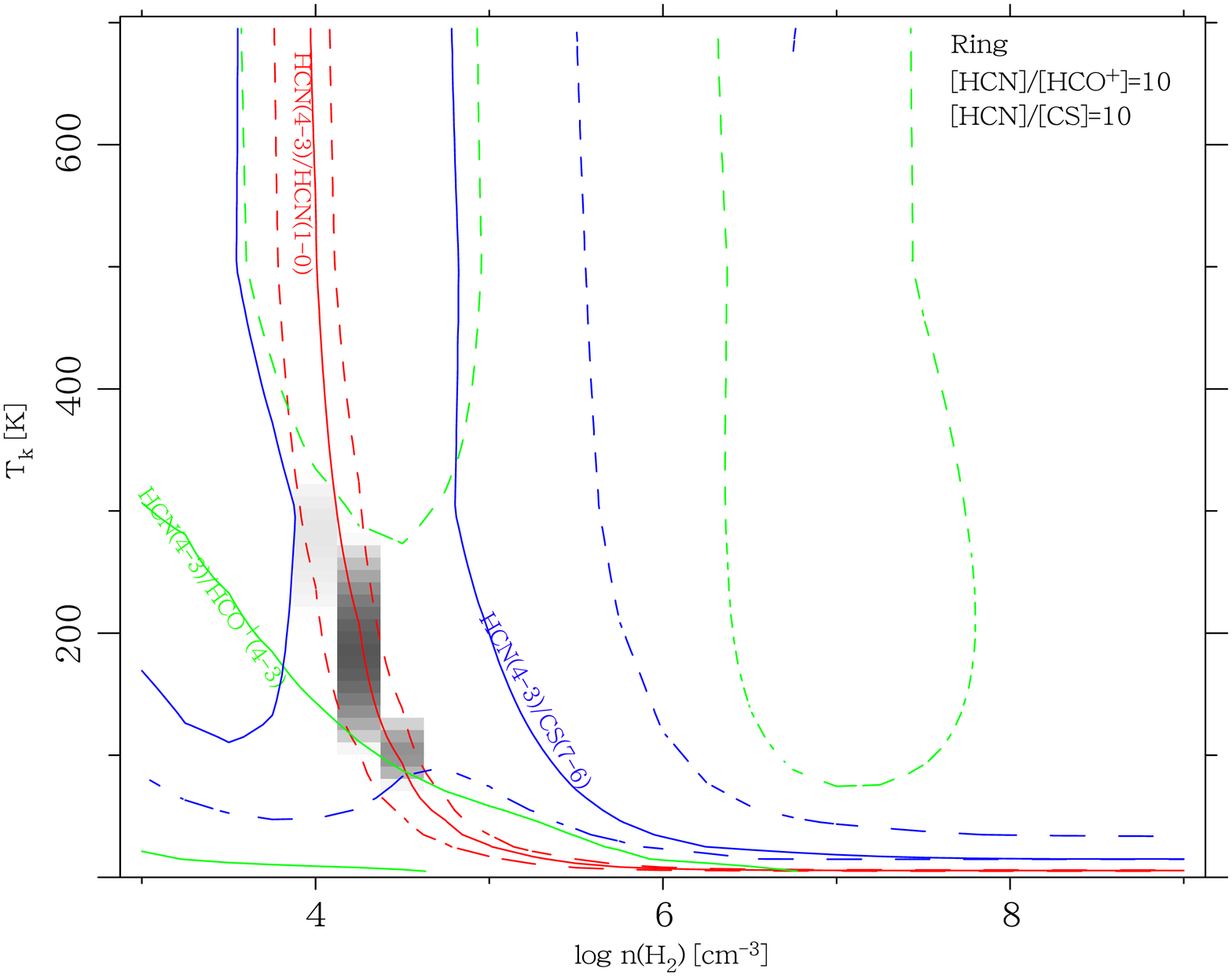}
	 \end{center}
	 \end{minipage}\\
 \end{tabular}
\caption{
Representative results of the $\chi^2$ test with RADEX to search for the best parameter sets to explain the observed ratios of HCN(4--3)/HCN(1--0) (red), HCN(4--3)/HCO$^+$(4--3) (green), and HCN(4--3)/CS(7--6) (blue) in (a) the CND and  (b) the star forming ring. 
The  parameters of the molecular hydrogen column density of $N_{\rm H_2} = 10^{22}$~cm$^{-2}$, 
 abundance ratios of HCO$^+$ and CS to H$_2$  
${\rm [HCO^+]/[H_2]}$ (and  
${\rm [CS]/[H_2]) =5.0\times10^{-9}}$, 
and abundance ratios of HCN to HCO$^+$ and CS 
${\rm [HCN]/[HCO^+]}$ (and  
${\rm [HCN]/[CS]})=10$ 
 were adopted both in the CND and ring. 
The dashed lines are the $1\sigma$ error of each track. 
The background gray scale indicates the  value of $\chi^2<1$. 
The best-fitted parameters achieved using these three tracks are 
$(n_{\rm H_2}, T_{\rm kin}) = (\sim10^{4.5}$~cm$^{-3}$, 350--550~K) at the  CND and 
$(n_{\rm H_2}, T_{\rm kin}) = (\sim10^{4-4.5}$~cm$^{-3}$, 80--300~K) in the star-forming ring.
}
 \label{fig:radex}
\end{figure}

Under these conditions, we applied RADEX and conducted a $\chi^2$ test, 
that is, $\chi^2=[\{R^{\rm mol}_{J_{\rm u}J_{\rm l}/J'_{\rm u}J'_{\rm l}}(\rm obs)-R^{\rm mol}_{J_{\rm u}J_{\rm l}/J'_{\rm u}J'_{\rm l}}(\rm model)\}/\sigma R^{\rm mol}_{J_{\rm u}J_{\rm l}/J'_{\rm u}J'_{\rm l}}(\rm obs)]^2$, where 
$\sigma R^{\rm mol}_{J_{\rm u}J_{\rm l}/J'_{\rm u}J'_{\rm l}}(\rm obs)$ is the standard deviation of the observed  intensity ratio to search for the best parameters to explain the observed intensities.
We confirmed that reasonable solutions based on the $\chi^2$ test could not be obtained when $T_{\rm bg}\geq10$~K.
Figures~\ref{fig:radex}(a) and (b) show the resultant best-fitted parameters in the CND and ring, respectively.
The dashed lines in figure~\ref{fig:radex} are the $\pm1\sigma$ error of each ratio, and 
the dark regions indicate the regions where $\chi^2<1$.
The best-fitted parameters with $\chi^2\lesssim1$ at the CND 
were obtained to be $T_{\rm k}=$350--550~K, 
$n_{\rm H_2}\sim10^{4.5}$~cm$^{-3}$, 
and $N_{H_2}=10^{22}$~cm$^{-2}$ 
when ${\rm [HCN]/[HCO^+]=10}$ and ${\rm[HCN]/[CS]=10}$. 
In the ring, the best-fitted parameters 
with $\chi^2\lesssim1$ were
$T_{\rm k}=80$--300~K, 
$n_{\rm H_2}\sim10^{4.5-5}$~cm$^{-3}$,
and $N_{H_2}=10^{22}$~cm$^{-2}$ 
when ${\rm [HCN]/[HCO^+]=10}$ and ${\rm [HCN]/[CS]=10}$. 
The derived column densities are comparable  to that derived from the CO(3--2) flux: $N_{\rm H_2}=0.9\times10^{22}$~cm$^{-2}$. 
The discrepancy in the abundance ratios of HCO$^+$ and CS with respect to HCN between the LTE analysis (${\rm[HCN]/[HCO^+]\approx[HCN]/[CS]}\sim3$) and non-LTE analysis (${\rm[HCN]/[HCO^+]=[HCN]/[CS]}=10$) can be caused by the subthermal emission for high transition lines, 
i.e., the density $n_{\rm H_2}\sim10^{4-4.5}$~cm$^{-3}$ is not high enough to thermalize these lines ($n^{\rm crit}_{\rm H_2}\sim10^{6-7}$~cm$^{-3}$).
Therefore, the high abundance ratios ${\rm[HCN]/[HCO^+]}$ and ${\rm [HCN]/[CS]}$ are needed to satisfy the observed line ratios.
In addition,  NGC~1097,  
which shows the similar discrepancy between the abundance ratio estimated through LTE  and non-LTE analyses, has been reported (\cite{martin2015},  \cite{izumi2016}).
By adopting the intensity ratios based on HCO$^+$,  
i.e., $R^{\rm HCO+}_{43/10}=0.60\pm0.10$ (CND) and $0.15\pm0.06$ (ring),  
$R^{\rm HCO+/HCN}_{43/43}=0.71\pm0.12$ (CND) and $1.33\pm0.78$ (ring), and 
$R^{\rm HCO+/CS}_{43/76}=4.04\pm0.70$ (CND) and $6.26\pm3.20$ (ring),  
in the analysis, 
we derived the best-fitted parameters $T_{\rm k}\sim$300--400~K, 
$n_{\rm H_2}\sim10^{4.5-5}$~cm$^{-3}$, 
and $N_{H_2}=10^{22}$~cm$^{-2}$ in the CND 
and 
$T_{\rm k}\sim$100--300~K, 
$n_{\rm H_2}\sim10^{4-4.5}$~cm$^{-3}$, 
and $N_{H_2}=10^{22}$~cm$^{-2}$ in the ring; 
these are consistent with the previous results based on HCN.
Although the  parameters in the ring could not be determined rigidly because of low S/N of the ratios, 
the kinematic temperature in the CND ($>300$~K) was clearly higher than that in the ring ($\sim80$--300~K). 
In addition, these results suggest that the variations of intensity ratios HCN/HCO$^+$ and HCN/CS in the CND and ring (section~\ref{sec:ratio})  mainly depend on the kinetic temperature  rather than on $n_{\rm H_2}$ and on the enhancement of the abundance ratio of HCO$^+$ and CS relative to HCN. 

\onecolumn
\twocolumn




  \onecolumn
\begin{table*}
  \tbl{Surface densities of molecular gas, star-formation rates (SFR) and star-formation efficiencies (SFE) at the nucleus and the spots in the ring.}{%
  \begin{tabular}{cccc}
\hline
Positions & $\Sigma_{\rm H_2}$~[\MO~pc$^{-2}$] & SFR~[\MO~yr$^{-1}$]\footnotemark[$*$] &SFE~[$\times 10^{-8}$~yr$^{-1}$]\\
\hline
Nucleus (CND)	&	$5.1\times10^2$ &	0.02 &	0.7	\\
Spot~1	&	$2.0\times10^2$ &	0.03 &	2.7	\\
Spot~2	&	$1.0\times10^2$ &	0.04 &	7.1	\\
Spot~3	&	$2.1\times10^2$ &	0.13 &	10.9	\\
Spot~4	&	$2.1\times10^2$ &	0.10 &	8.4	\\
Spot~5	&	$1.2\times10^2$ &	0.10 &	14.7	\\
Spot~6	&	$1.6\times10^2$ &	0.08 &	8.8	\\
Spot~7	&	$1.8\times10^2$ &	0.06 &	5.9	\\
\hline
\end{tabular}}
\label{tab:sfe}
\begin{tabnote}
\footnotemark[$*$]~\citet{falcon}
\end{tabnote}
  \end{table*}
\begin{table*}
  \tbl{Fitted parameters of gas in the north side of the CND}{%
  \begin{tabular}{ccccccc}
\hline
Positions 	 &Component	 & $F_{\rm CO32}$~(mJy) & Velocity~(km~s$^{-1}$) & $\Delta v_{\rm FWHM}$~(km~s$^{-1}$)&$S_{\rm CO32}$~(Jy~km~s$^{-1}$) &$M_{\rm H_2}$~($10^4~$\MO)\footnotemark[$*$] \\
\hline
N~(\timeform{0"}, 0\farcs{5})&Blue 		 &$38.6\pm2.5$	 &$1390.2\pm0.8$	&$39.8\pm2.4$	&	$1.6\pm0.1$	&	$5.5\pm0.5$\\
~					&Wide		 &$23.7\pm4.8$	 &$1468.5\pm4.4$	&$224.8\pm20.2$&	$5.7\pm1.3$	&	$19.1\pm4.2$\\
~					&Narrow		 &$112.7\pm4.5$	 &$1482.1\pm0.4$	&$76.8\pm2.1$	&	$9.2\pm0.4$	&	$31.0\pm1.5$\\
\hline
N~(\timeform{0"}, 0\farcs{75})	&Blue	 &$26.0\pm1.6$	 &$1401.7\pm0.9 $	&$37.2\pm2.9$	&	$1.0\pm0.1$	&	$3.5\pm0.3$\\
~						&Wide	 &$34.6\pm1.7$	 &$1467.5\pm2.3$	&$163.6\pm4.3$&	$6.0\pm0.3$	&	$20.3\pm1.1$\\
~						&Narrow	 &$16.7\pm1.8 $	 &$1474.2\pm1.2$	&$31.0\pm4.3$	&	$0.6\pm0.1$	&	$1.9\pm0.3$\\
\hline
NW~(-0\farcs{5}, 0\farcs{5})	&Red	 &$86.5\pm8.5$	 &$1526.5\pm3.7$	&$53.0\pm4.7$&	$4.9\pm0.6$	&	$16.4\pm2.2$\\
~						&Wide	 &$35.4\pm3.1$	 &$1477.1\pm3.4$	&$169.9\pm7.1$&	$6.4\pm0.6$	&	$21.6\pm2.1$\\
~						&Narrow	 &$54.2\pm11.3$	 &$1483.8\pm5.3$	&$50.2\pm6.2$	&	$2.9\pm0.7$	&	$9.8\pm2.4$\\
\hline
NW~(-0\farcs{5}, 0\farcs{75})	&Red	 &$33.5\pm2.1$	 &$1528.5\pm2.7$	&$73.8\pm7.7$	&	$2.6\pm0.3$	&	$8.9\pm1.1$\\
~						&Wide	 &$24.9\pm1.7$	 &$1470.9\pm5.8$	&$193.8\pm7.6$&	$5.1\pm0.4$	&	$17.3\pm1.4$\\
~						&Narrow	 &$28.6\pm3.2$	 &$1479.6\pm1.0$	&$28.9\pm3.1$	&	$0.9\pm0.1$	&	$3.0\pm0.5$\\
\hline
\end{tabular}}
\label{tab:out}
\begin{tabnote}
\footnotemark[$*$]~Molecular gas mass is derived from equation~(\ref{eq:mass}).
\end{tabnote}
  \end{table*}
\twocolumn

\subsection{Activity of the central region of NGC~613}
\begin{figure*}[th]
 \begin{center}
  \includegraphics[width=\linewidth]{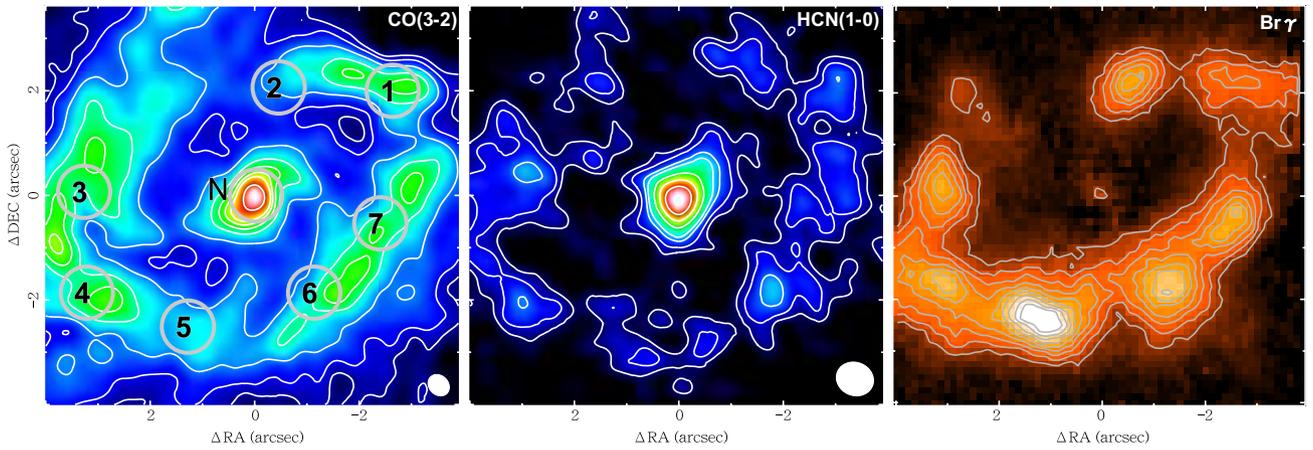}
 \end{center}
 \caption{Enlarged integrated intensity maps of  our (a) CO(3--2) and (b) HCN(1--0) to match (c) Br$\gamma$ image with VLT  \citep{falcon}.
 The circles mark the positions of the nucleus and spots~1 thorough 7 denoted by \citet{falcon}. 
 }
 \label{fig:brg}
\end{figure*}
\subsubsection{Low star-formation efficiency at CND}
We found abundant  gas in the CND relative to the ring, 
while \citet{falcon} showed lower star-formation rate (SFR) in the CND, 0.02~\MO~yr$^{-1}$, than those at the spots in the ring, 0.03--0.13~\MO~yr$^{-1}$, 
by using Br$\gamma$  with SINFONI on the Very Large Telescope. 
Figure~\ref{fig:brg} shows that in the CND, the molecular line intensities are strong, 
but the Br$\gamma$ intensity is rather weak, 
where Br$\gamma$ is produced through photoionization in the Str\"{o}mgren spheres around O- or B-type stars 
and its intensity is an indicator of SFR.
Table~\ref{tab:sfe} shows the star-formation efficiency (SFE) defined as the ratio of the SFR surface density to the surface mass density of molecular hydrogen at each position, 
where $R_{32/10}=2$ is adopted.
The SFE in the CND is almost an order of magnitude lower than those at the spots in the star-forming ring. 
By applying $R_{32/10}=0.25$ in the ring because  $R_{32/10}$ decreases with radius (section~\ref{sec:mass}), the difference in the SFEs between the CND and ring is increased further.
The weak intensity of C$_2$H, which is a tracer of massive star formation, also indicates inactive star formation in the CND (section~\ref{sec:band3}).

The much lower SFE in the CND cannot be caused by the underestimation of the SFR or uncertainty of the estimated surface density of molecular gas.
Br$\gamma$ is affected by dust extinction,
although lesser than H$\alpha$.
\citet{falcon} determined the amount of extinction at the spots in the ring from the Brackett decrement Br$\gamma /$Br10 
to be the visual extinction of $A_{\rm V}=6.50$--15.27 at the spots.   
In contrast, the extinction at the CND $A_{\rm V}=3.62$ was estimated using the color excess $E(H-K)$ and CO line strength index  due to the low S/N of the Br lines.
Even if the difference of the two measurements is considered, generally 2--5, 
the SFE  at the CND remains lower than those at the spots.
The uncertainty of the surface density of molecular gas is mainly caused by  $R_{32/10}$ and $X_{\rm CO}=N({\rm H_2})/I_{\rm CO}$.
$R_{32/10}(=2)$  may be overestimated in the ring and $X_{\rm CO}[=0.5\times10^{20}$~cm$^{-2}$~(K~km~s$^{-1})^{-1}]$ is expected to increase with the radius (e.g., \cite{nakai}, \cite{bolatto2013}). 
Given that 
$R_{32/10}$ in the CND is four times higher than that in the ring (e.g., \cite{tsai2012}) 
and $X_{\rm CO}$ in the center is $\sim0.3$~dex lower than that in the galactic disk (e.g., \cite{sandstorm}), 
SFEs in the CND and the spots change from the values in table~\ref{tab:sfe} to $7.8\times10^{-5}$~yr$^{-1}$ and $1.2\pm0.5\times10^{-4}$~yr$^{-1}$, respectively.
SFE at the CND  is still lower than those at the spots.

\subsubsection{Heating source of molecular gas in CND}
We determined that the temperature of the molecular gas in the CND ($T_{\rm rot}\sim12$~K, $T_{\rm k}>300$~K) is higher than that in the ring ($T_{\rm rot}\sim8$~K, $T_{\rm k}\sim80-300$~K), 
and the ratios of the HCN(4--3)/HCO$^+$(4--3) and HCN(4--3)/CS(7--6) in the CND  locate at the AGN domain (figure~\ref{fig:izumi}) except the starburst.
In addition to the findings that Br$\gamma$ emission in the CND is weaker than that in the ring  \citep{falcon},
\citet{asmus}  demonstrated that 
the nuclear MIR flux of NGC~613 is considerably lower ($\sim6\%$) than that in the ring, 
marginally resolving the nucleus  with T--ReCS in the N-filter mounted on Gemini-South  ($\theta_{\rm FWHM}\sim\timeform{0.39"}$).
These results indicate that the star-formation activity is an unlikely dominant heating source for the hot gas in the CND.

\citet{goulding} identified that NGC~613 hosts AGN, based on the detection of an emission line  [Ne\emissiontype{V}]~${\rm \lambda14.32\mu m}$
by using $Spitzer/$IRS ($\timeform{4.7"}\times\timeform{11.3"}$ for $\lambda\sim$9.9--19.6$\mu$m).
However, by using the ratios among emission lines, 
e.g., [Ne\emissiontype{V}],  [Ne\emissiontype{II}], and [O \emissiontype{III}], and IR luminosity, 
they suggested that the activity of the central region of NGC~613 is dominated by star formation (AGN contribution $<5\%$).
\citet{falcon} discussed about the heating source of molecular gas at the CND,  
and concluded that the gas cannot be excited through photoionization by UV and soft X-ray radiation but through shock by the nuclear jets, 
by comparing the flux ratios of [Fe\emissiontype{II}]/Br$\gamma$, H$_2$(1--0)~S(1)/H$_2$(2--1)~S(1), and H$_2$(1--0)~S(2)/H$_2$(1--0)~S(0) in the CND with those in the ring and theoretical model.
The distribution of the  steep spectral index derived from 4.9 and 95~GHz around the center $(\alpha\lesssim-0.6$, figure~\ref{fig:cont}) and the feature of outflowing  gas around the CND in the PV-diagram (figure~\ref{fig:pv_diagram}) are compatible with the mechanical heading by the jets.

\subsubsection{Outflow of molecular gas in CND}
\label{sec:outflow}
\begin{figure}[h]
\begin{center}
\begin{tabular}{ccc}
	 \begin{minipage}{0.4\hsize}
\par
{\bf (a)}
		\FigureFile(60mm,150mm){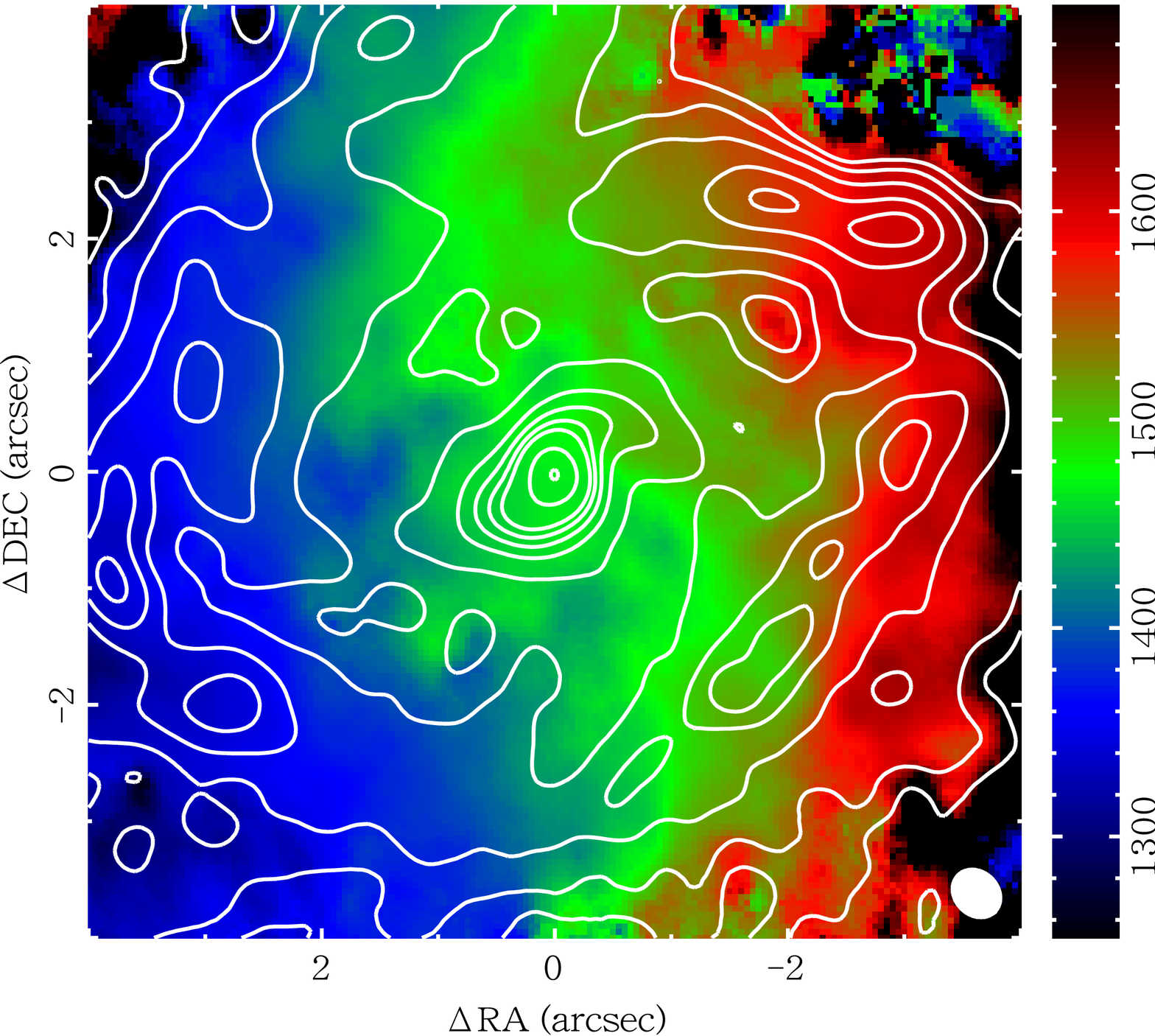}
	 \end{minipage}
	 \begin{minipage}{0.4\hsize}
\par
{\bf (b)}
		\FigureFile(60mm,150mm){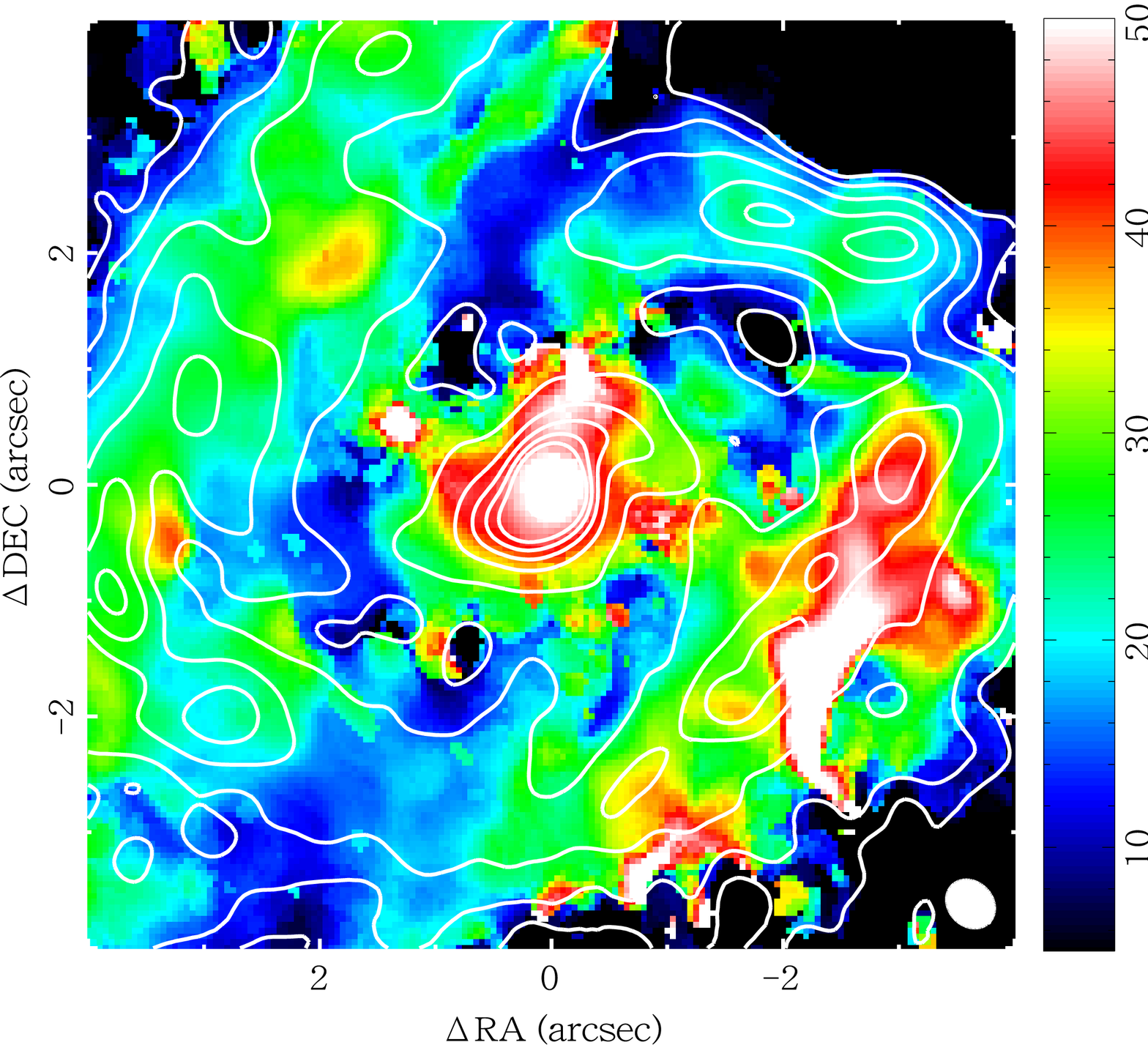}
	 \end{minipage}\\
	 \begin{minipage}{0.4\hsize}
\par
{\bf (c)}
		\FigureFile(60mm,150mm){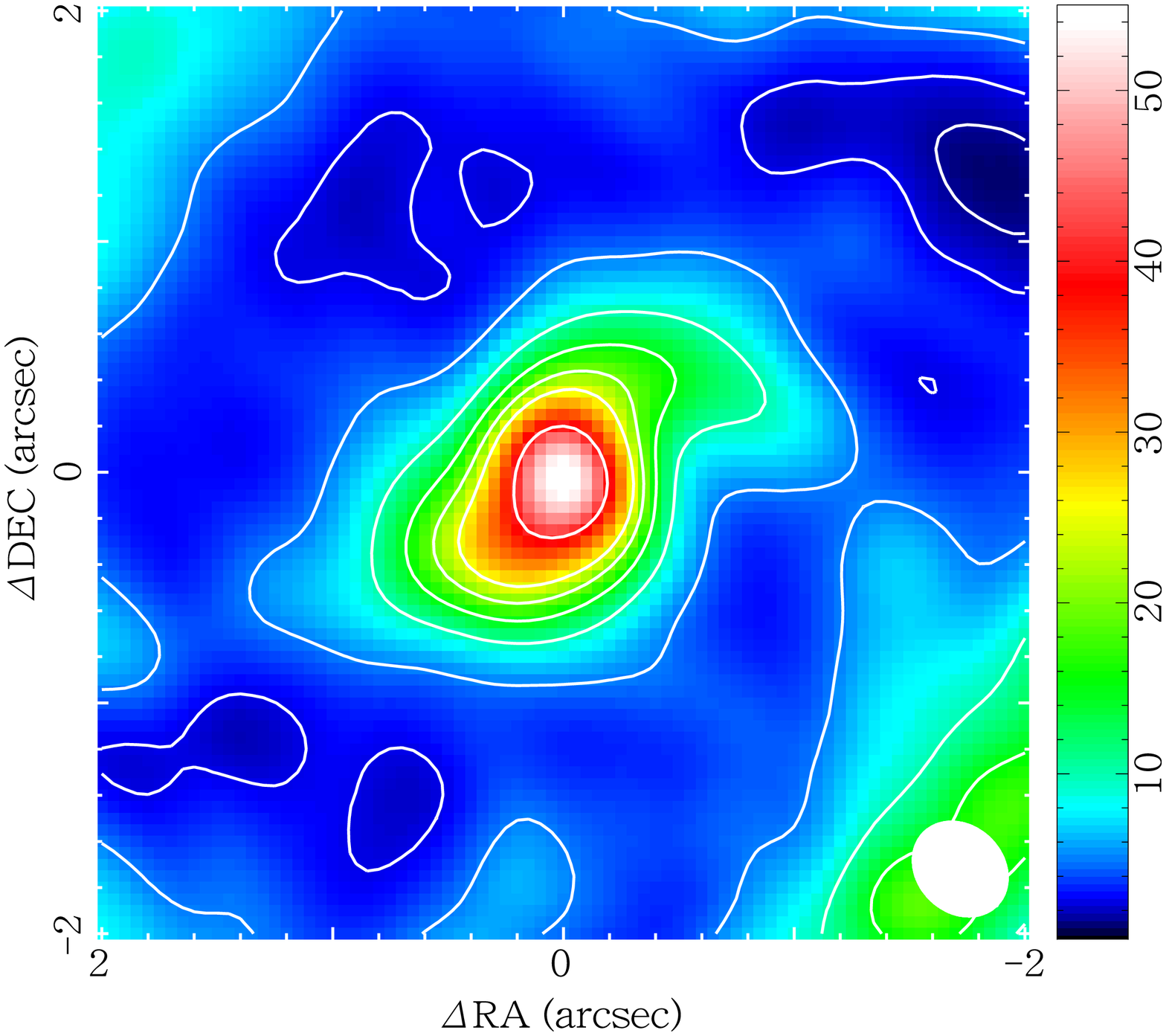}
	 \end{minipage}
	 \begin{minipage}{0.4\hsize}
\par
{\bf (d)}
		\FigureFile(60mm,150mm){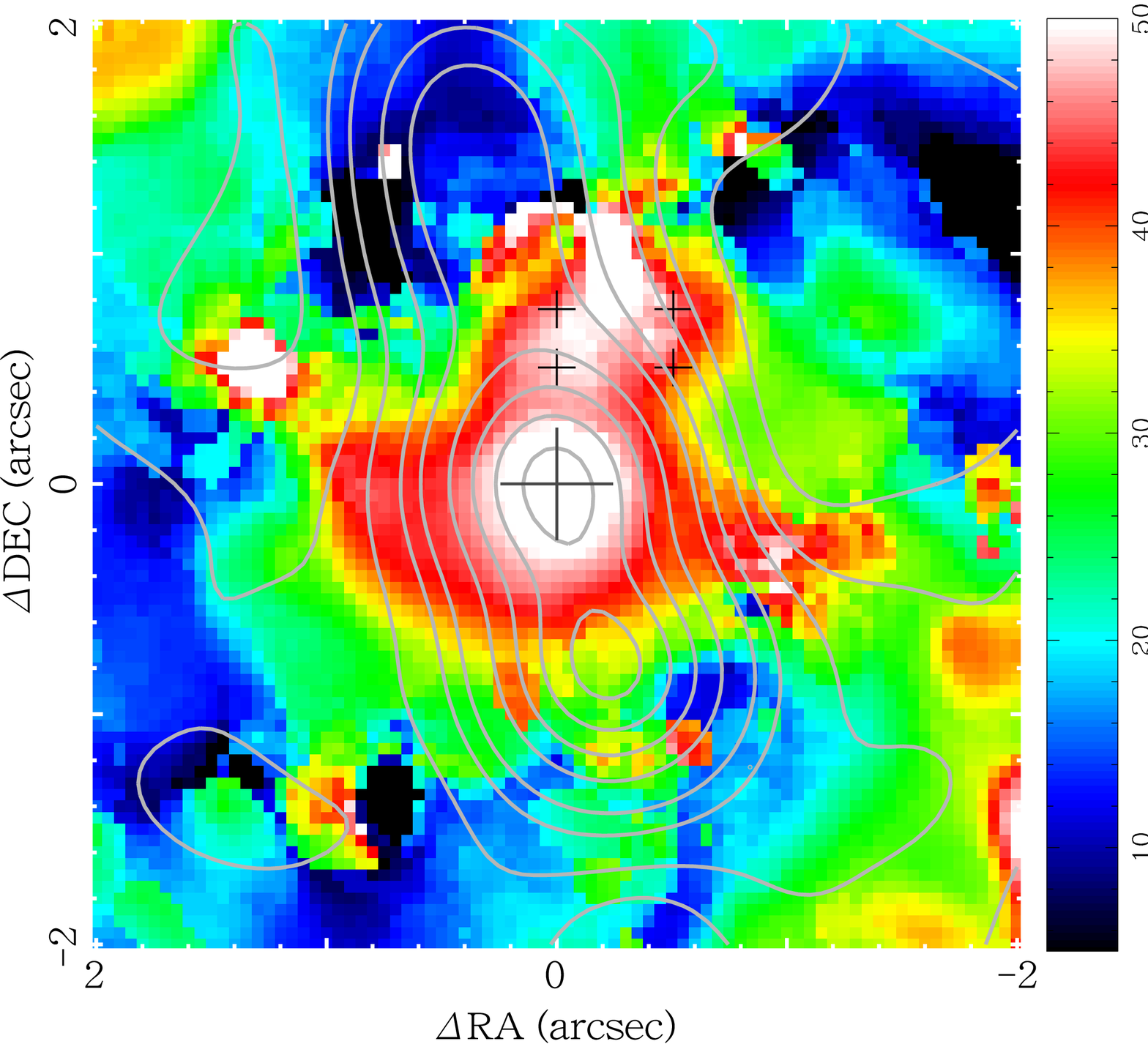}
	 \end{minipage}\\
	 \begin{minipage}{0.8\hsize}
\par
{\bf (e)}
		\FigureFile(120mm,150mm){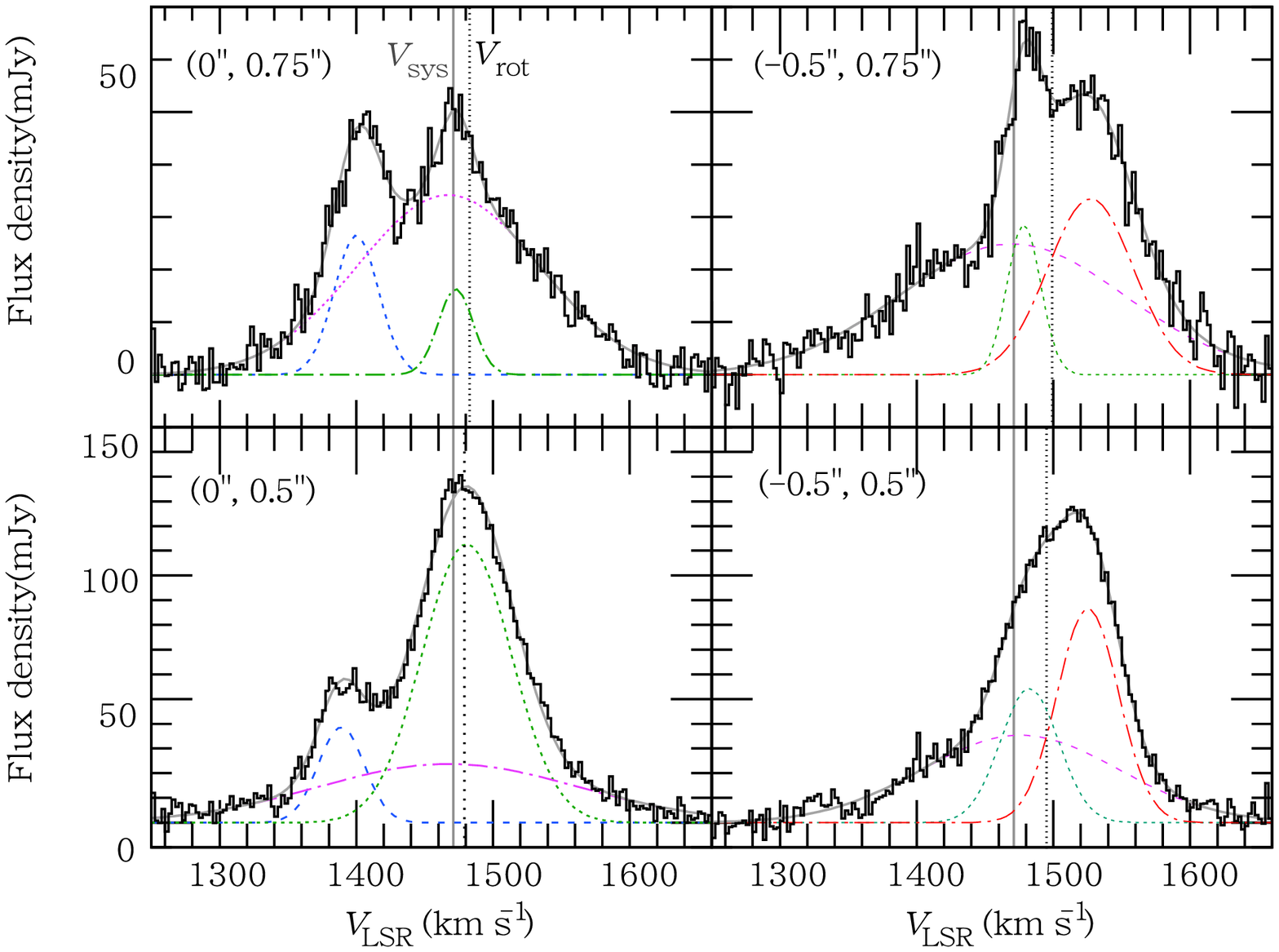}
	 \end{minipage}
 \end{tabular}
\end{center}
\caption{
(a) CO(3--2) velocity field map derived from intensity-weighted mean velocities.
(b) Intensity-weighted CO velocity dispersion map. 
Enlarged view of (c) the integrated intensity maps of CO(3--2) 
and (d) intensity-weighted CO velocity dispersion map.
The contours in (d) are same as the 4.9~GHz continuum map presented in figure~\ref{fig:cont} .
(e) Representative CO(3--2) spectra in the north side, $(\Delta RA, \Delta DEC)=$(\timeform{0"}, 0\farcs{5}) and (\timeform{0"}, 0\farcs{75}), and the northwest side, ($-0\farcs{5}$, 0\farcs{5}) and ($-0\farcs{5}$, 0\farcs{75}).
The positions are shown in (d).
The three Gaussian fits to the lines are overlaid with dotted lines.
}
 \label{fig:out}
\end{figure}

Figures~\ref{fig:out}~(a) and (b) show maps of 
the  intensity-weighted CO velocity field ($\bar{v}=\sum(T v) / \sum T$) and intensity-weighted  CO velocity dispersion $\left(\Delta v=\sqrt[]{\sum[T(v-\bar{v})^2]/\Sigma T}\right)$, where $T$ is the intensity per channel.
The velocity dispersion at the  west-side of the ring ($\Delta v\approx50$~km~s$^{-1}$) is comparable to that in the CND 
because the velocity components along the galactic bar are overlapped in the beam 
and the dispersion may be overestimated because the baseline range in a spectrum is not enough for high velocity (section~\ref{sec:observation}). 
The northern CND  shows high velocity dispersion  ($\Delta v \gtrsim 50$~km~s$^{-1}$)  in agreement with that of [Fe\emissiontype{II}] (\cite{boker}, \cite{falcon}); 
however their distribution is different, i.e., $\theta_{\rm PA}\sim\timeform{-20D}$ for CO and $\theta_{\rm PA}\sim\timeform{30D}$ for [Fe\emissiontype{II}]. 
In addition, the large dispersion extends towards north-west and north-east [figure~\ref{fig:out}(d)] regardless of the distribution of molecular gas  [figure~\ref{fig:out}(c)]. 
Figure~\ref{fig:out}(e) shows the representative spectra in the north and northwest regions of the CND.
The spectra are fitted by three Gaussians, and the resultant parameters are summarized in table~\ref{tab:out}.
In addition to the broad emission features of $\Delta v_{\rm FWHM}> 150$~km~s$^{-1}$ and narrow features of $\Delta v_{\rm FWHM}\sim 30$--80~km~s$^{-1}$ 
close to the rotational and systemic velocities, which probably reflect the gas in the disk, 
the blueshifted and redshifted features are clearly decomposed in the north and northwest regions, respectively.
The molecular gas mass of each component is estimated to be of $\sim10^5$~\MO (table~\ref{tab:out}).
We calculated the virial mass 
given by $M_{\rm vir}=1040~\sigma_{v}~R$, where $\sigma_v$ is the cloud velocity dispersion in km~s$^{-1}$ 
and $R$ is the  cloud radius in pc \citep{solomon}, 
assuming a spherical cloud with a density profile $\rho\propto r^{-1}$ and size of $\sim30$~pc being comparable with the beam size for simplicity.
The virial mass ($\gtrsim10^6$~\MO) is obviously larger than the CO luminosity mass, suggesting that the molecular gas is not a gravitationally bound state.

Both the blueshifted and redshifted components may not be coplanar gases because the systematic vertical gas motion to the disk is unlikely.
Considering that the northern part of the galactic disk is the far side, 
the blueshifted and redshifted components are regarded as the outflowing gas from the disk 
and the gas dragged by the jets in the disk, respectively.
The observed distribution and kinematics of the molecular gas could be explained by the effect of the radio jets.
The simulations of \citet{wagner2011} and \citet{wagner2012} have shown 
that the jets create a spherical bubble as propagating through  ISM and a cocoon of shocked material accelerates molecular gas to high velocities and  over a wide range of directions, preventing star formation in the CND, i.e., negative feedback. 
In the future, we will discuss the relation between molecular gas and the NGC~613 jets in detail.
However, to discuss the relation observations of CO(1--0) with higher angular resolution ($\theta_{\rm FWHM}<0\farcs{1}$) are needed (e.g., \cite{cicone}).
 
\section{Summary}
\label{sec:summary}
In this study, we made observations of molecular lines and continuum emission toward NGC~613 by using ALMA Bands~3 and 7.
The main conclusions are summarized as follows: 
\begin{enumerate}

\item Radio continuum emissions were detected at 95 and 350~GHz from both the CND and star-forming ring. 
The 95~GHz continuum extends from the center at a position angle of $\timeform{20D} \pm \timeform{8D}$.
The archival 4.9 GHz data and our 95 GHz data show spectral indices of $\alpha\lesssim-0.6$ and $-0.2$ along the jet and in the star-forming ring; these can be produced by synchrotron and free--free emissions, respectively. 

\item The emission of CO(3-2), HCN(1-0), HCN(4-3), HCO$^+$(1-0), HCO$^+$(4-3), CS(2-1), and CS(7-6)  was detected in both the CND and ring. 
While SiO(2--1) was detected marginally at the edge of the radio jets, 
C$_2$H(1--0, 3/2--1/2) was detected unexpectedly in the east side of the ring.

\item The basic parameters of NGC~613 were derived to be the systemic velocity $V_{\rm sys}=1471\pm3$~km~s$^{-1}$, 
position angle $\theta_{\rm maj}=\timeform{118D}\pm\timeform{4D}$,  
and inclination angle $i=\timeform{46D}\pm\timeform{1D}$  in the region $0\farcs2 \gtrsim r \gtrsim 4\farcs5$.
The velocity field shows rigid rotation as a function of $V_{\rm rot}=(49\pm1)$~[km~s$^{-1}$~arcsec$^{-1}$]~$\times~r$ at $r>0\farcs5$; 
however, velocity increases steeply in the central region ($r < 0\farcs5 \approx 42$~pc).
Next, a velocity lower than $V_{\rm sys}$ is featured, bridging between the CND and southern part of the ring, 
and may be a sign of the outflow from the CND.

\item The molecular gas mass in the region $r\leq\timeform{5"}$ is $M_{\rm H_2}=7.2\times10^7-5.8\times10^8$~$\MO$ for $R_{32/10}=0.25-2$.
The radial distribution of  the surface density of the molecular gas $\Sigma_{\rm H_2}$ derived from CO(3--2) shows an exponential decrease  of $\Sigma_{\rm H_2}=1330\exp(-r/r_{\rm e})$~[\MO~pc$^{-2}$], where $r_{\rm e}=0\farcs48(=41$~pc) at $r\leq100$~pc.

\item 
The rotational temperatures and column densities of the molecules derived 
from $J=1-0$ and $4-3$ lines of HCN and HCO$^+$ and  $J=2-1$ and $7-6$ of CS in the CND are as follows: 
$T_{\rm rot}=11.6\pm1.8$~K, $N{\rm (HCN)}=7.7\pm0.9\times10^{13}$~[cm$^{-2}$] for HCN, 
$T_{\rm rot}=12.9\pm2.6$~K, $N{\rm (HCO^+)}=2.2\pm0.3\times10^{13}$~[cm$^{-2}$] for HCO$^+$, and 
$T_{\rm rot}=17.5\pm2.9$~K, $N{\rm (CS)}=2.2\pm0.3\times10^{13}$~[cm$^{-2}$] for CS.
The rotational temperature decreased with distance from the center.

\item 
The variations of the HCN(1--0)/HCO$^+$(1--0) and HCN(1--0)/CS(2--1) ratios from the CND through the ring correspond to the variations of the abundance ratio of HCO$^+$ and CS with respect to HCN.
The ratios of HCN(4--3)/HCO$^+$(4--3) and HCN(4--3)/CS(7--6) in the CND and ring are consistent with the ratios in the AGN and starburst domains already reported, respectively. 
The HCN(4--3)/HCO$^+$(4--3) and HCN(4--3)/CS(7--6) ratios  likely depend on the kinetic temperature rather than the volume density of H$_2$ and the abundance ratio of HCO$^+$ and CS with respect to HCN.

\item 
The physical parameters  in the inner CND, utilizing the intensity ratios of  HCN, HCO$^+$, and CS and assuming non-LTE radiative transfer are $T_{\rm k}=$350--550~K, 
$n_{\rm H_2}\sim10^{4.5}$~cm$^{-3}$, 
and $N_{H_2}=10^{22}$~cm$^{-2}$ 
with $\chi^2<1$ 
for ${\rm [HCN]/[HCO^+]=10}$ and ${\rm[HCN]/[CS]=10}$, and
$T_{\rm k}=80$--300~K, 
$n_{\rm H_2}=10^{4-4.5}$~cm$^{-3}$, and 
$N_{H_2}=10^{22}$~cm$^{-2}$ 
for ${\rm [HCN]/[HCO^+]=10}$ and ${\rm [HCN]/[CS]=10}$ in the ring.
However, the parameters could not be determined rigidly because of the low S/N.

\item
The SFE in the CND is almost an order of magnitude lower than those at the spots in the star-forming ring, 
even though  the dominant activity of the central region is the star formation rather than AGN.
The large velocity dispersion  of CO extending toward the north side of the CND 
and decomposing into blueshifted and redshifted features are probably explained by the effect of the radio jets.
These results strongly suggest that  the jets heat gas in the CND where the feedback prevents star formation.

\end{enumerate}

\begin{ack}
We are grateful to the anonymous referee for careful and constructive comments that improved this paper.
We thank J. Falc\'{o}n--Barroso for providing us the data with VLT (figure~\ref{fig:h2} and \ref{fig:brg}) 
and T. Minamidani for useful discussions.
This paper makes use of the following ALMA data:
ADS/JAO.ALMA\#2013.1.01329.S. ALMA is a partnership of ESO (representing its member states), NSF (USA) and NINS (Japan), together with NRC (Canada) and NSC and ASIAA (Taiwan) and KASI (Republic of Korea), in cooperation with the Republic of Chile. The Joint ALMA Observatory is  operated by ESO, AUI/NRAO and NAOJ.
We referred a script developed by Y. Tamura for this calculation. 
This research has made use of the NASA/IPAC Extragalactic Database (NED) which is operated by the Jet Propulsion Laboratory, California Institute of Technology, under contract with the National Aeronautics and Space Administration.
The National Radio Astronomy Observatory is a facility of the National Science Foundation operated under cooperative agreement by Associated Universities, Inc.
Data analysis was in part carried out on the open use data analysis computer system at the Astronomy Data Center, ADC, of the National Astronomical Observatory of Japan.
\end{ack}

\onecolumn
\twocolumn

\end{document}